\documentclass[aps,prf,preprint,longbibliography]{revtex4-1}

\usepackage{textgreek}
\usepackage{graphicx}
\usepackage{mathptmx}      
\usepackage{latexsym}
\usepackage{xcolor}
\usepackage{hyperref}

\graphicspath{{./Figures/}}

\newcommand{\sagar}[1]{\textcolor{black}{#1}}

\begin{document}


\title{Buoyant finite-size particles in turbulent duct flow}


\author{Sagar Zade}
\email[Email address: ]{{zade@mech.kth.se}}
\author{Walter Fornari}
\author{Fredrik Lundell}
\author{Luca Brandt}
\affiliation{Linn\'e Flow Centre and SeRC (Swedish e-Science Research Centre), \\ KTH Mechanics, SE 100 44 Stockholm, Sweden}


\date{\today}

\begin{abstract}
Particle Image Velocimetry (PIV) and Particle Tracking Velocimetry (PTV) have been employed to investigate the dynamics of finite-size spherical particles, slightly heavier than the carrier fluid, in a horizontal turbulent square duct flow. 
Interface resolved Direct Numerical Simulations (DNS) have also been performed with the Immersed Boundary Method (IBM) at the same experimental conditions, bulk Reynolds number $Re_{2H}$ = 5600, duct height to particle size ratio $2H/d_p$ = 14.5, particle volume fraction $\Phi$ = 1\% and particle to fluid density ratio $\rho_p/\rho_f$ = 1.0035. A good agreement has been observed between experiments and simulations in terms of the overall pressure drop, concentration distribution and turbulent statistics of the two phases.
Additional experimental results considering two particle sizes, $2H/d_p$ = 14.5 and 9 and multiple $\Phi$ = 1, 2, 3, 4 and 5\% are reported at the same $Re_{2H}$. The pressure drop monotonically increases with the volume fraction, almost linearly and nearly independently of the particle size for the above parameters. However, despite the similar pressure drop, the microscopic picture, the fluid velocity statistics, differs significantly with the particle size. 
This one-to-one comparison between simulations and experiments extends the validity of interface resolved DNS in complex turbulent multiphase flows and highlights the ability of experiments to investigate such flows in considerable details, even in regions where the local volume fraction is relatively high.
\end{abstract}


\maketitle

\section{Introduction}
\label{intro}
Particle-laden flows are widely encountered in environmental problems and industrial applications such as carriage of silt by rivers, drifting of snow, sorting of crushed materials, transportation of nuclear waste, etc. Since settling effects due to gravity are generally non-negligible in these flows, the understanding of the turbulence modulation mechanisms during particle transport under sedimentation is of practical importance. Among the many factors affecting turbulence modulation, particle size and density, volume fraction, fluid inertia and particle-particle/particle-wall interaction are of extreme importance \citep{crowe2011multiphase, balachandar2010turbulent}. In this study, we focus on the specific case of finite-size particle transport in a fully developed turbulent square duct flow. It also represents a reasonably complex wall-bounded case, for comparing simulations and experiments due to the presence of a mean secondary flow and the resulting non-homogeneity in the duct cross-section.

Point-particle simulations are often employed to study particle laden flows. For the case of small heavy particles in a highly turbulent horizontal square duct flow, \citet{yao2010inertial} found that particle resuspension is promoted by the drag force arising from the secondary flows as well as shear induced lift forces. Previously, \citet{winkler2004preferential} simulated point particles in a vertical square duct and found that, in general, particles accumulate in regions of high compressional strain and low swirling strength. However, near the wall, the tendency of particles to accumulate in regions of high vorticity increases with particle response time. \citet{sharma2006turbulent} showed that while passive tracers tend to remain within the secondary flow, high inertia particles accumulate close to the walls in a square duct. For a review on particle deposition and entrainment mechanism from the wall in a turbulent flow, the reader is referred to \citet{soldati2009physics}. 

The effect of finite-size sedimenting particles in a square duct was investigated in \citet{lin2017effects} for different Shield's number $Sh={\tau_{w}}/{(\rho_p-\rho_f)d_pg}$, which signifies the relative strength of shear forces $\tau_{w}d_p^2$ to buoyancy forces $(\rho_p-\rho_f)d_p^3g$ ($g$ being the acceleration due to gravity). These authors observed that the presence of particles increases the secondary flow circulation which, in-turn, causes particles to accumulate preferentially at the face center of the bottom wall. Also, at constant $\Phi$ and pressure gradient, the flow rate reduced as the particle settling effect increased (lower $Sh$). \sagar{It is also worth to note the recent simulations of \cite{fornari2017suspensions} for neutrally buoyant particles in a square duct up to a volume fraction of 20\%. These authors found that for $\Phi \leq$ 10\%, particles preferentially accumulate on the corner bisectors and turbulence production is enhanced, whereas at $\Phi$ = 20\% particles migrate towards the core region and turbulence production decreases below the values for $\Phi$ = 5\%.} \citet{shao2012fully}, in their channel flow simulations with settling particles, observed that when settling is significant enough, particles form a sediment layer that acts like a rough wall. Vortex structures shedding from this region ascend into the core and substantially increase the turbulence intensity there. Also, the effects of smaller particles on the turbulence are found to be stronger than those of larger particles at the same $\Phi$. \citet{kidanemariam2013direct} simulated turbulent open channel flow with low volume fractions of finite-size, heavy particles. They found that particles showed strong preferential concentration in the low-speed streaks due to quasi-streamwise vortices, which resulted in particles moving, on an average, slower than the average velocity of the fluid phase.

Experimental measurement of velocity and concentration field in suspensions are quite difficult, and this has therefore limited our understanding of the interactions between the turbulent fluid and the near-mobile-bed region. Optical measurement techniques like PIV in suspension flows of finite-size particles relies on the use of particles that are transparent i.e. the refractive indices of the fluid and particles are nearly the same with respect to the wavelength of the light used for illumination. However, the available particles are composed of materials like plastic, metal, glass etc., which are usually opaque thus, a volume fraction of 0.5\% in a domain of 5 to 10 cm has been indicated as the limit \citep{poelma2006turbulence}. Some limited options available for Refractive-Index-Matching (RIM) fluids to enable the use of PIV can be found in \citet{wiederseiner2011refractive}. However, they are often difficult to scale-up due to issues related to long-time properties of the suspending solution, thermal stability, handling and cost. Super-absorbent hydrogel particles in water, used in this study, can be advantageously used for performing RIM-PIV as previously shown in \citet{klein2012simultaneous}, \citet{byron2013refractive} and \citet{harshani2016experimental}. The mechanical properties of commercially available spherical hydrogel particles are discussed in \citet{dijksman2017refractive}.

\sagar{By performing fully resolved DNS simulations at the same bulk Reynolds number, particle size, density and volume fraction as in experiments, we cross-validate both tools and demonstrate the suitability of hydrogel particles to be used as rigid spheres in PIV experiments. Good agreement between experiments and simulations under the above conditions also indicate that the statistics in turbulent flow may not be very sensitive to the exact value of collision and friction parameter required in simulations. Additional experiments for two particle sizes are presented here to estimate the turbulence modulation and particle dynamics as a function of the particle volume fraction.}

\section{Experimental method}
\label{Experimental method}

\begin{figure}
 (a) \includegraphics[width=0.6\linewidth]{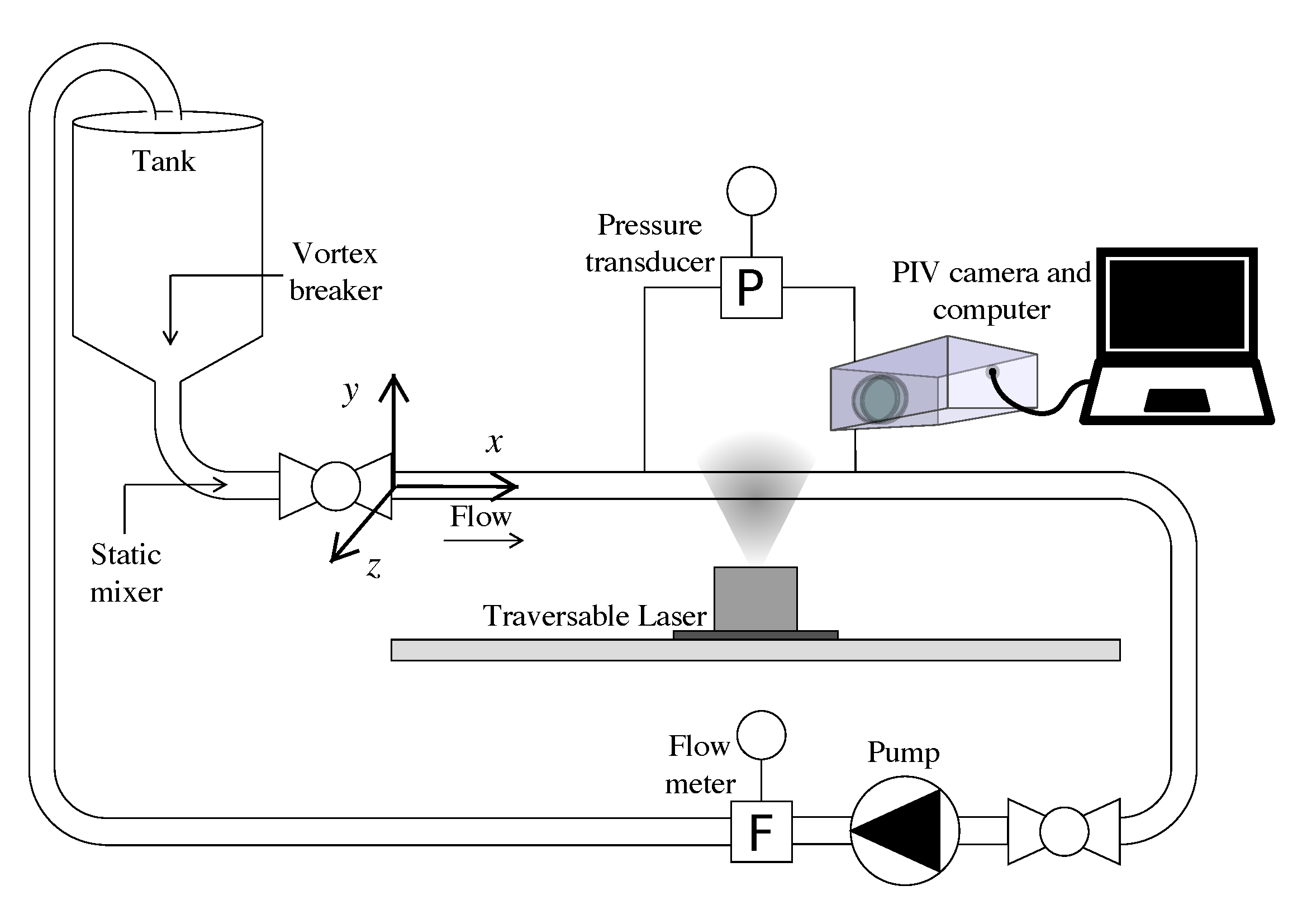}
 (b) \includegraphics[width=0.3\linewidth]{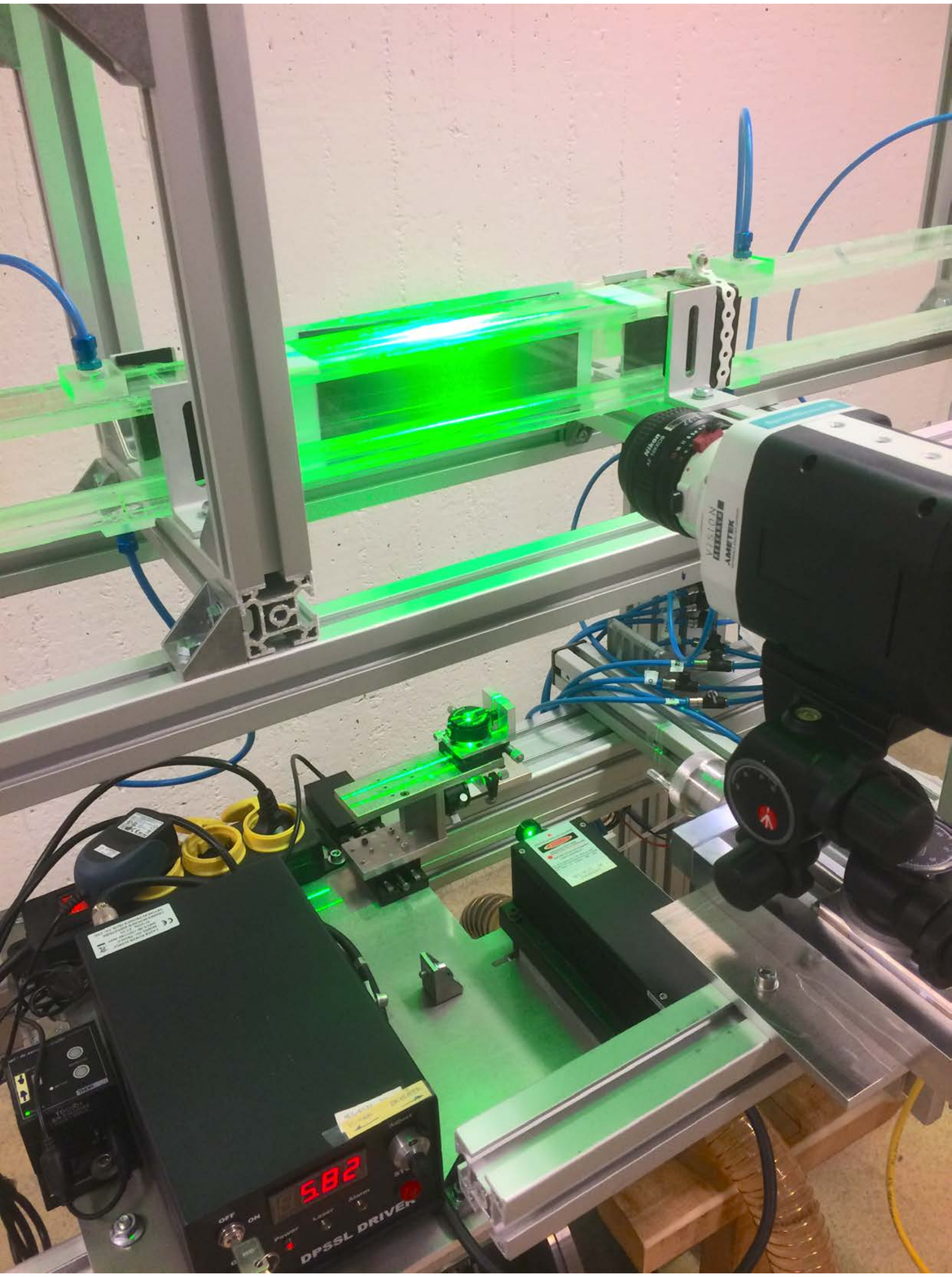}
\caption{(a) Schematic of the flow-loop (b) Photo of the section where PIV is performed.}
 \label{fig:Set-up schematic}  
\end{figure}

The experiments are performed in a 5 m long square duct with 50 mm x 50 mm cross section. The entire duct is made up of transparent acrylic permitting visualization throughout its length. Figure \ref{fig:Set-up schematic} (a) shows a schematic of the flow loop. The fluid is recirculated through a closed loop consisting of a tank that is open to the atmosphere, where the particle-fluid mixture can be introduced. A static mixer is mounted close to the inlet of the duct to neutralize any swirling motions that may arise from the gradual 90$^\circ$ bend at the exit of the tank. It is followed by a section providing a smooth transition from a circular to a square cross-section. A tripping tape is installed at the entrance of the duct to trigger turbulence. The temperature is maintained at nearly 20$^\circ$C by means of an immersed-coil heat-exchanger in the tank. In order to minimize mechanical breakage of the particles, a very gentle disc pump (Discflo Corporations, CA, USA) has been chosen.

An electromagnetic flowmeter (Krohne Optiflux 1000 with IFC 300 signal converter, Krohne Messtechnik GmbH, Germany) is used to measure the volume flow rate of the particle-fluid mixture. The Reynolds number $Re_{2H}$, used hereafter, is based on the average or bulk velocity $U_{Bulk} = $, the kinematic viscosity of the fluid $\nu_f$ and the duct height $2H$. Pressure drop is measured at a streamwise distance of nearly $140H$ from the inlet (the turbulent flow is seen to be fully developed at this entry length), across a length of $54H$ using a differential pressure transducer (0 - 1 kPa, Model: FKC11, Fuji Electric France, S.A.S.). Data acquisition from the camera, flow meter and pressure transducer is performed using a National Instruments NI-6215 DAQ card using Labview\textsuperscript{TM} software. Additional details can be found in \citet{zade2018experimental}.

\subsection{Particle properties}
\label{Particle properties}

The particles are commercially procured super-absorbent (polyacrylamide based) hydrogel spheres which are deliverd in dry condition. They are graded into different sizes using a range of sieves from which two sizes are selected for these experiments. Once mixed with tap water and left submerged for around one day, they grow to two equilibrium sizes: 3.5$\pm$0.8 mm (3 times standard deviation) and 5.60$\pm$0.9 mm yielding a duct height to particle diameter ratio $2H/d_p$ of 14.5 and 9 respectively. These two particles would be hereby refered to as smaller particles SP, and larger particles LP, respectively. The particle size was determined both by a digital imaging system and from the PIV images of particles in flow and a small spread in the particle diameter was observed. The fact that a Gaussian like particle size distribution has small effect on the flow statistics has been shown in \citet{fornari2018effect}. To retain the matching of refractive indices of the particle and fluid, and at the same time facilitate the detection of particles in the PIV images, a small amount of fluorescent Rhodamine (in ppm) was added to the water in which the particles expand. This enhances the contrast of the particles in the PIV images (shown later). 

The density of the particles was determined using 2 methods: (i) by measuring the volume displacement by a known mass of particles and (ii) by determining the terminal settling velocity in a long liquid settling column. In the first method, a known mass of fully expanded particles was put in a water-filled container of uniform diameter. The rise in the level of water due to the particles was measured using a very precise laser distance meter (optoNTDC 1710, Micro-Epsilon Messtechnik GmbH, resolution: 0.5\textmugreek m). In the second method, a single particle with a known diameter was gently dropped in a long vertical pipe filled with water, wide enough so as to minimize the wall effects, and the settling velocity is determined (after it has reached steady state). The relation for drag force $F$ on a settling particle in \citet{crowe2011multiphase},
\begin{equation}
  \frac{F}{\rho_{f}U_{T}^2A}=\frac{12}{Re_{p}}(1+0.15Re_{p}^{0.687})
  \label{eqn:Particle density}
\end{equation} applicable in the transitional regime: 1$<$Re$<$750, 
was used to relate the particle diameter $d_p$ and terminal velocity $U_T$ to the unknown particle density $\rho_{p}$ . Here, $A$ is the projected area of the particle in the falling direction. In equation \ref{eqn:Particle density}, $Re_p$ is the particle Reynolds number given by $\rho_{p}U_{T}d_{p}/\mu_f$ where, ${\mu_f}$ is the dynamic viscosity of the liquid. Both the above tests were performed at a room temperature of around 20$^\circ$C and yielded nearly the same value of density ratio $\rho_{p}/\rho_{f}$ of 1.0035$\pm$0.0003. The latter method was preferred as it was less sensitive to measurement uncertainties. \sagar{The restitution coefficient of the hydrogel spheres is found to be around 0.9, when a stationary particle is made to fall on a flat acrylic sheet from a height of 0.1 m in air.}

\subsection{Velocity measurement: PIV + PTV}
\label{Velocity measurement: PIV + PTV}

\begin{figure}
 (a) \includegraphics[width=0.35\linewidth]{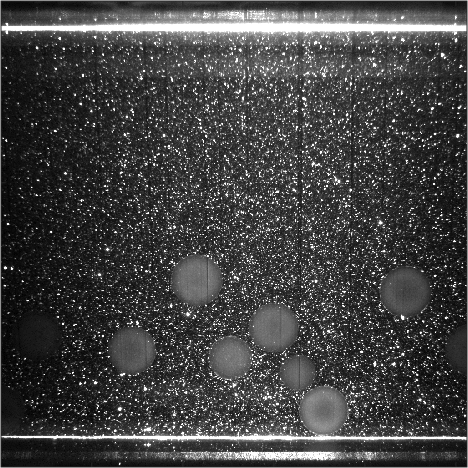}
 (b) \includegraphics[width=0.35\linewidth]{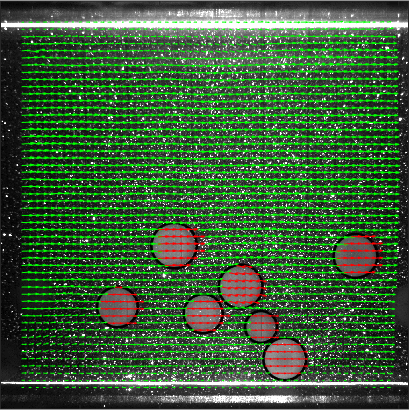}
\caption{(a) Raw Image and (b) PIV (green, in fluid phase) + PTV (red, in particle phase) velocity vectors. The above example corresponds to large particles LP at $\Phi$ = 3\%.}
\label{fig:PIV+PTV}  
\end{figure}

The coordinate system used in this study is indicated in figure \ref{fig:Set-up schematic} (a) with $x$ the streamwise, $y$ the wall-normal and $z$ the spanwise directions. The velocity field is measured using 2D Particle Image Velocimetry (2D-PIV) in three spanwise planes: $z/H$ = 0, 0.4 and 0.8. These measurements are performed at a streamwise distance of $x/H\approx$ 150 from the entrance of the duct. 

A continuous wave laser (wavelength = 532 nm, power = 2 W) and a high-speed camera (Phantom Miro 120, Vision Research, NJ, USA) are used to capture successive image pairs. The thickness of the laser light-sheet is 1 mm. Figure \ref{fig:Set-up schematic} (b) shows the PIV set-up. 

For imaging the full height of the duct, a resolution of approximately 60 mm/1024 pixels was chosen. The frame rate (acquisition frequency) was chosen so that the maximum pixel displacement did not exceed a quarter of the size of the final interrogation window IW \citep{raffel2013particle}. Images were processed using an in-house, three-step, FFT-based, cross-correlation algorithm \citep{kawata2014velocity}. The final size of the IW is 32 $\times$ 32 pixel. The degree of overlap can be estimated from the fact that the corresponding final resolution is 1 mm x 1 mm per IW. Additional near wall measurement are conducted by zooming the camera on a small region close to the wall with a resolution that is $\approx$ 3 times higher. Each experiment has been repeated at least 3 times and 500 image pairs have been observed to be sufficient to ensure statistically converged results.

Figure \ref{fig:PIV+PTV} (a) depicts one image from a typical PIV sequence for particle-laden flow. As mentioned earlier, the contrast between particle and fluid was enhanced by using a small quantity of Rhodamine. Raw images captured during the experiment were saved in groups of two different intensity levels. The first group of images (an example being figure \ref{fig:PIV+PTV} (a)) was used for regular PIV processing according to the algorithm mentioned above. The second group of images were contrast-enhanced, e.g. they were sharpened and their intensity adjusted, and used for detecting the particles only, using a circular Hough transform \citep{yuen1990comparative}. From the detected particles in image A and B of the PIV sequence, a nearest neighbor approach was used to determine their translational motion. Particles that were detected only in one image of the pair were, thus, eliminated by the PTV algorithm. For the Eulerian PIV velocity field, we define a mask matrix, which assumes the value 1 if the point lies inside the particle and 0, if it lies outside. The fluid phase velocity is thus determined on a fixed mesh. The particle velocity is determined using PTV at its center, which is assigned to the grid points inside the particle (mask = 1). The velocity field of the particle-phase is, now, available at the same grid points as that of the fluid and the ensemble averaging, reported later, are phase averaged statistics. Figure \ref{fig:PIV+PTV} (b) shows the combined fluid (PIV) and particle (PTV) velocity field.

\section{Numerical set-up and methodology}
\label{Numerical set-up and methodology}

\begin{figure}
\centering
 \includegraphics[width=0.6\linewidth]{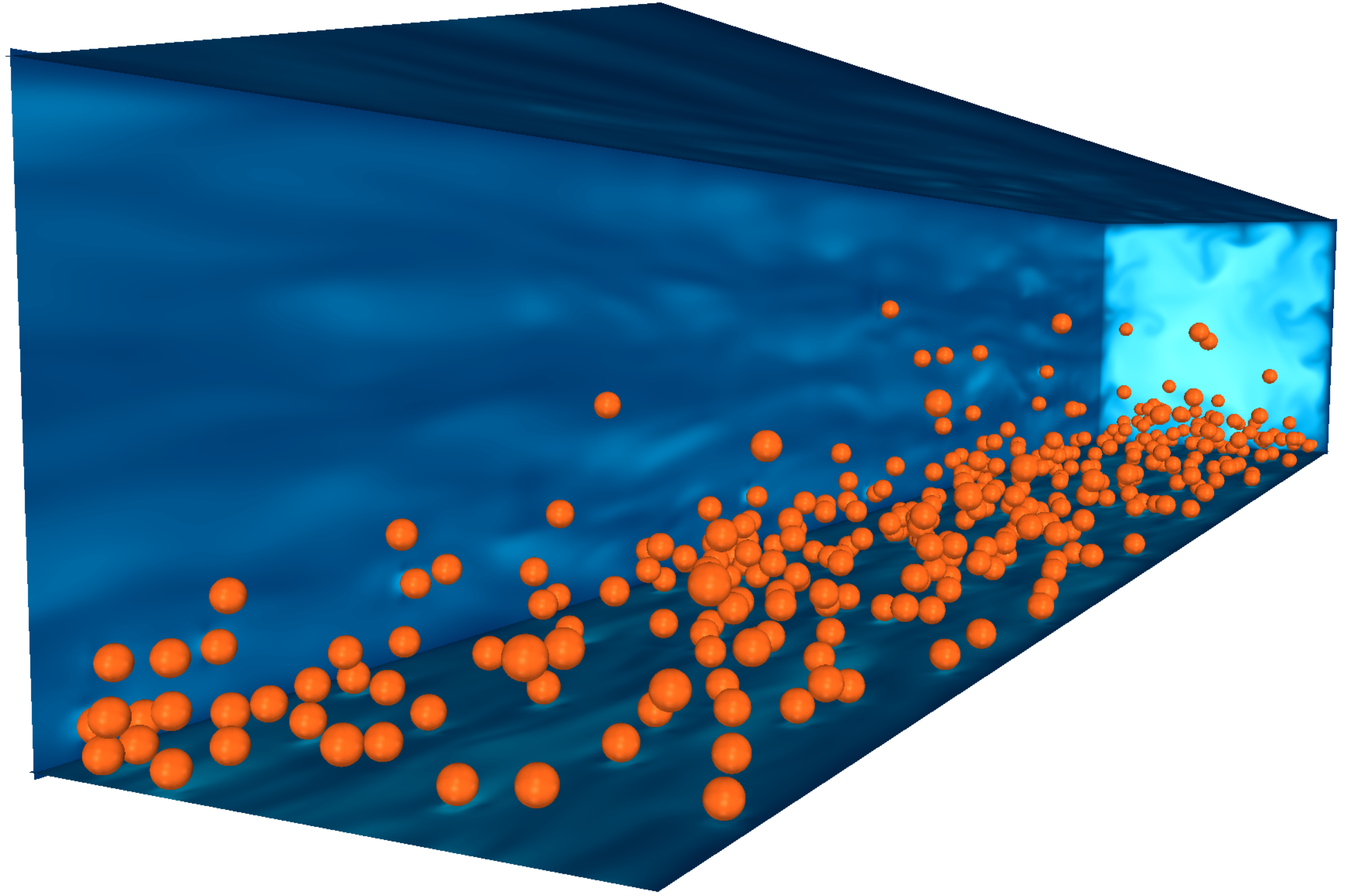}
 \caption{Instantaneous snapshot of the magnitude of the near-wall fluid streamwise velocity together with the particles. The solid volume fraction $\Phi = 1\%$.}
  \label{fig:DNS_3d_view}  
\end{figure}

The direct numerical simulations DNS have been performed using the immersed boundary method IBM, originally developed by \citet{breugem2012second}, which fully models the coupling between the solid and fluid phases. The flow is evolved according to the incompressible Navier--Stokes equations, whereas the particle motion is governed by the Newton--Euler Lagrangian equations for the particle linear and angular velocities. Using the IBM, the boundary condition at the moving fluid/solid interfaces is modelled by an additional force on the right-hand side of the Navier--Stokes equations, making it possible to discretize the computational domain with a fixed staggered mesh on which the fluid phase is evolved using a second-order finite-difference scheme. Time integration is performed by a third-order Runge--Kutta scheme combined with pressure correction at each sub-step. When the distance between two particles becomes smaller than twice the mesh size, lubrication models based on Brenner's asymptotic solution \citep{brenner1961slow} are used to correctly reproduce the interaction between the particles. A soft-sphere collision model is used to account for collisions between particles. An almost elastic rebound is ensured with a restitution coefficient set at 0.97. More details and validations of the numerical code are provided in previous publications \citep{picano2015turbulent, fornari2017suspensions}. 

The simulations are performed in a Cartesian computational domain of size $L_x = 12H$, $L_y = 2H$ and $L_z = 2H$. The domain is discretized by a uniform mesh ($\Delta x = \Delta y = \Delta z$) of 2592$\times$432$\times$432 Eulerian grid points in the streamwise and cross-flow directions. The bulk velocity of the entire mixture is kept constant by adjusting the streamwise pressure gradient to achieve a constant bulk Reynolds number $Re_{2H}$ = 5600. The volume fraction $\Phi$ = 1\% corresponds to 353 particles. The number of Eulerian grid points per particle diameter is 30 ($\Delta x$ = 1/24). The gravitational forces acting on the particle with respect to fluid viscous forces are quantified by the Galileo number $Ga = \sqrt{((\rho_p/\rho_f-1)g(d_p^3)/\nu_f^2)}$ equal to 40 for the particles simulated ($2H/d_p$ = 14.5, $\rho_p/\rho_f$ = 1.0035). The particle Shield's number $Sh$ = 0.45, based on the shear stress for single-phase flow.

\sagar{
Figure \ref{fig:DNS_3d_view} shows an example of the instantaneous particle distribution in the computational domain. From the near-wall streamwise velocity, signature of near-wall streaks can be more prominently seen for walls other than the bottom wall where particle concentration is high. For the bottom wall, hot-spots of higher fluid streamwise velocity is seen below each particle. The statistics are collected after the initial transient phase of $490H/U_{Bulk}$, using an averaging period of $1040H/U_{Bulk}$.} 

\section{Results}
\label{Results}
 
We first compare the results of the DNS code with the experimental measurements for single phase flow at $Re_{2H}$ = 5600. Then we discuss the results from DNS and experiments for smaller particles SP at a volume fraction $\Phi$ = 1\%. Finally, we present purely experimental results for $\Phi$ = 1, 2, 3, 4 and 5\% for SP and LP.

\subsection{Single-phase flow validation}
\label{Single-phase flow validation}

\begin{figure}
  (a) \includegraphics[height=0.30\linewidth]{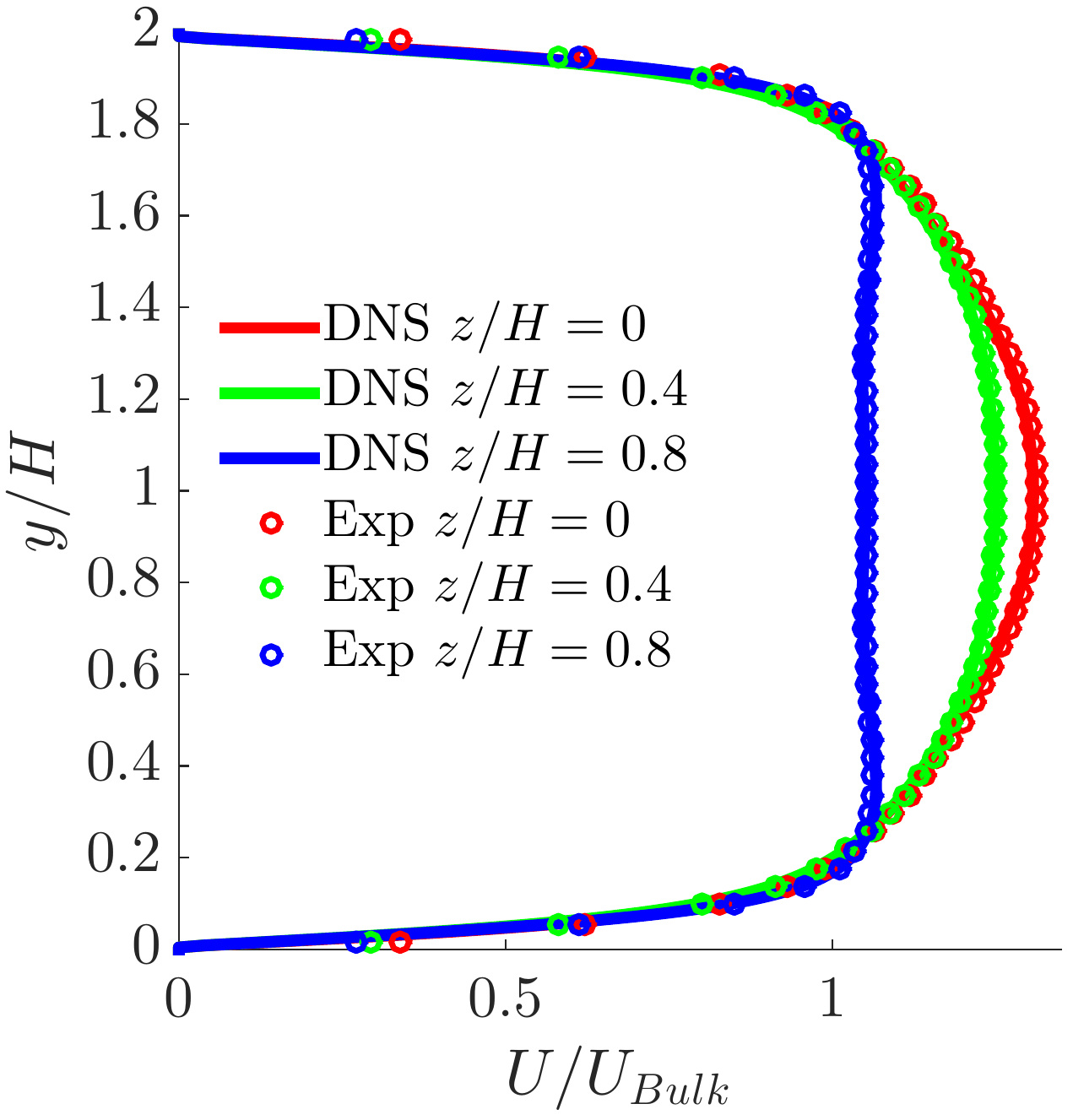}
  (b) \includegraphics[height=0.30\linewidth]{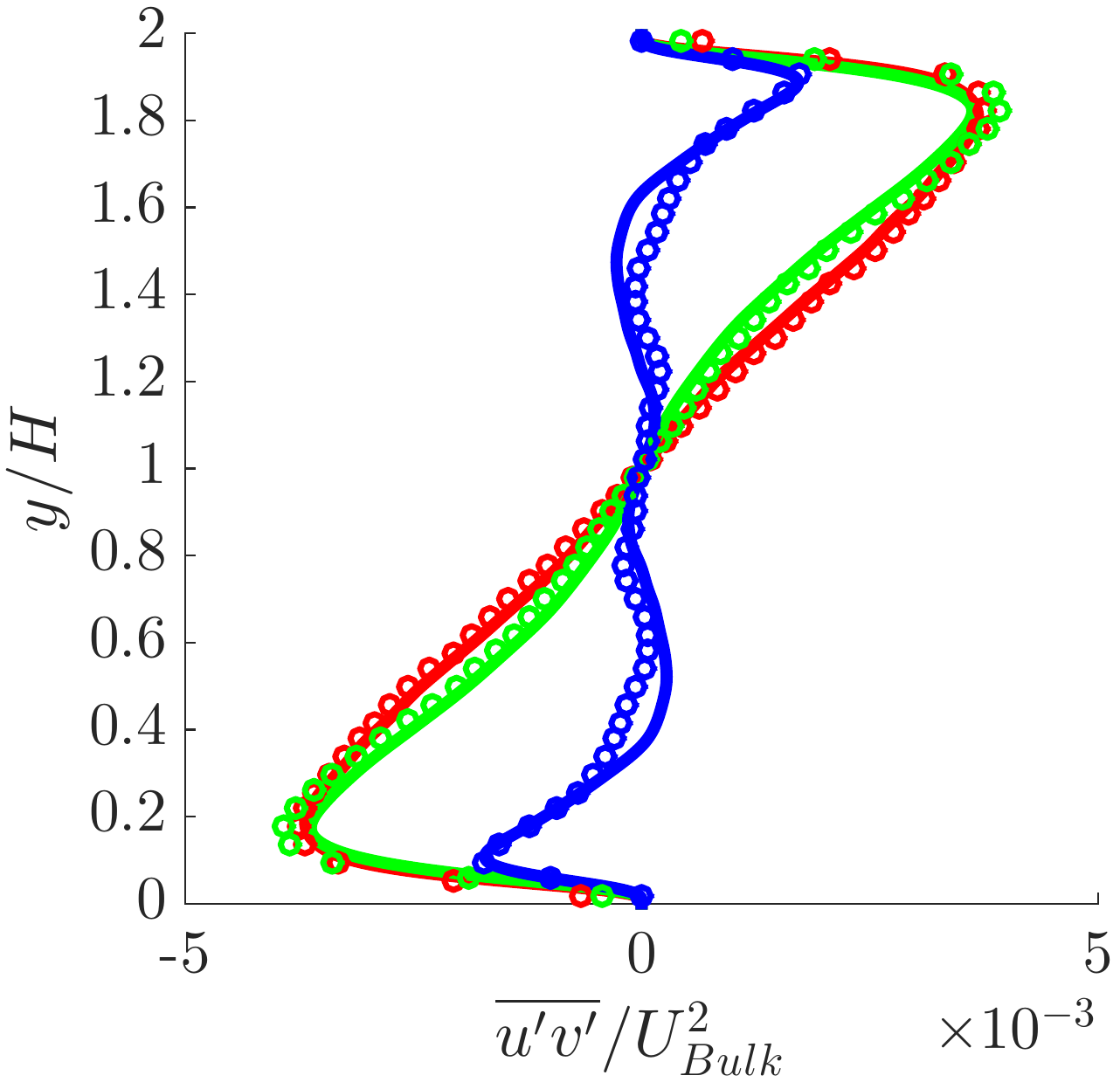}
  (c) \includegraphics[height=0.30\linewidth]{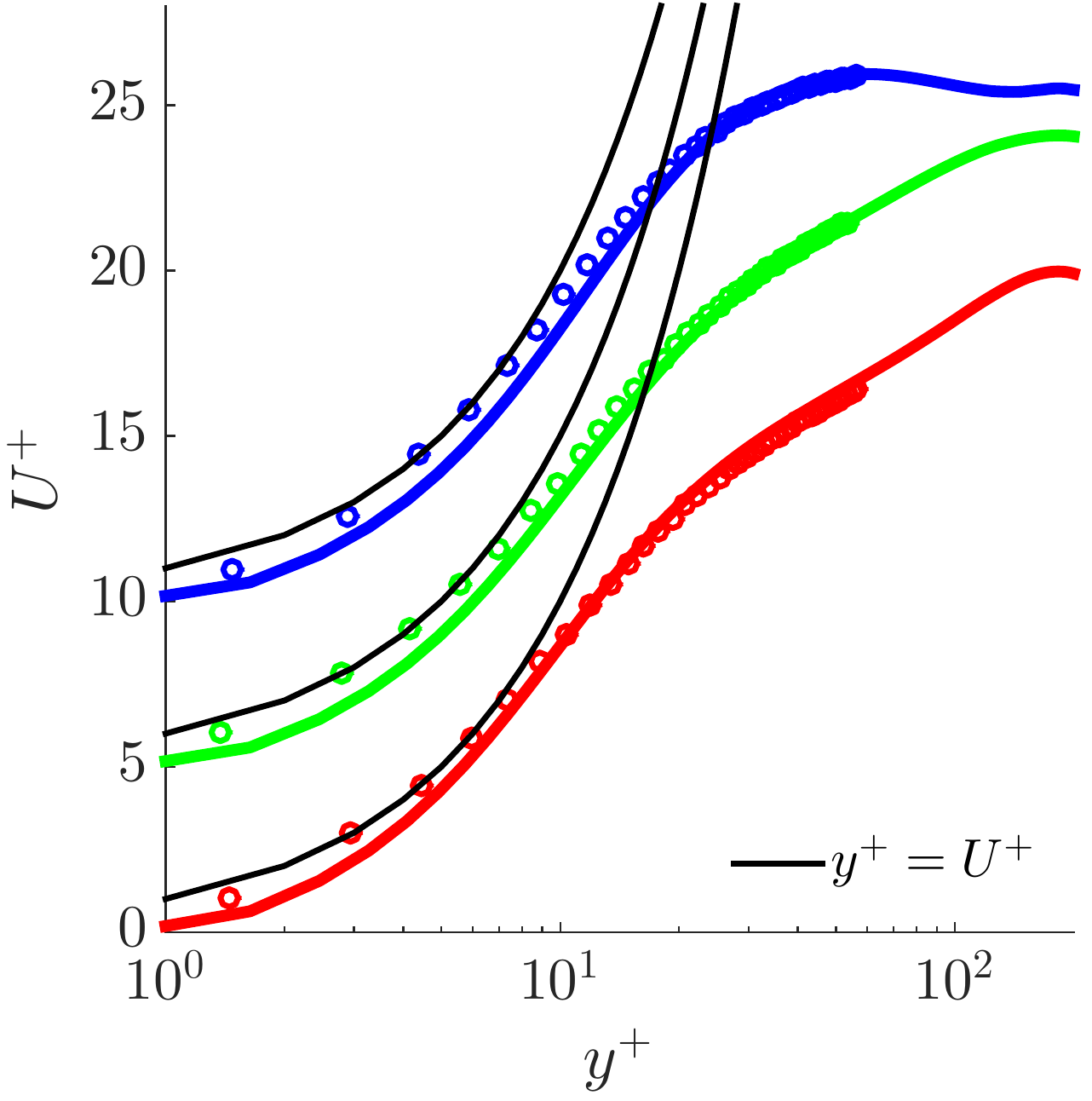}
\caption{Single phase velocity statistics: (a) Mean streamwise velocity, (b) primary Reynolds shear stress scaled in bulk units ($U_{Bulk}$ and $2H$) and (c) mean streamwise velocity scaled in inner units ($u_\tau$ and $\nu/u_\tau$). In (c), the plots for successive spanwise plane is shifted upwards by $5U^+$ units for better visualization).}
\label{fig:Single_phase_velocity_validation}  
\end{figure}

Figures \ref{fig:Single_phase_velocity_validation} (a) and (b) show the streamwise velocity and the corresponding Reynolds shear stress for the single-phase flow measured at three different spanwise planes: $z/H$ = 0, 0.4 and 0.8. It is well documented that the gradient in the Reynolds stresses leads to a secondary flow of Prandtl's second kind in the duct (see \citet{gavrilakis1992numerical}). This secondary motion appears in the form of 4 pairs of counter-rotating vortices near the duct corners, driving high speed fluid from the center of the duct towards the corners and low speed fluid from the wall towards the center. This redistribution of momentum is reflected in the streamwise velocity profiles of figure \ref{fig:Single_phase_velocity_validation} (a). The measured mean streamwise and fluctuating velocity statistics are in good agreement with those of the simulations. The maximum deviation appears in the second order turbulent statistics in the $z/H$  = 0.8 plane due to the proximity of this plane to the side wall ($z/H$ = 1). This is because, in this near-wall region, the shear rate and turbulence intensity are relatively higher, leading to larger out-of-plane motion and hence, larger uncertainties in PIV measurements. Figure \ref{fig:Single_phase_velocity_validation} (c) shows, again, the mean streamwise velocity close to the bottom wall, now scaled in inner units, from separate measurements with higher resolution, where it was possible to measure velocity statistics from distances as small as $y^+ \approx 1.5$.

\subsection{Comparing experiments and simulations: Smaller particles SP ($2H/d_p$ = 14.5) at $\Phi$ = 1\%}
\label{Comparing experiments and simulations: Smaller particles SP at volume fraction of 1}

\begin{figure}
  \includegraphics[height=0.33\linewidth]{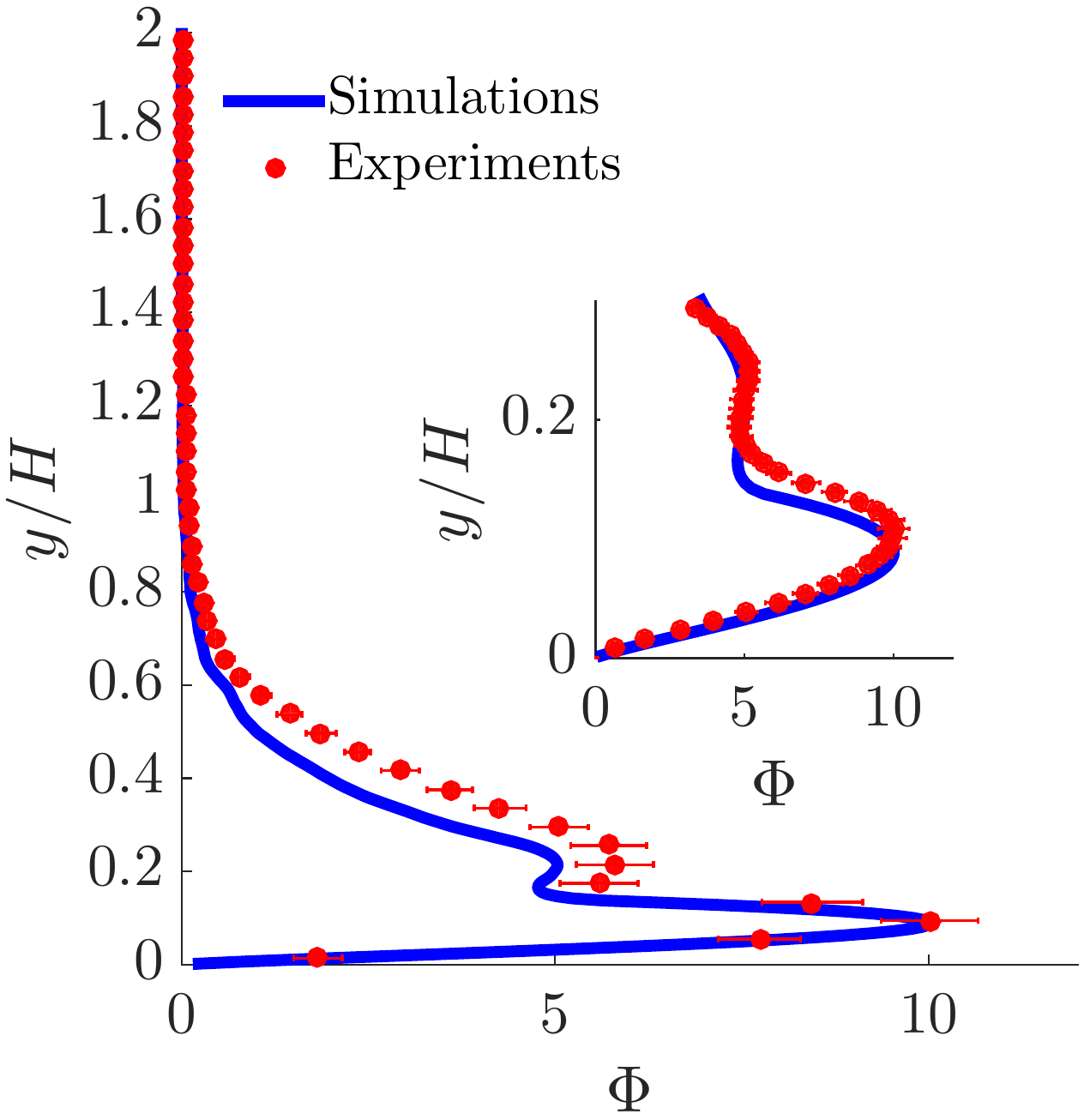}
  \includegraphics[height=0.33\linewidth]{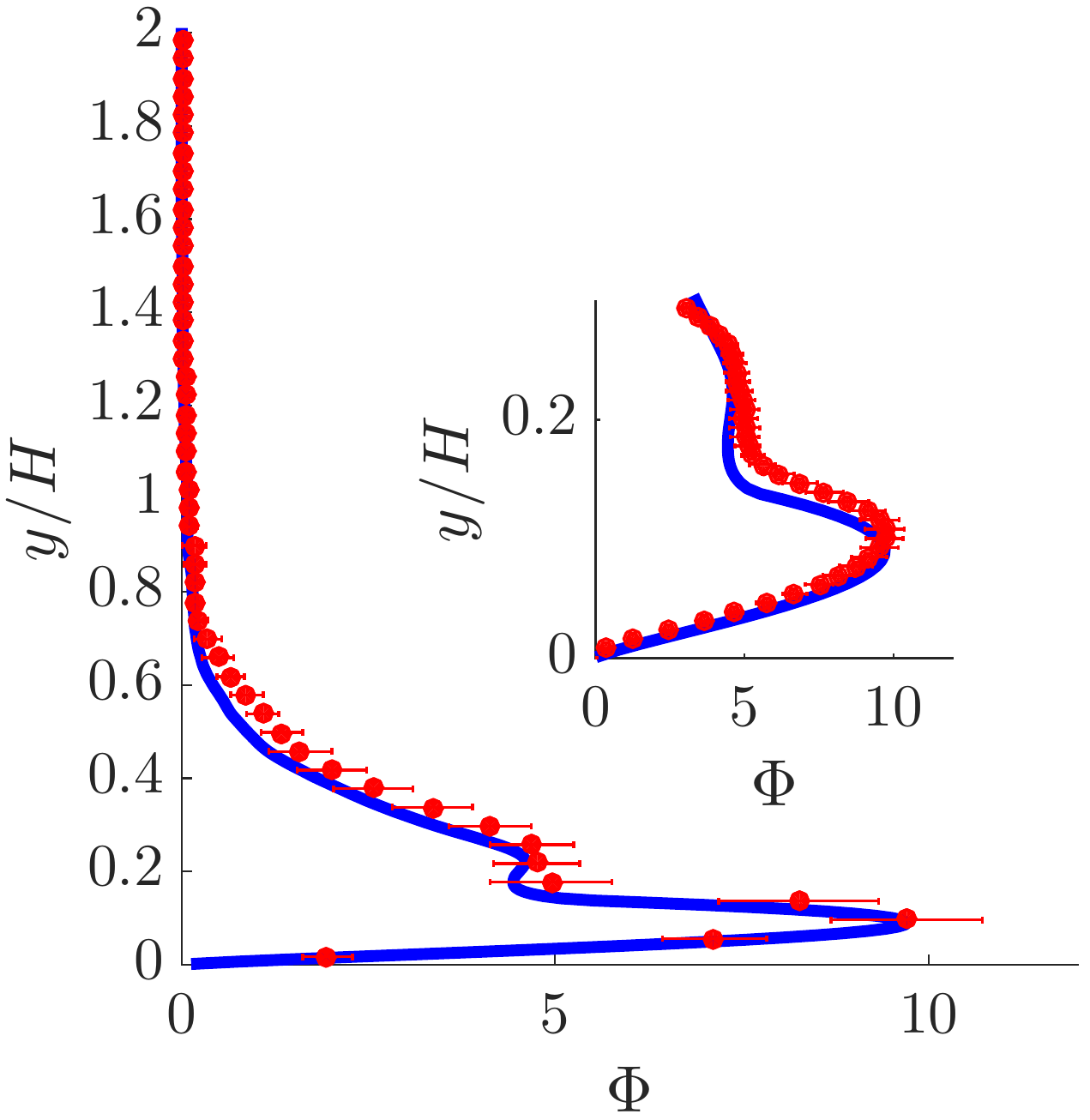}
  \includegraphics[height=0.33\linewidth]{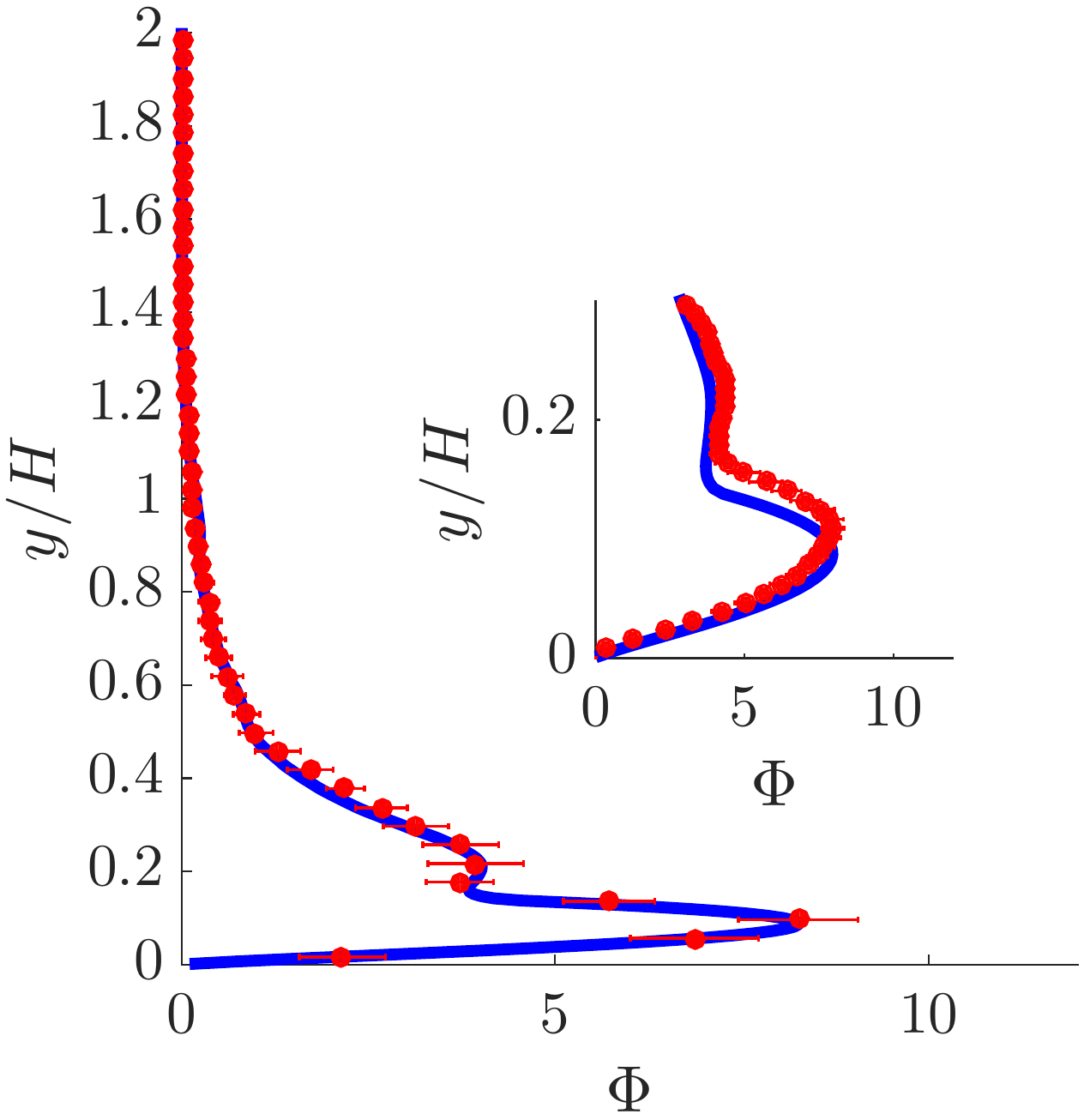}
  \includegraphics[height=0.33\linewidth]{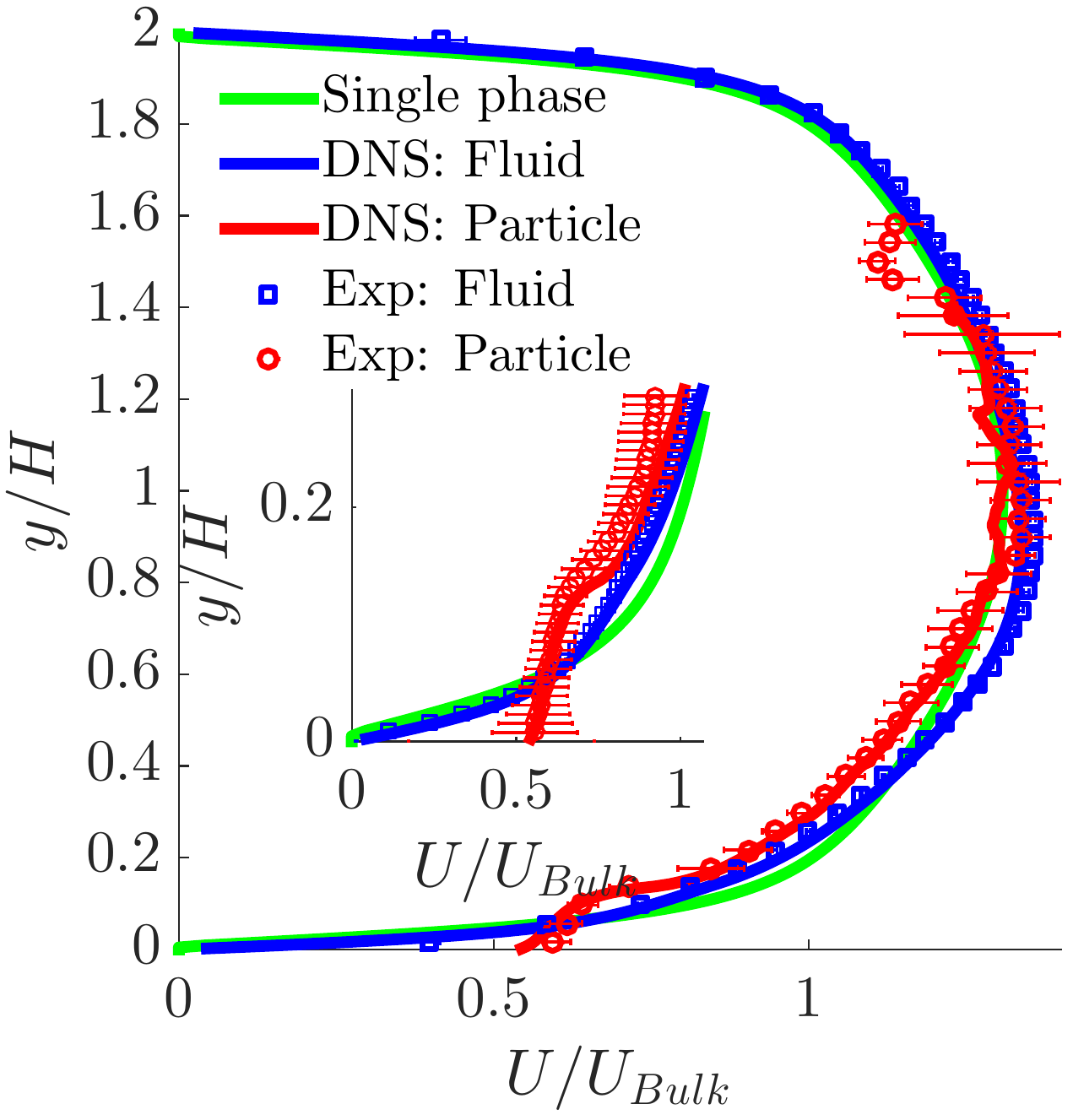}
  \includegraphics[height=0.33\linewidth]{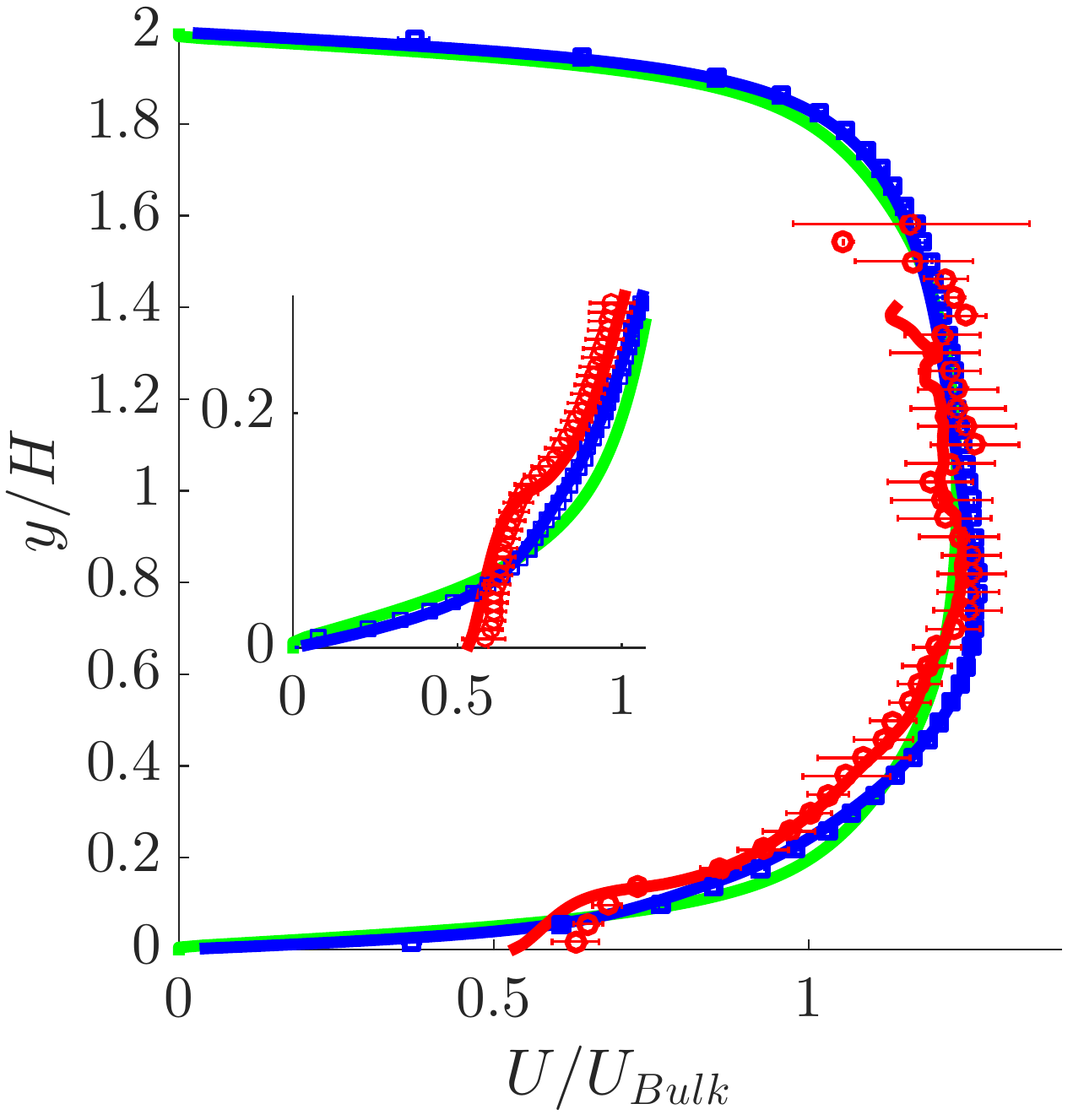}
  \includegraphics[height=0.33\linewidth]{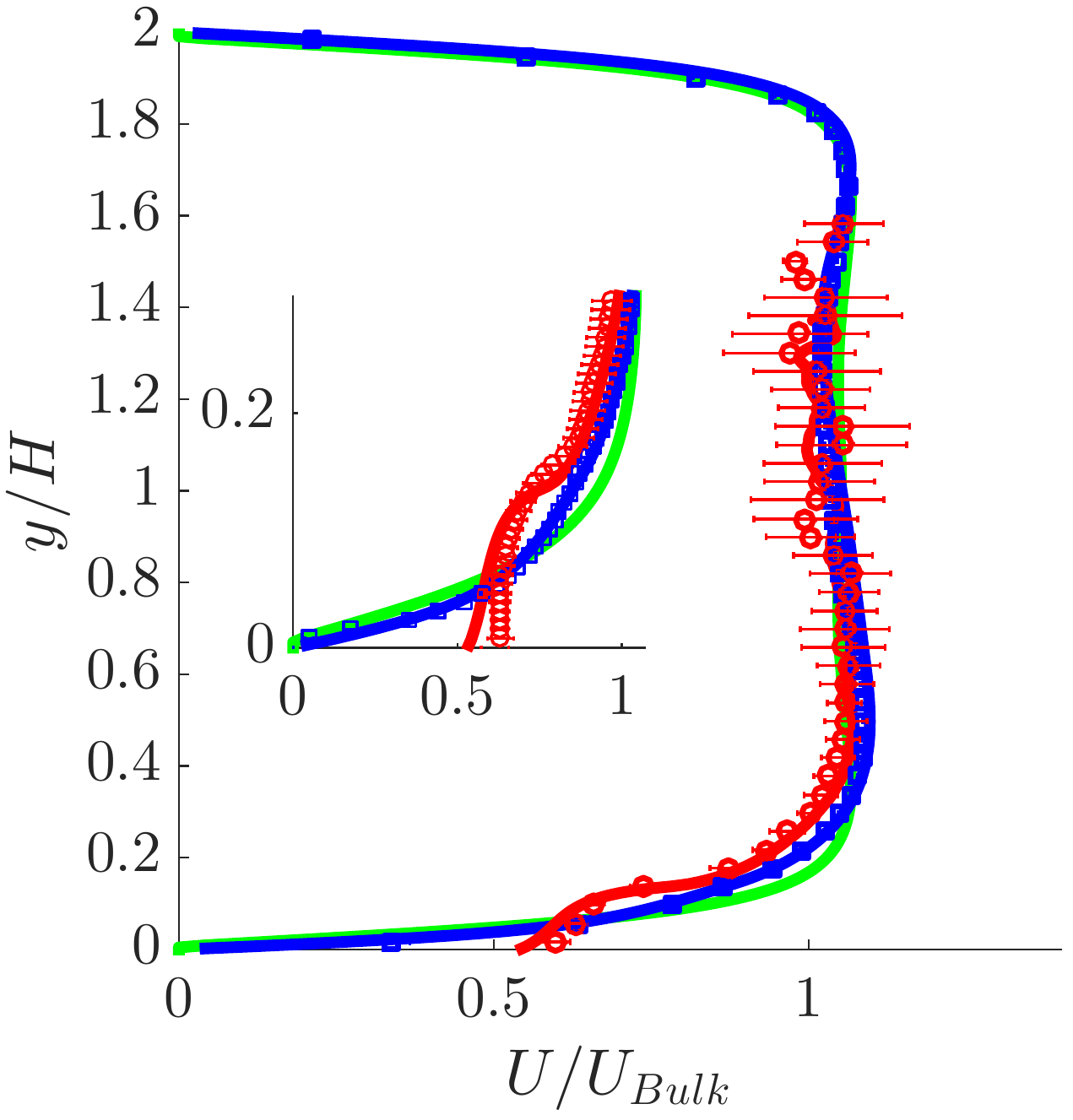}
  \includegraphics[height=0.32\linewidth]{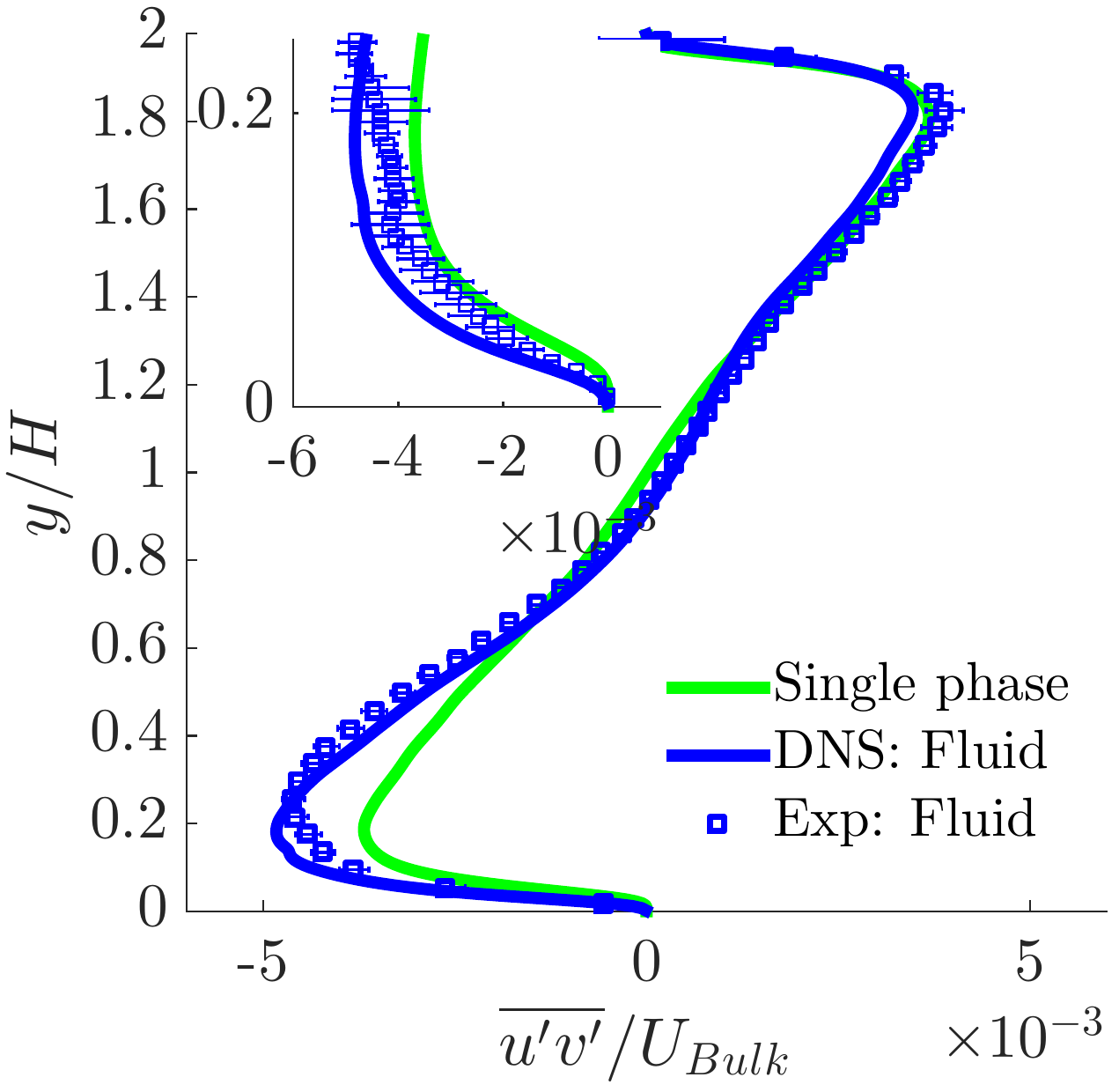}
  \includegraphics[height=0.32\linewidth]{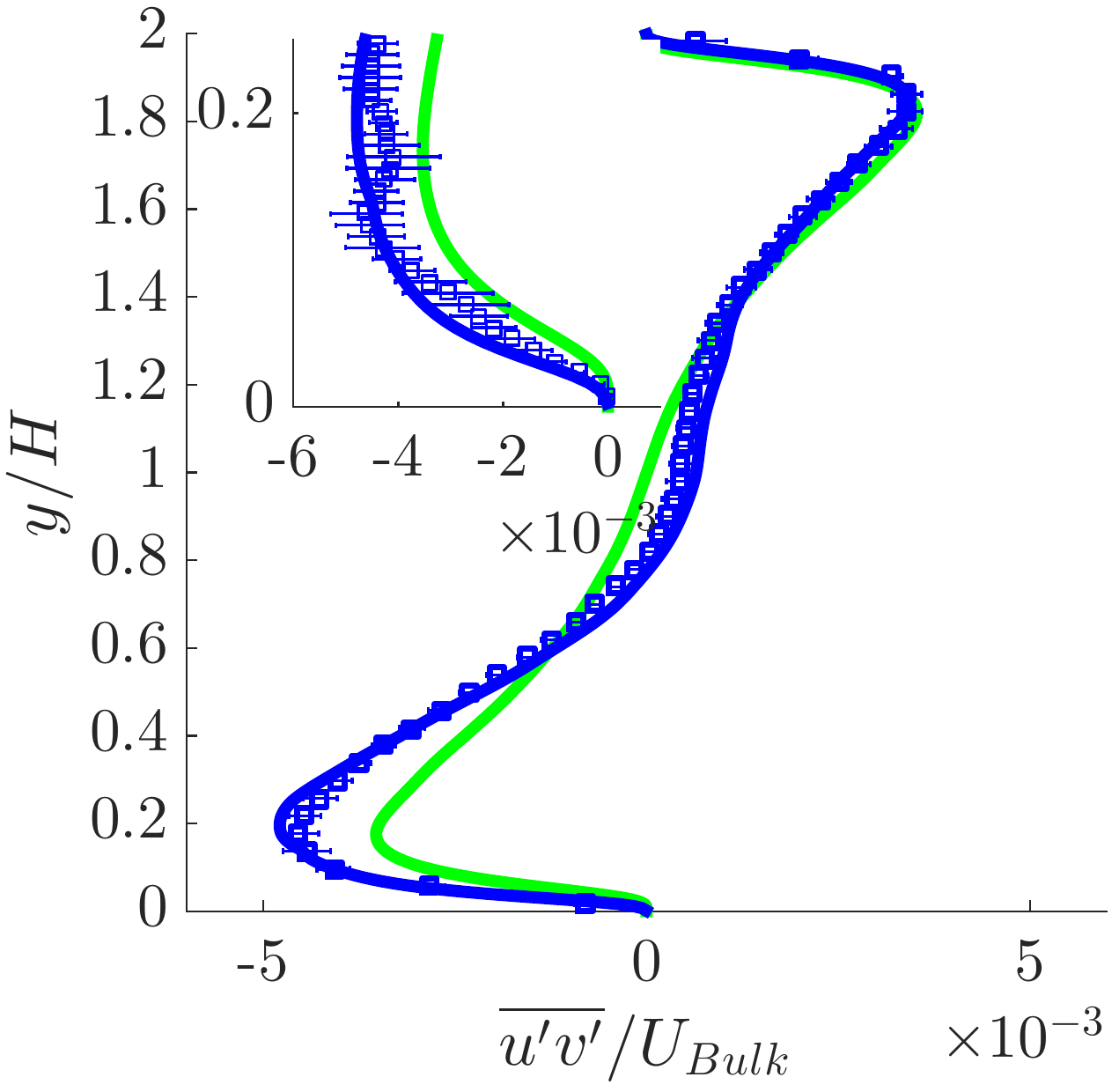}
  \includegraphics[height=0.32\linewidth]{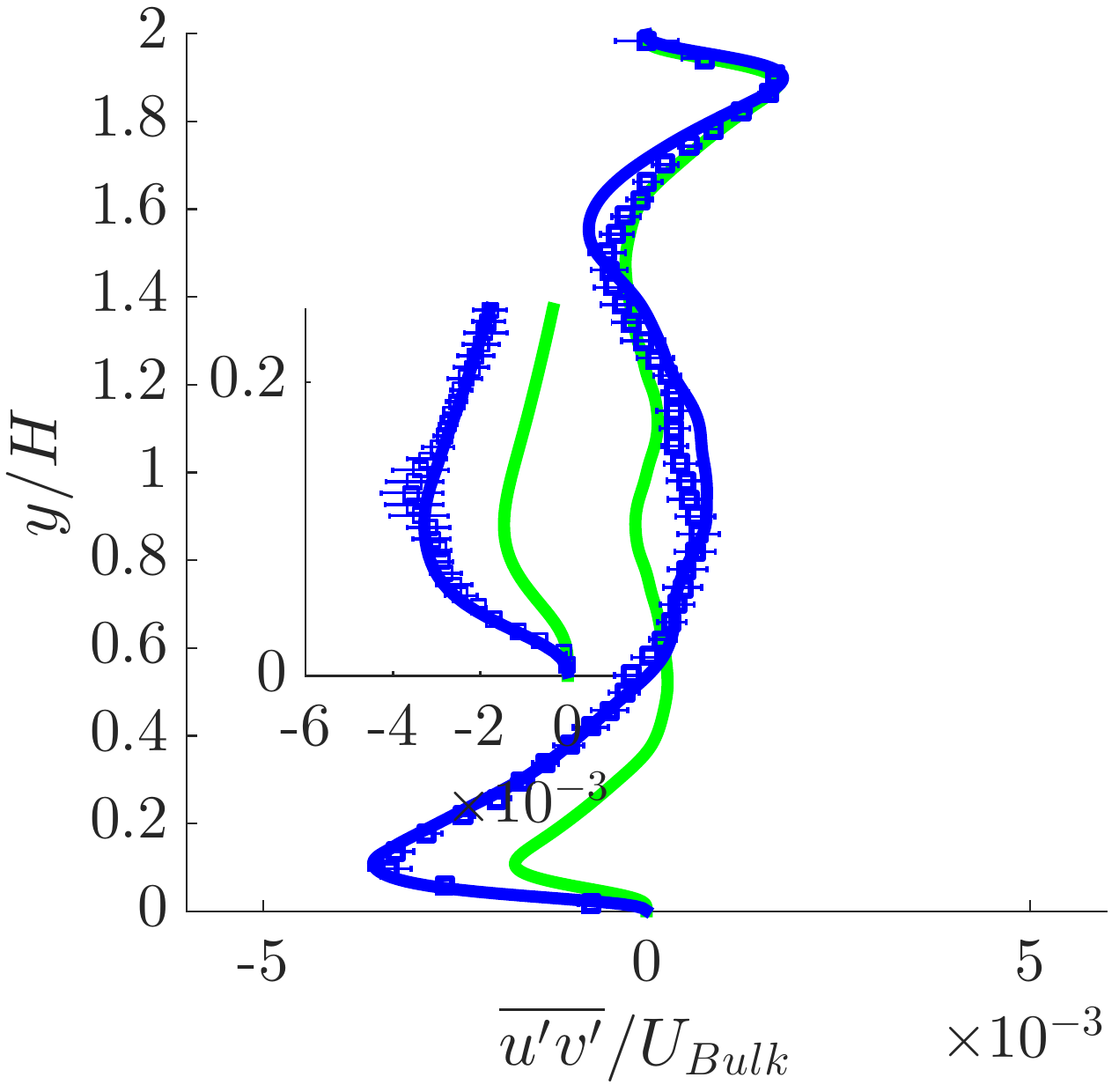}
\caption{Simulation and experimental results for SP at $\Phi$ = 1\%: mean particle concentration \textit{(top-row)}, mean streamwise velocity, \textit{(middle-row)} and primary Reynolds shear stress \textit{(bottom-row)} for 3 span-wise planes namely $z/H = 0$ \textit{(left-column)}, $z/H = 0.4$ \textit{(middle-column)} and $z/H = 0.8$ \textit{(right-column)}.}
\label{fig:Comparing simulations and experiments}  
\end{figure}

\begin{figure}
  (a) \includegraphics[height=0.35\linewidth]{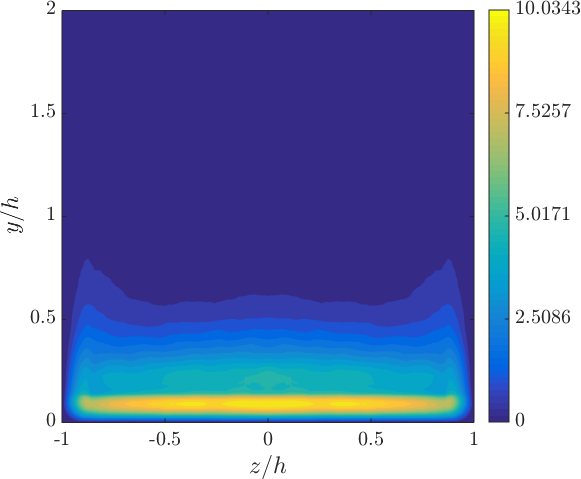}
  (b) \includegraphics[height=0.35\linewidth]{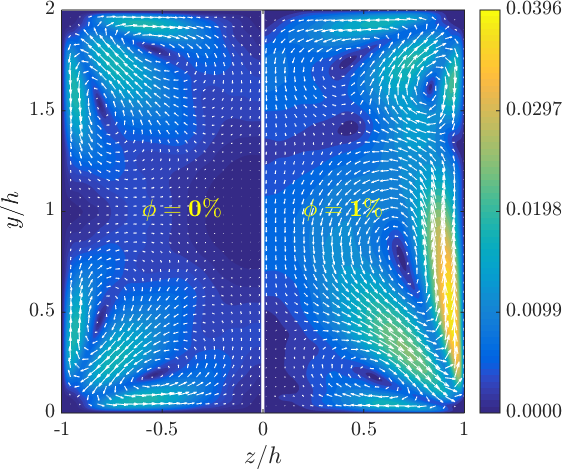}
\caption{(a) Particle concentration distribution (in percentage) and (b) mean secondary fluid velocity $\sqrt{V^2+W^2}/U_{Bulk}$ from simulations. In figure (b), the left panel corresponds to single phase flow $\Phi = 0\%$ and right panel corresponds to $\Phi = 1\%$.}
\label{fig:Concentration and secondary flow}  
\end{figure}

The finite-size particles, while being carried forward by the surrounding fluid also sediment due to their relatively higher density. They are often resuspended upwards by instantaneous turbulent structures, local shear-induced lift and collisions with neighboring particles. Thus, particles mostly occupy the bottom half of the duct and undergo a saltation type of motion \citep{bagnold1956flow} i.e. particles removed from the bottom wall are carried by the fluid, before being transported back to the wall.

Figure \ref{fig:Comparing simulations and experiments} shows the mean particle concentration (first-row), fluid and particle mean streamwise velocity (second-row) and fluid Reynolds shear stress (third-row) for smaller particles SP at a volume fraction $\Phi$ = 1\%. Each column corresponds to a spanwise plane starting from the plane of the wall-bisector $z/H = 0$ (left column) to a plane near the side-wall $z/H = 0.8$ (right-column). 

Particles, on average, form a one diameter thick layer of high concentration in contact with the wall, as seen in the mean concentration profiles (first-row of figure \ref{fig:Comparing simulations and experiments}). Reduced concentration appears above the bottom layer. The concentration is maximum (peak value of $\approx$10\%) in the plane of the wall-bisector $z/H = 0$ and drops towards the side-wall $z/H = 1$. With increasing bulk velocity, the concentration in the bottom layer will reduce until, at sufficiently higher velocities, all the particles will be homogeneously distributed (see \citet{zade2018experimental}). The concentration distribution measured experimentally is the area concentration in a two-dimensional slice whose thickness is equal to the thickness of the laser light-sheet; the particle is recognized as the largest projection of a spherical section cut by the light-sheet. Hence, the different apparent diameter of particles in figure \ref{fig:PIV+PTV} (a). The conversion from such an area-concentration to volume-concentration is non-trivial and hence to maintain consistency between simulations and experiments, we normalize the experimental area-concentration so that its maximum value equals the maximum of the volume-concentration in DNS. The insets in the figure \ref{fig:Comparing simulations and experiments} (first-row) show the details of the high particle concentration region obtained by separate near-wall measurements. 

The mean streamwise velocity of the fluid phase is skewed due to the asymmetry in the particle concentration when compared to the profile for the single-phase flow (second-row of figure \ref{fig:Comparing simulations and experiments}). The fluid velocity is reduced by the particle drag in the particle-rich region and this is compensated, in order to maintain same volume-flow rate, by a proportional increase in the upper regions. On average, particles move slowly compared to the fluid nearly in the whole cross section, except close to the walls, where particles are not constrained by the no-slip condition limiting the fluid velocity to zero at the walls. The particle velocity is displayed without considering the particle rotation as it was not possible to measure particle rotation for our RIM particles. Early trials with tracer-embedded RIM particles as in \citet{byron2013refractive} showed that even a small quantity of tracers severely reduced the optical visibility of the suspension and limited the applicability of PIV+PTV in regions of high concentration. The apparent slip velocity is expected to be lower if rotation is also included while calculating particle velocity. 

The fluid Reynolds shear stress is seen to increase when compared to the single-phase case in particle dominated areas (third-row of figure \ref{fig:Comparing simulations and experiments}) in all three spanwise planes. The enhanced fluid Reynolds shear stress is indicative of increased mixing of the fluid phase in this region. For this low volume fraction $\Phi$ = 1\%, the Reynolds shear stress approaches the values of the single-phase flow in the particle-free upper region of the duct. The mean velocity profile in this region also remains similar to the single-phase case. 

Overall, we find reasonable agreement between numerical and experimental results in terms of particle concentration and velocity statistics. Also, the pressure drop, shown later, is seen to agree quite well. 

Additional details regarding the flow can be obtained from figure \ref{fig:Concentration and secondary flow} where we present the mean particle distribution and the mean secondary motion for the fluid phase, from simulations, for the entire cross-section of the duct. As shown in figure \ref{fig:Concentration and secondary flow} (a), particles suspend to higher elevations near the two side-walls ($z/H$ = -1 and 1), most likely owing to an enhanced secondary flow as shown in figure \ref{fig:Concentration and secondary flow} (b). In the single phase flow, the peak magnitude of this secondary motion $\sqrt{V^2+W^2}\approx0.02U_{Bulk}$. In the presence of particles, this peak value increases to $\approx0.04U_{Bulk}$ and occurs closer to the side-walls, thus causing particles to rise upwards in that region. Due to the higher concentration of particles near the bottom-wall, the secondary flow is no longer symmetric along the vertical direction, as also previously observed in \citet{lin2017effects}. 

\subsection{Effect of particle size and volume fraction}
\label{Effect of particle size and volume fraction}

\begin{center}
\begin{table}[b]
\caption{Cases investigated}
\label{tab:Parameter}
\begin{tabular}{llll}
\hline\noalign{\smallskip}
${2H}/d_p$ & $\Phi \%$ & Studies performed \\ 
\noalign{\smallskip}\hline\noalign{\smallskip} \\
$14.5$ & $1\pm0.1$ & PIV+PTV \& DNS \\
\noalign{\smallskip}
$14.5$ & $1,2,3,4,5\pm0.1$ & PIV+PTV\\
\noalign{\smallskip}
$9$ & $1,2,3,4,5\pm0.1$ & PIV+PTV\\
\noalign{\smallskip}\hline
\end{tabular}
\end{table}
\end{center}

The summary of the different experiments performed is listed in table \ref{tab:Parameter}. The $Re_{2H}  = U_{Bulk}2H/\nu_f$ in experiments is maintained at $5660\pm150$ across all cases. The ratio of the particle terminal velocity $U_{T}$ to the bulk velocity $U_{Bulk}$ is 0.08 and 0.13 for SP and LP respectively. To quantify the role of gravity on the mode of particle transport, the Rouse number $Ro={U_T}/{\kappa u_\tau}$ \citep{rouse1937} is often used. Here $\kappa$ is the von K\'arm\'an constant and $u_\tau = \sqrt{\tau_{w}/\rho_f}$ is the friction velocity for single phase flow with $\tau_{w}$ the average wall-shear stress estimated from the streamwise pressure gradient: $\tau_w = (dP/dx)(H/2)$. In our experiments, $Ro$ is around $\approx$ 3 and 5 for SP and LP, which corresponds to bed load transport i.e. particles transported along the bed ($Ro\geq2.5$) \citep{fredsoe1992mechanics}. The particle Stokes number $St={(\rho_p{d_p}^2/{18\rho_f\nu_f}})/(H/U_{Bulk})$ 
based on the fluid bulk time scale is $\approx$ 3 and 8 for SP and LP, respectively. Finally, the particle size in inner length scales of the single phase flow for the small particles is $\approx$ 25$\delta_\nu$ (where $\delta_\nu = \nu/u_{\tau}$ is the viscous length scale) and $\approx$ 40$\delta_\nu$ for the large particles. So, the SP are already around 5 times larger than the thickness of the viscous sub-layer.

\subsubsection{Pressure drop}
\label{Pressure drop}

\begin{figure}
  \includegraphics[height=0.45\linewidth]{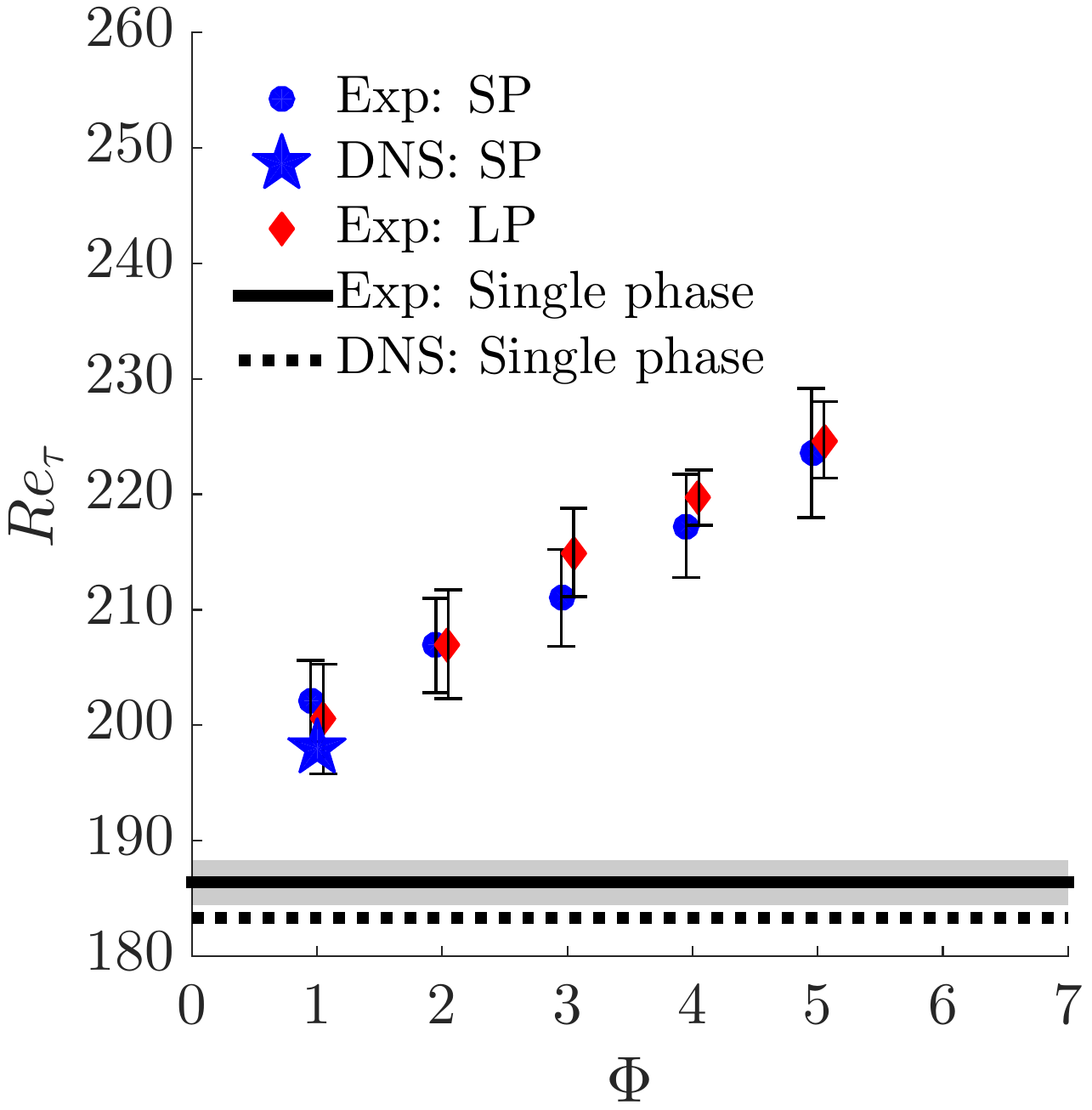}
\caption{Friction Reynolds number at bulk $Re_{2H}$=5600 as a function of the particle size and volume fraction.}
\label{fig:Re_tau_phi} 
\end{figure}

The pressure drop, expressed as the friction Reynolds number $Re_{\tau} = u_{\tau}H/\nu_f$, is plotted in figure \ref{fig:Re_tau_phi} as a function of particle volume fraction $\Phi$. As expected, $Re_{\tau}$ increases with increasing $\Phi$. Within the limits of the errorbars, particles of both sizes result in similar values of the pressure drop. The pressure drop predicted by DNS for $\Phi$ = 1\% of SP (star symbol in figure \ref{fig:Re_tau_phi}) is also close to the experimentally measured value. The small offset between DNS and experiments with particle-laden case is of the same order as the difference in the single-phase case i.e. $<2\%$. 


\subsubsection{Concentration and velocity distribution}
\label{Concentration and velocity distribution}

\begin{figure}
  \includegraphics[height=0.33\linewidth]{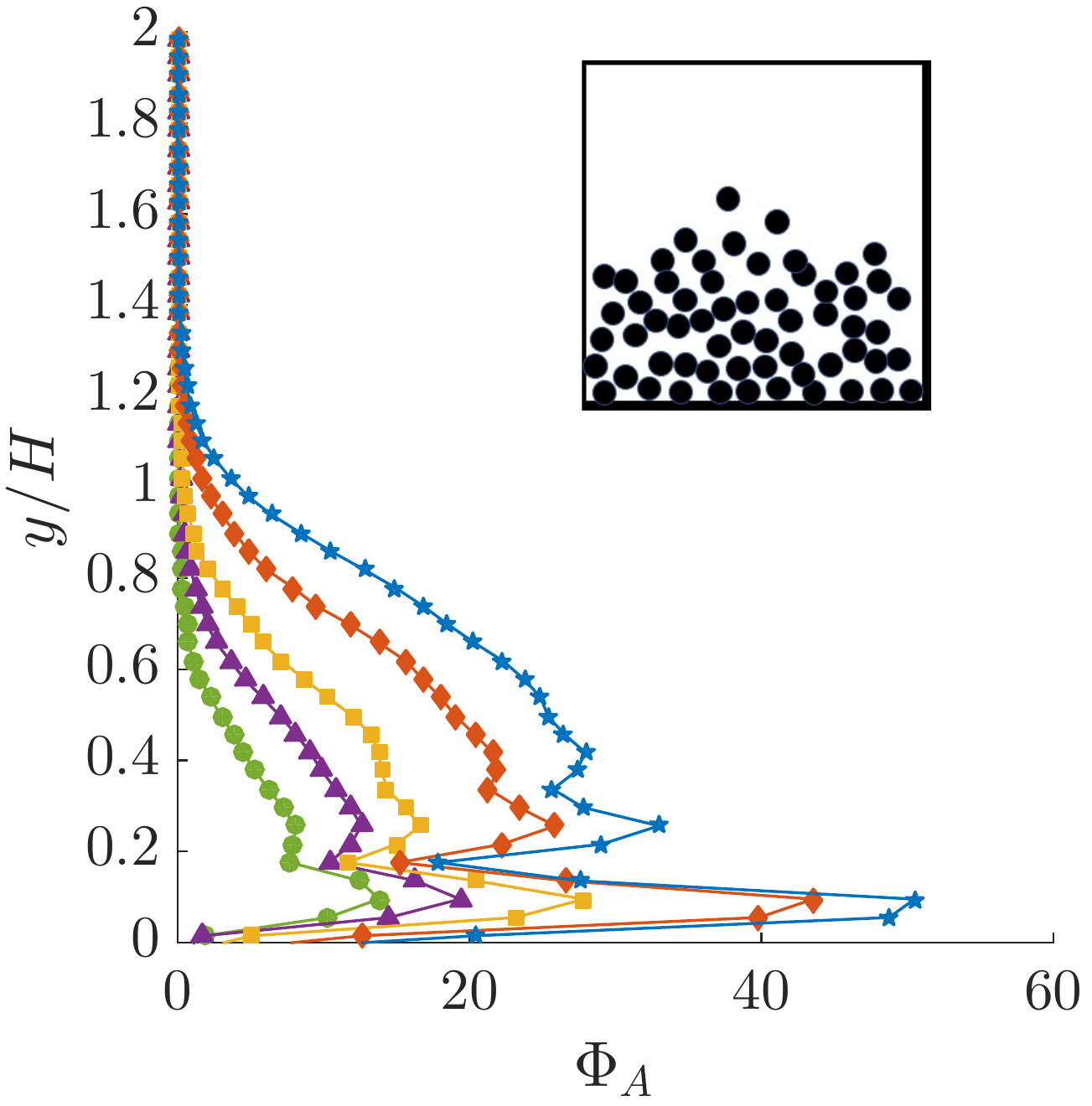}
  \includegraphics[height=0.33\linewidth]{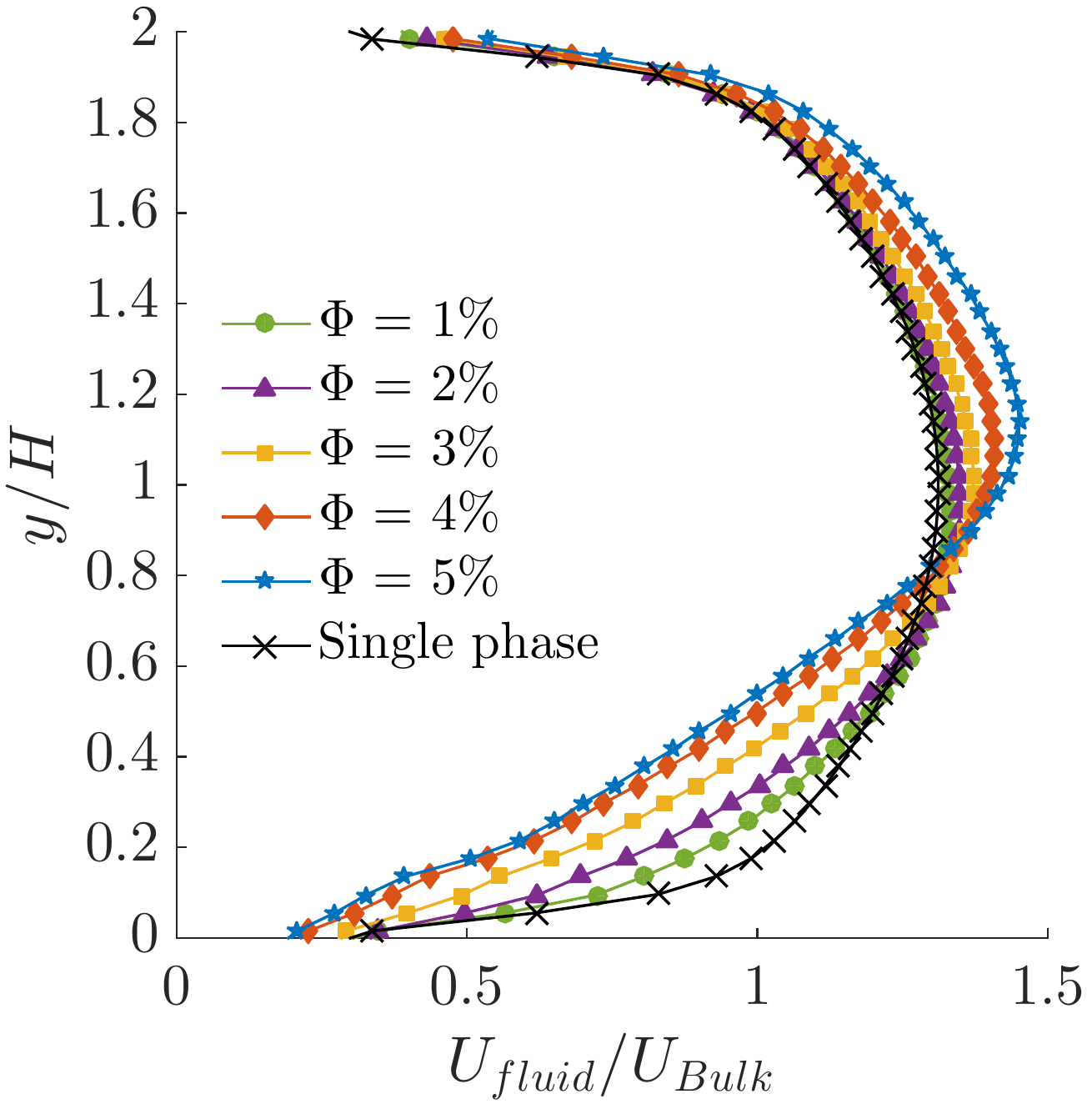}
  \includegraphics[height=0.33\linewidth]{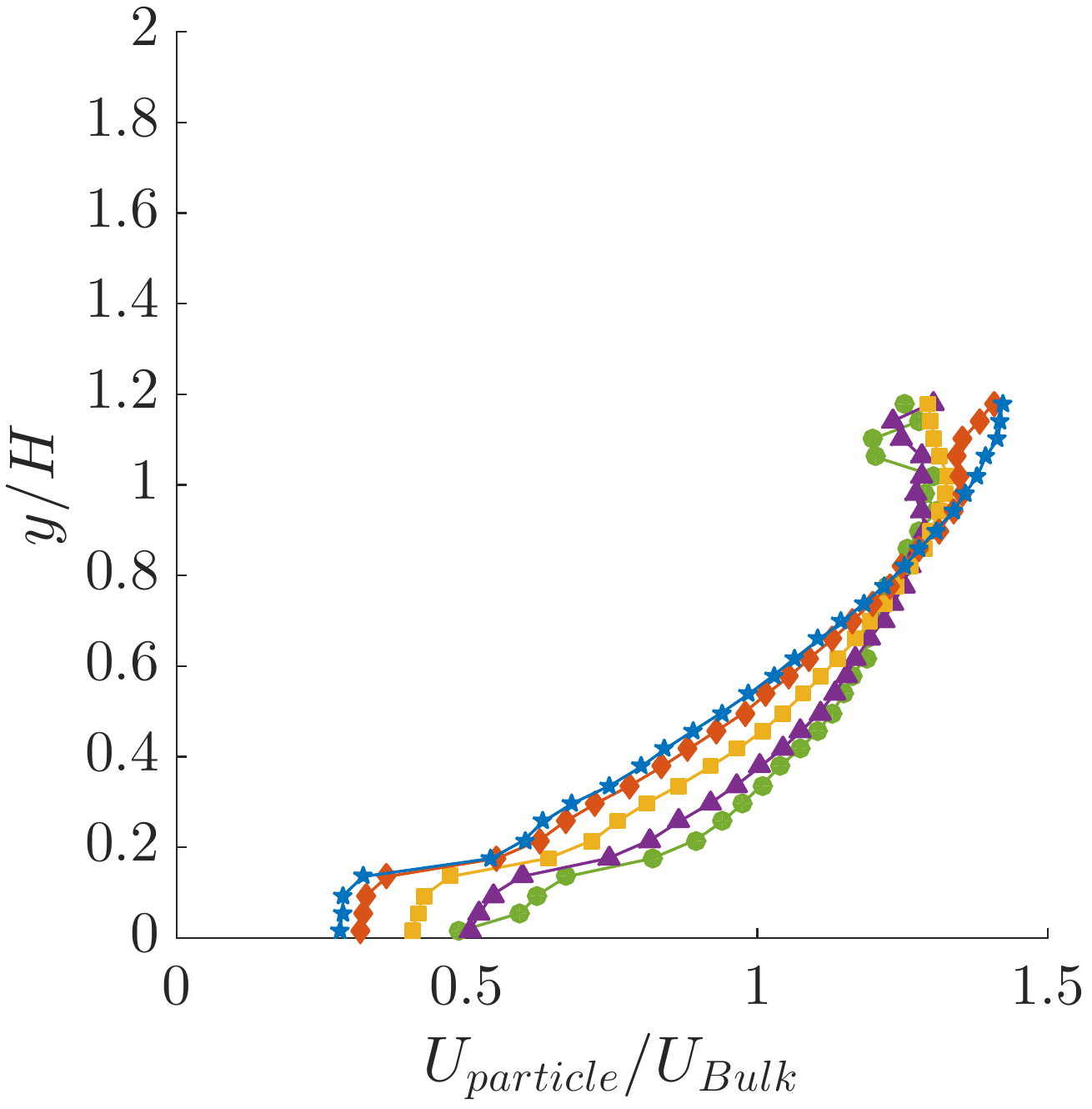}
  \includegraphics[height=0.33\linewidth]{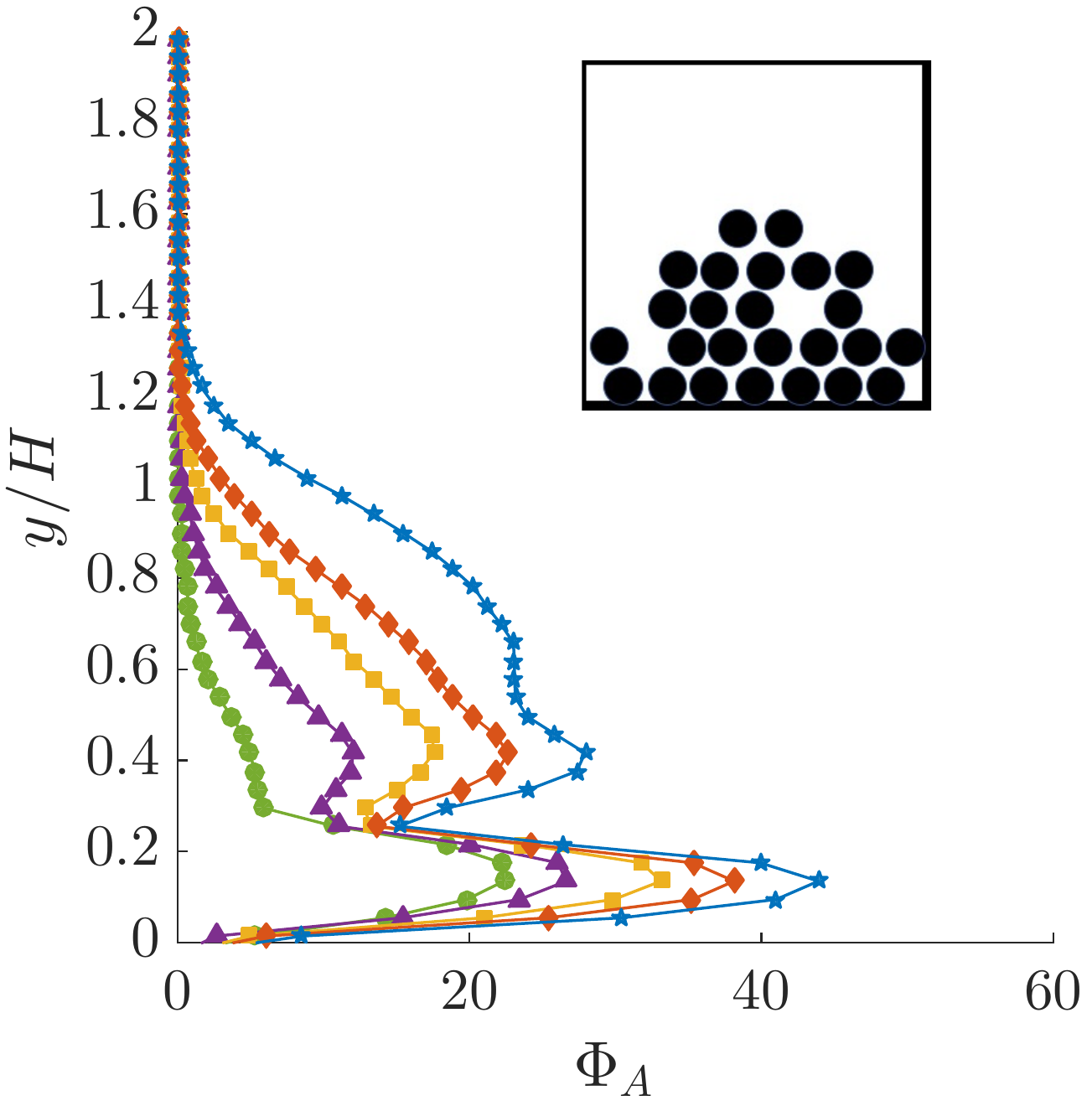}
  \includegraphics[height=0.33\linewidth]{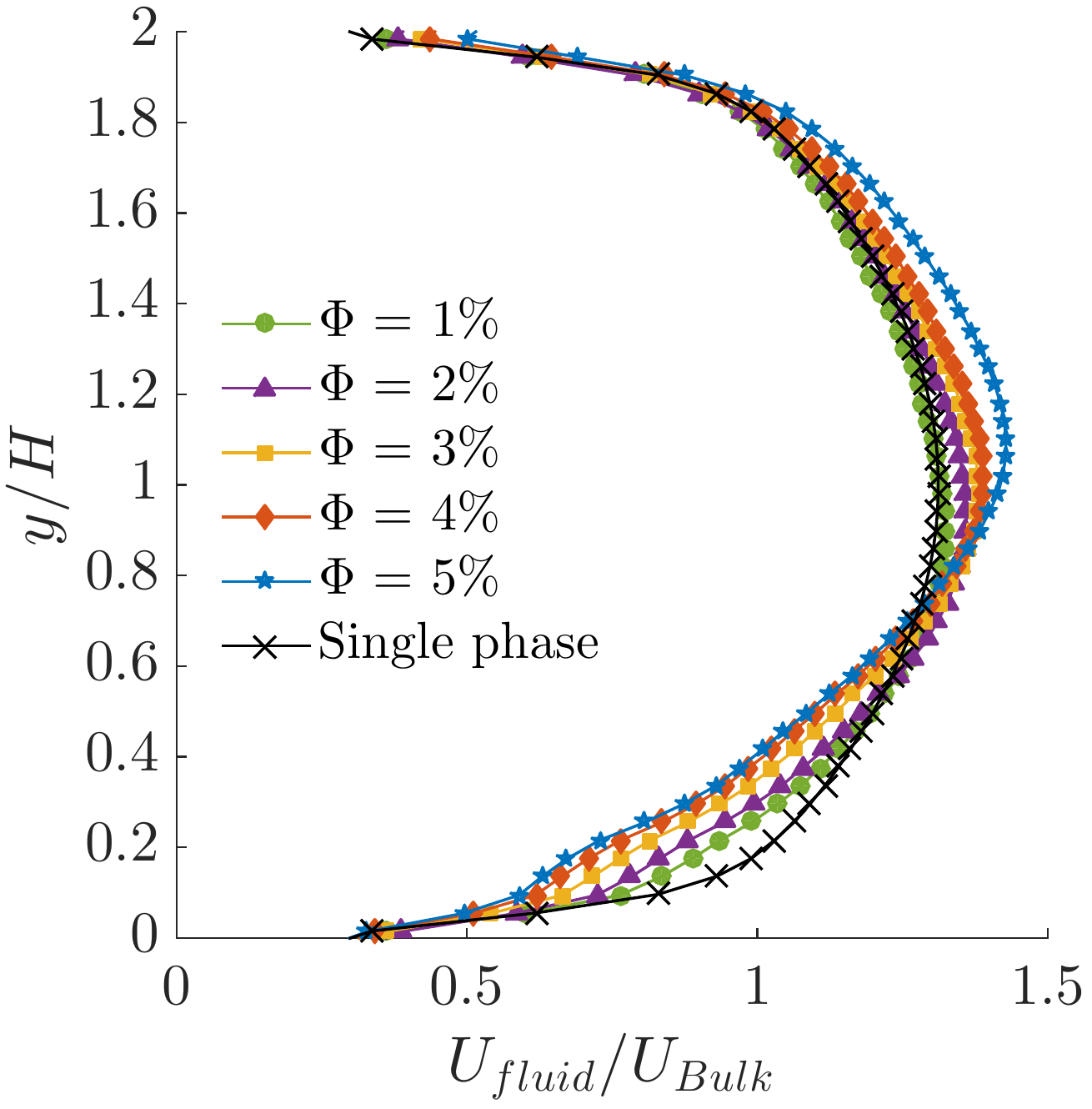}
  \includegraphics[height=0.33\linewidth]{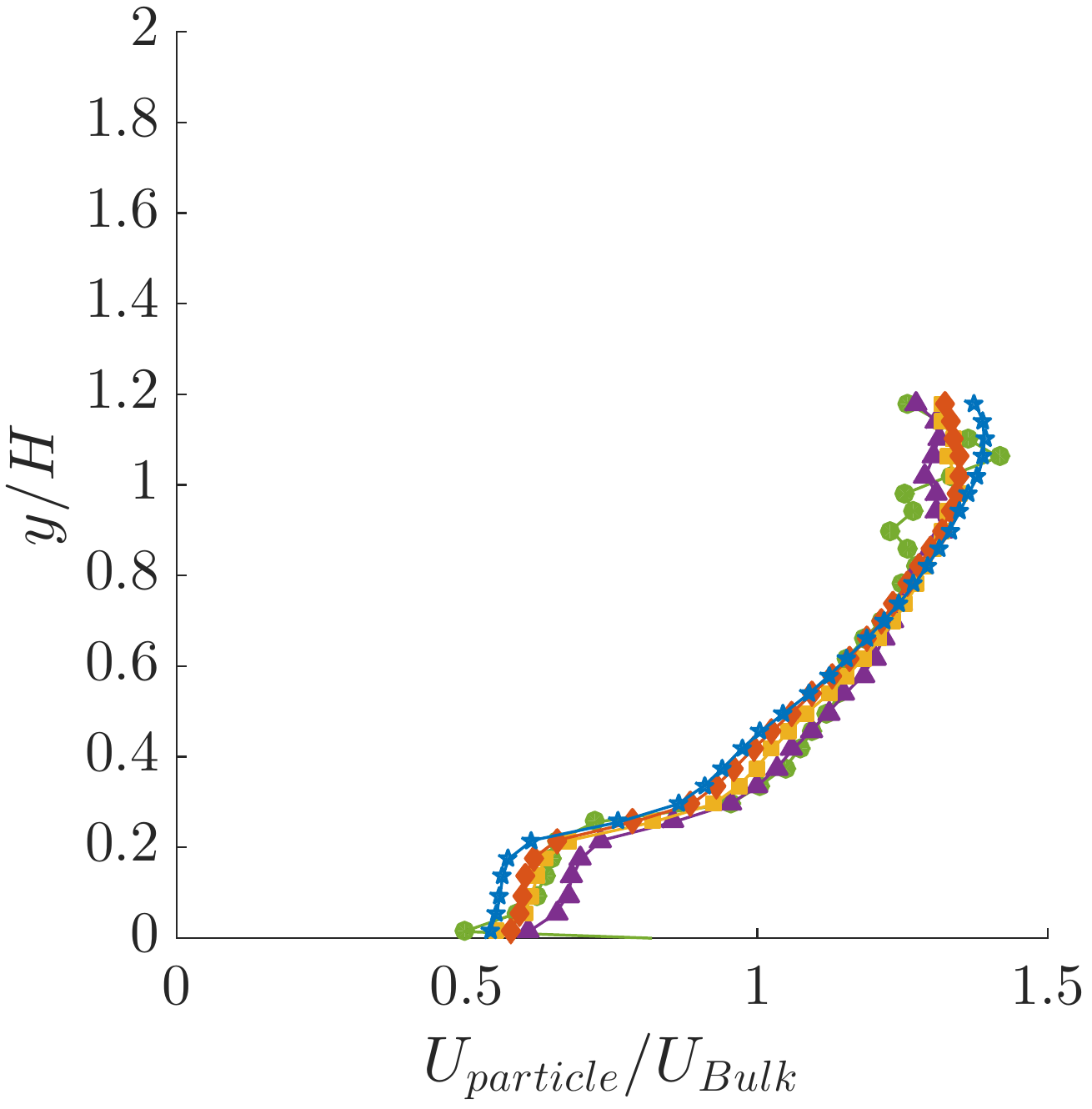}
\caption{Particle concentration and mean velocity at $z/H = 0$: mean particle concentration \textit{(left-column)}, mean fluid streamwise velocity \textit{(middle-column)} and  mean particle streamwise velocity \textit{(right-column)}. The \textit{(top-row)} corresponds to smaller particles and the \textit{(bottom-row)} to larger particles. The relative size and distribution of the particles is illustrated by the cartoon in the corresponding inset of the figures in the \textit{left-column}.}
\label{fig:Effect of particle size and volume fraction Umean_phi}  
\end{figure}

Even though the overall pressure drop is very similar for the two particle sizes and seems mainly to be a function of the concentration, noticeable differences are seen in the turbulent velocity statistics of the fluid phase. 

Figure \ref{fig:Effect of particle size and volume fraction Umean_phi} shows the mean particle area-concentration $\Phi_A$, mean fluid and particle streamwise velocity for SP (top-row) and LP (bottom-row) in the plane of the wall-bisector $z/H = 0$. The concentration distribution is similar for both the particle sizes for all bulk volume fractions (first-column of figure \ref{fig:Effect of particle size and volume fraction Umean_phi}). The thickness of the particle layer is higher for LP due to their larger size. With an increase in nominal volume fraction, a second layer of high particle concentration appears above the first one and and at the highest concentration $\Phi$ = 5\%, there is a tendency for a third layer, more so for SP due to their larger number $N\propto\Phi/d_p^3$. The maximum particle concentration reaches values above 40\% and yet using RIM-PIV, it was possible to measure velocity inside the flow i.e. beyond 6-7 particle diameters.

Compared to the single-phase case, the mean fluid streamwise velocity is monotonically reduced by increasing $\Phi$ in the lower region where the particle concentration is high (second-column of figure \ref{fig:Effect of particle size and volume fraction Umean_phi}). In this region, the mean velocity approaches a linear profile, or constant shear rate, for increasing $\Phi$. To maintain the same flow rate, the upper region experiences, therefore, higher velocity and the peak in the velocity profile is displaced upwards with increasing $\Phi$. The above modification is more pronounced for SP than LP. Particles, on average, move slower than the fluid phase (third-column of figure \ref{fig:Effect of particle size and volume fraction Umean_phi}) except close to the wall. For SP, the apparent slip velocity in the near-wall region is lower than LP and reduces further with increasing $\Phi$. The lower velocity of SP as compared to LP is due to the fact that the bottom layer of SP is exposed to lower fluid velocity. This reduction of particle velocity with $\Phi$ conforms with the reduced fluid velocity for the corresponding $\Phi$.

\begin{figure}
  \includegraphics[height=0.33\linewidth]{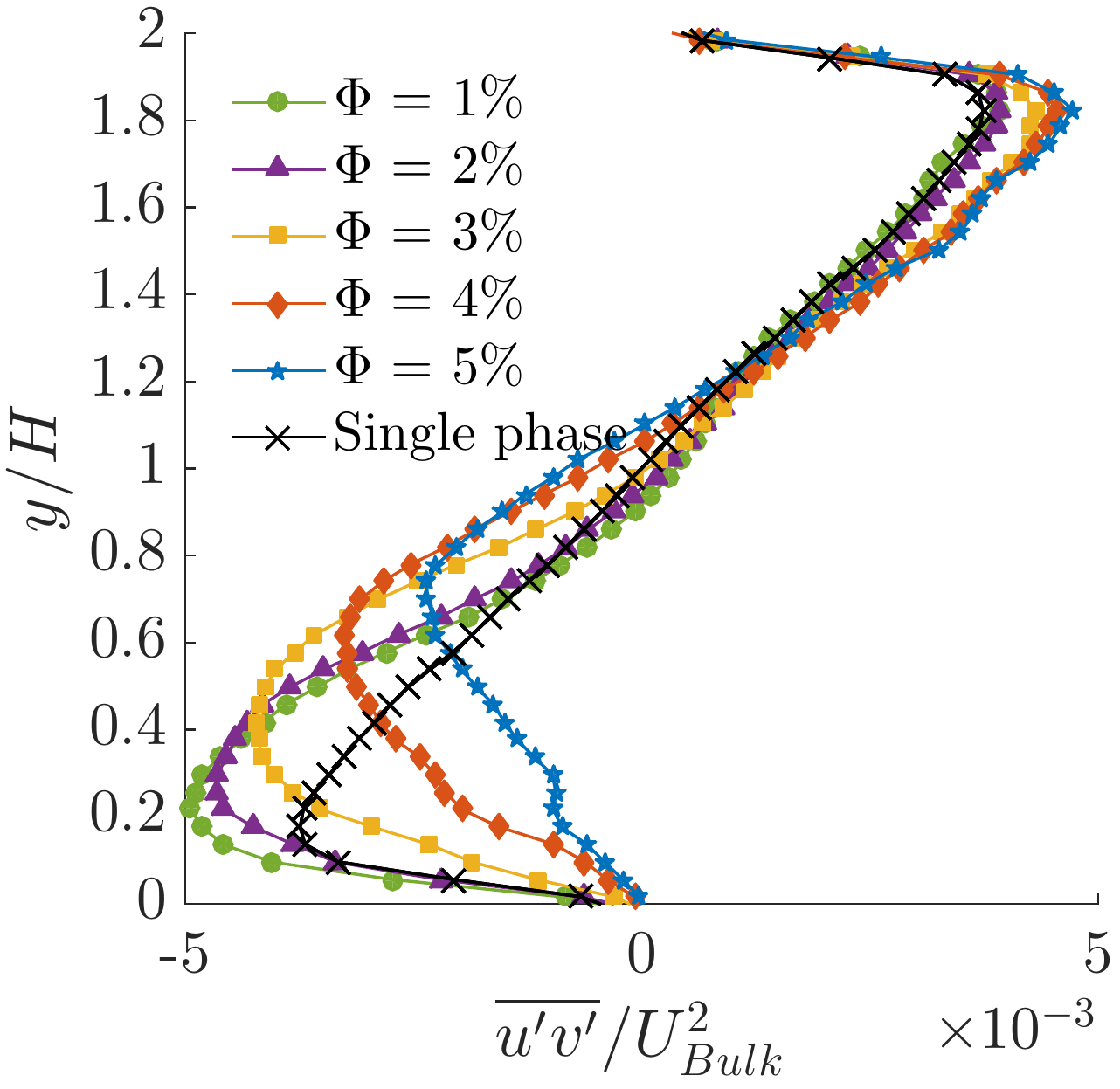}
  \includegraphics[height=0.33\linewidth]{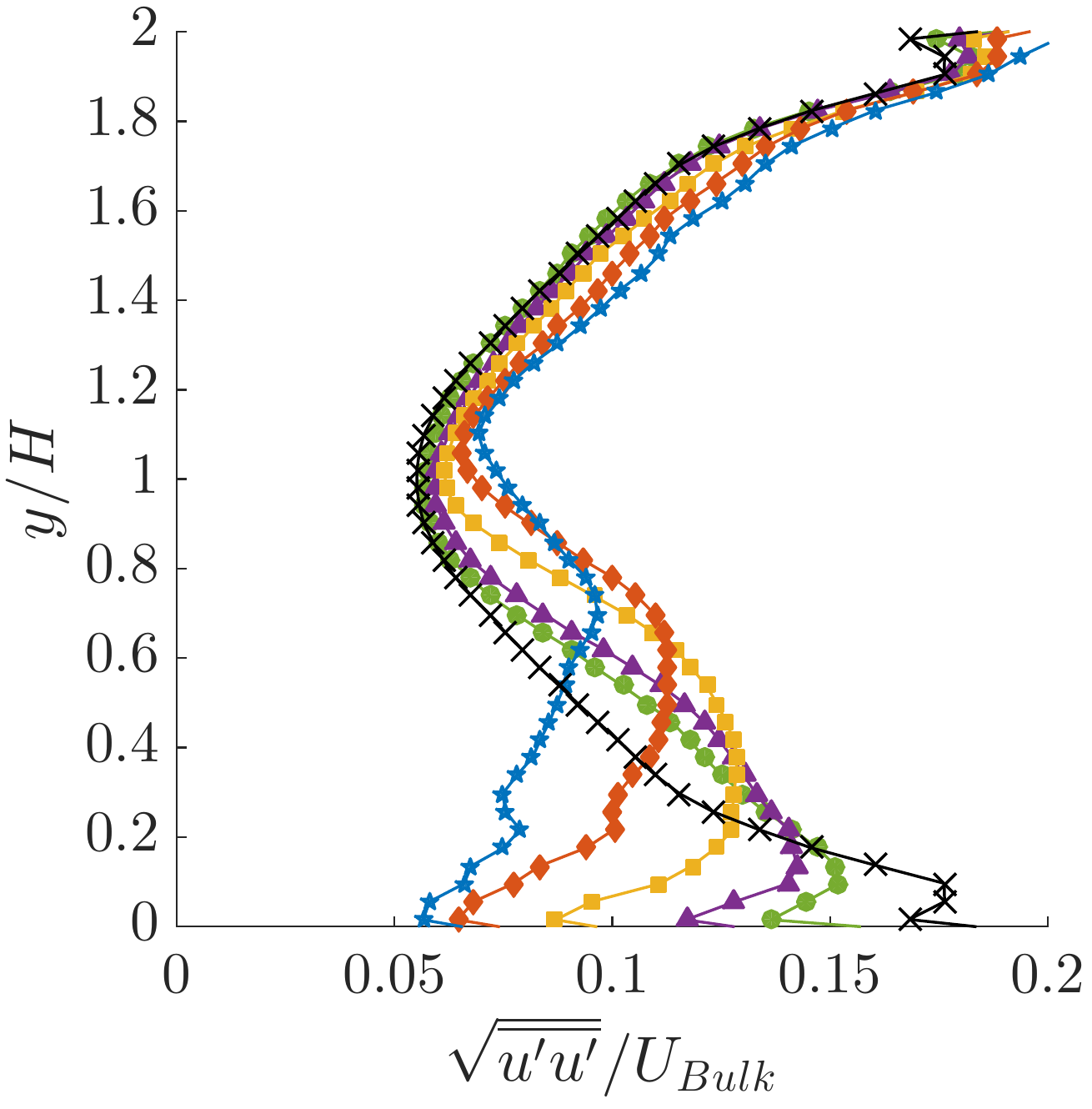}
  \includegraphics[height=0.33\linewidth]{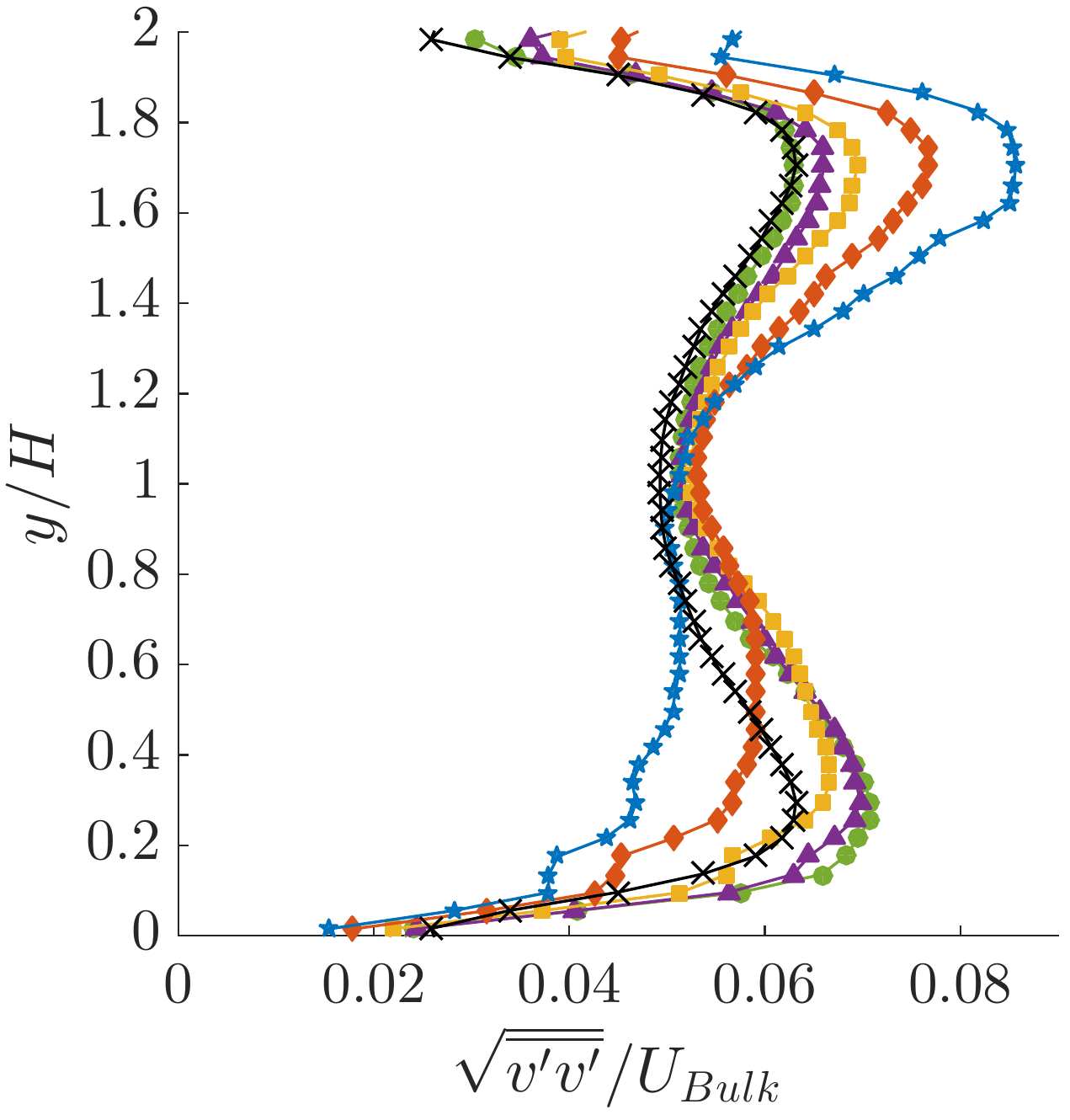}
  \includegraphics[height=0.33\linewidth]{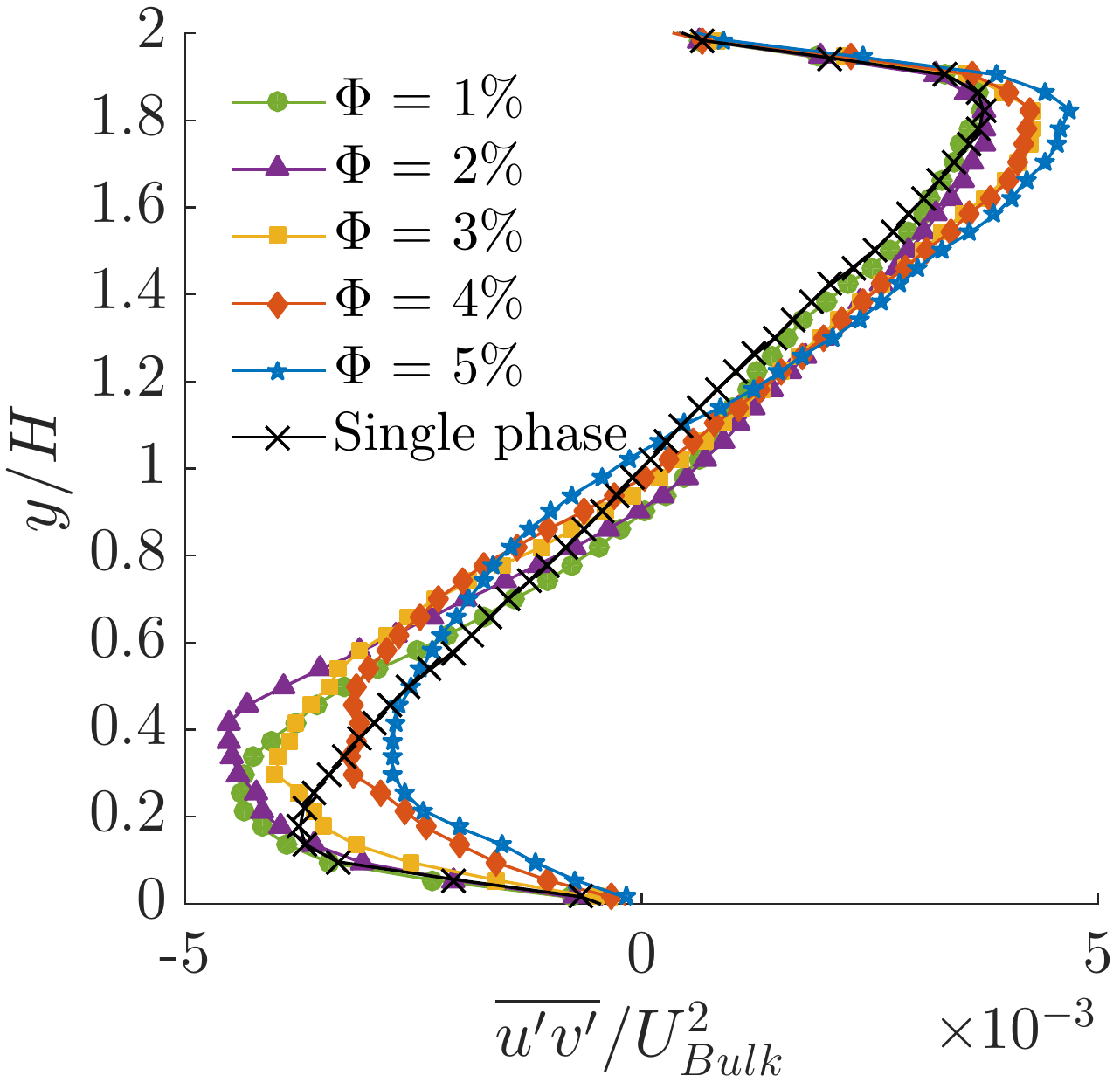}
  \includegraphics[height=0.33\linewidth]{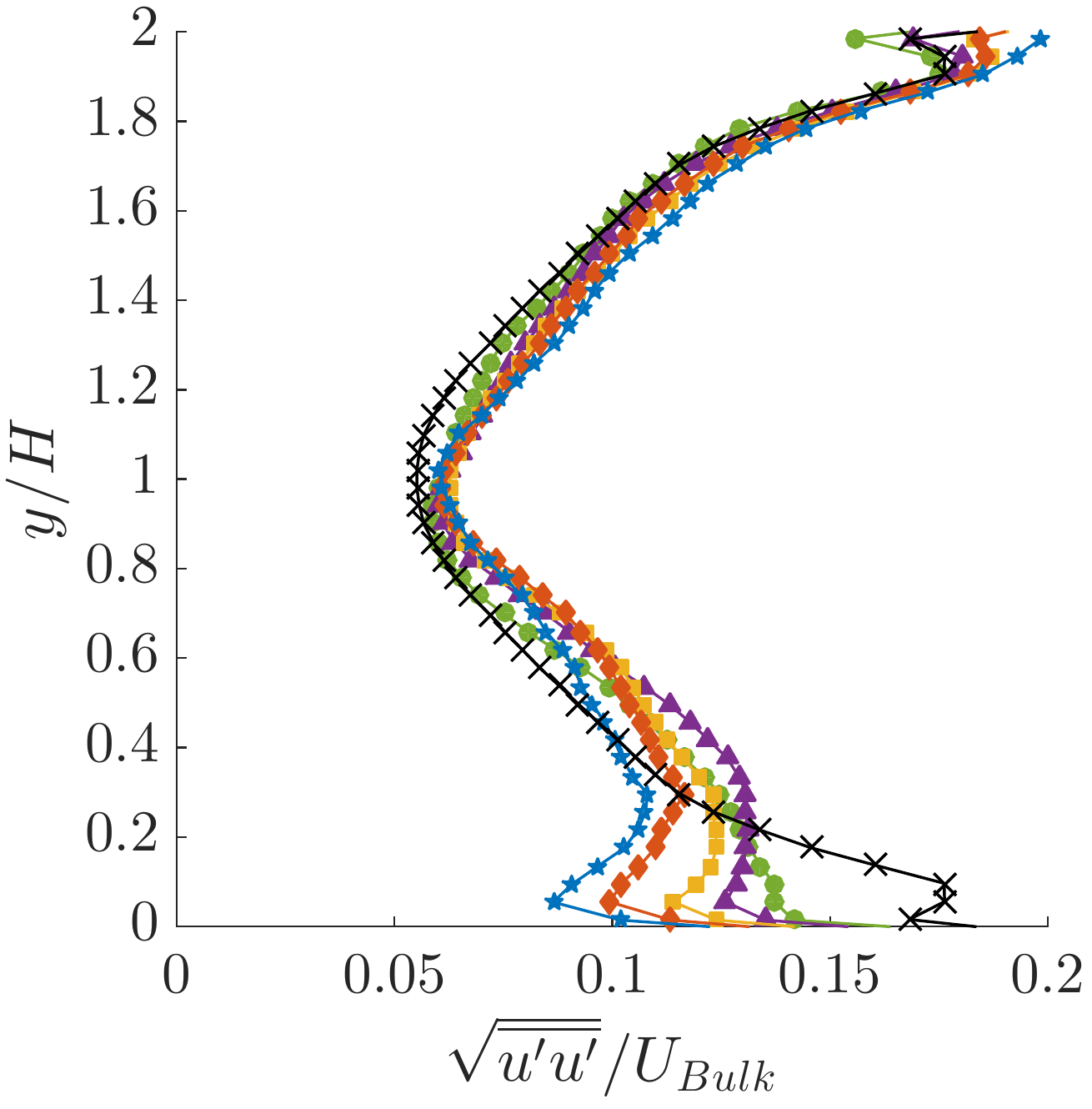}
  \includegraphics[height=0.33\linewidth]{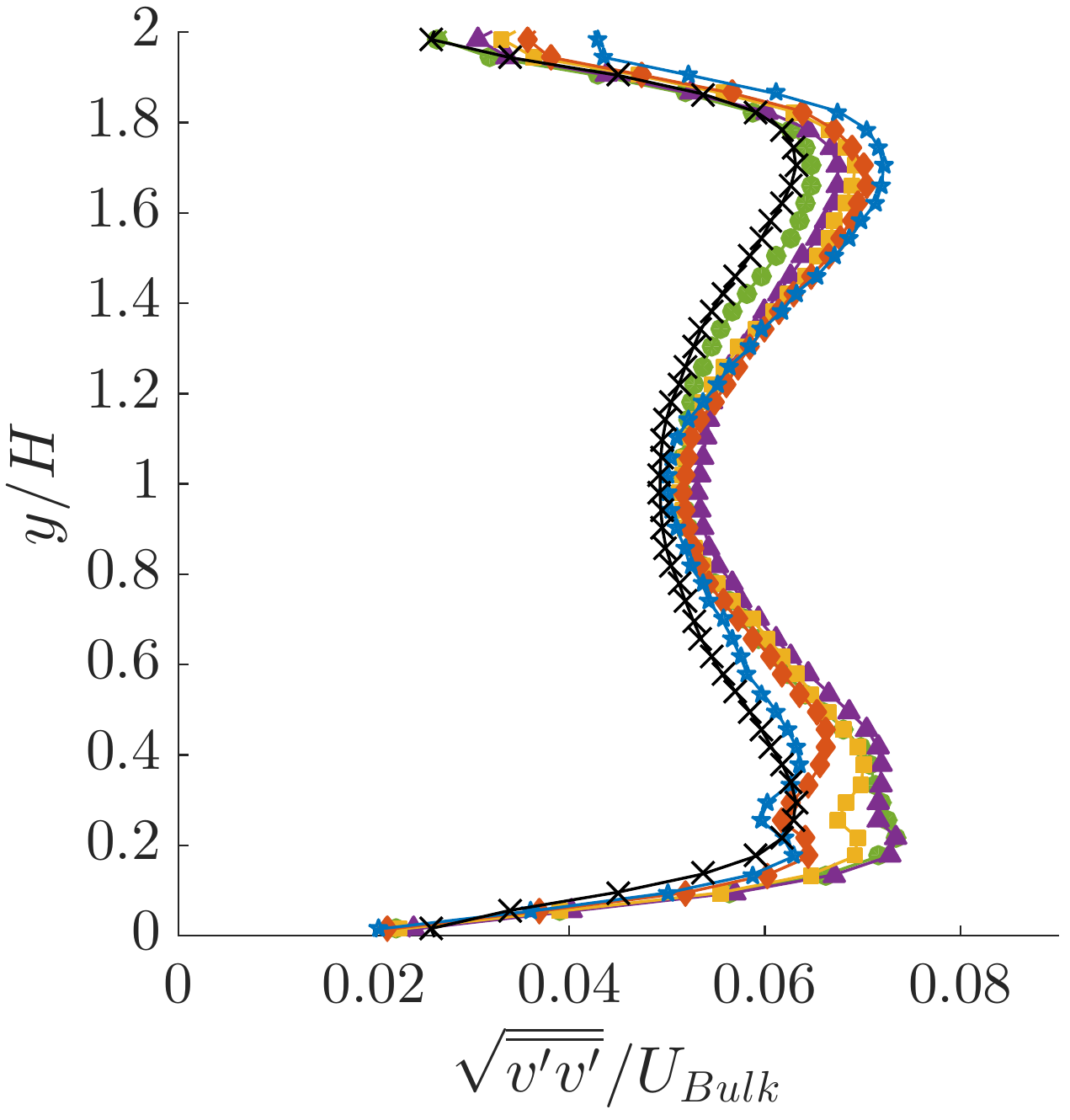}
\caption{Fluid velocity fluctuation statistics at $z/H = 0$:  primary Reynolds shear stress \textit{(left-column)}, RMS of streamwise fluctuating velocity \textit{(middle-column)} and RMS of wall-normal fluctuating velocity \textit{(right-column)}. The \textit{(top-row)} displays results for smaller particles SP and the \textit{(bottom-row)} for larger particles LP.}
\label{fig:Effect of particle size and volume fraction uv}  
\end{figure}

The fluid velocity fluctuation statistics are more significantly changed for SP than LP (see figure \ref{fig:Effect of particle size and volume fraction uv}) indicating that, for a given particle density and volume fraction, larger particle number has larger effects, as also observed by \citet{shao2012fully}. \sagar{These authors suggested that at same $\Phi$, the fluid inertial effects are more pronounced for particles with a larger number density i.e. smaller particles are more effective than larger particles.} Also the area on the wall occupied by SP ($\sim Nd_p^2 \propto \Phi/d_p$) is larger than LP, causing larger hindering effects for the near-wall fluid structures. 

For $\Phi$ = 1 and 2\%, the negative peak in the fluid Reynolds shear stress (first-column of figure \ref{fig:Effect of particle size and volume fraction uv}) increases. With increasing $\Phi$, the location of this peak is displaced away from the wall, again more for SP than LP. For $\Phi$ larger than 3\%, the Reynolds shear stress is increasingly suppressed in the lower (near-wall) part of the particle-rich region while it increases in the upper part of the duct until it approaches values similar to or slightly higher than the single-phase case in the particle-free region. 

A similar behavior is seen for the rms of the streamwise velocity fluctuations: these are increasingly suppressed in the lower part of the duct, more for SP than LP, followed by an increase before finally approaching values similar to or slightly higher than the single-phase case. The wall-normal velocity fluctuations systematically increase with $\Phi$ in the particle-free upper regions of the duct while displaying an increase at lower $\Phi$ and decrease at larger $\Phi$ in the lower regions of the duct. The above observations are qualitatively similar to the channel flow DNS simulations of \citet{shao2012fully}: In the near-wall region, the presence of particles disrupts the larger coherent flow structures, e.g. high/low speed streaks, and thus reduces the streamwise turbulence intensity (also see \citet{picano2015turbulent} and \citet{fornari2017suspensions}). On the other hand, particle-induced small-scale vortices in the near wall region increase the wall-normal (and spanwise, not measured here) velocity fluctuations. At larger volume fractions, sufficient to cover the bottom wall with particles (corresponding to $\Phi \approx$ 4.5\% for SP and 7\% for LP considering that all particles settle on the bottom wall at maximum packing, assumed here as 65\%), the particle layer behaves like a rough wall. Vortices that are shed as the fluid moves above this particle wall-layer are transported towards the core, leading to an enhancement of the fluid velocity fluctuations in that region, as evident in figure \ref{fig:Effect of particle size and volume fraction uv} \sagar{and also found in \citet{shao2012fully}.}

\subsubsection{Particle distribution and relative velocity field}
\label{Particle distribution and relative velocity field}

\begin{figure}
  \includegraphics[width=0.24\linewidth]{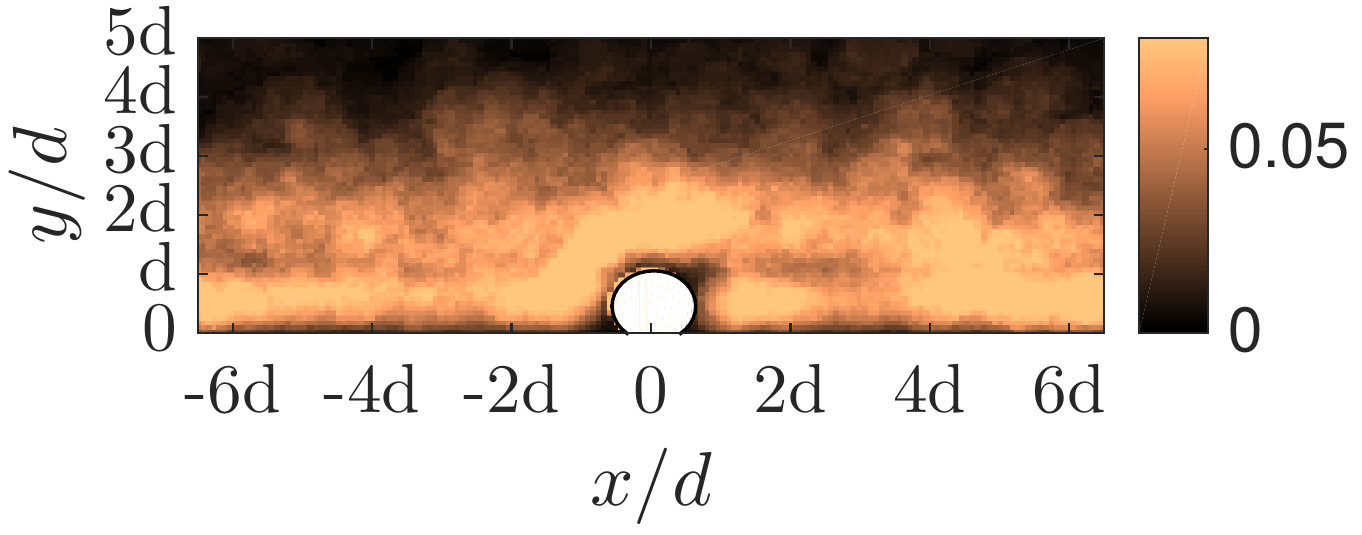}
    \includegraphics[width=0.24\linewidth]{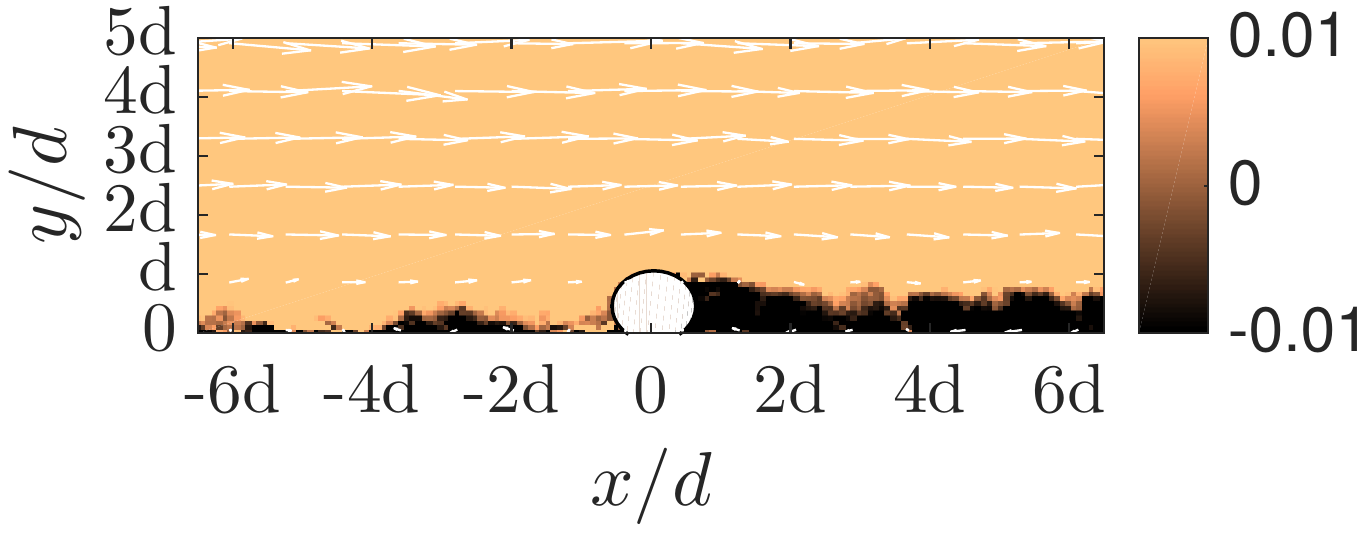}
  \includegraphics[width=0.24\linewidth]{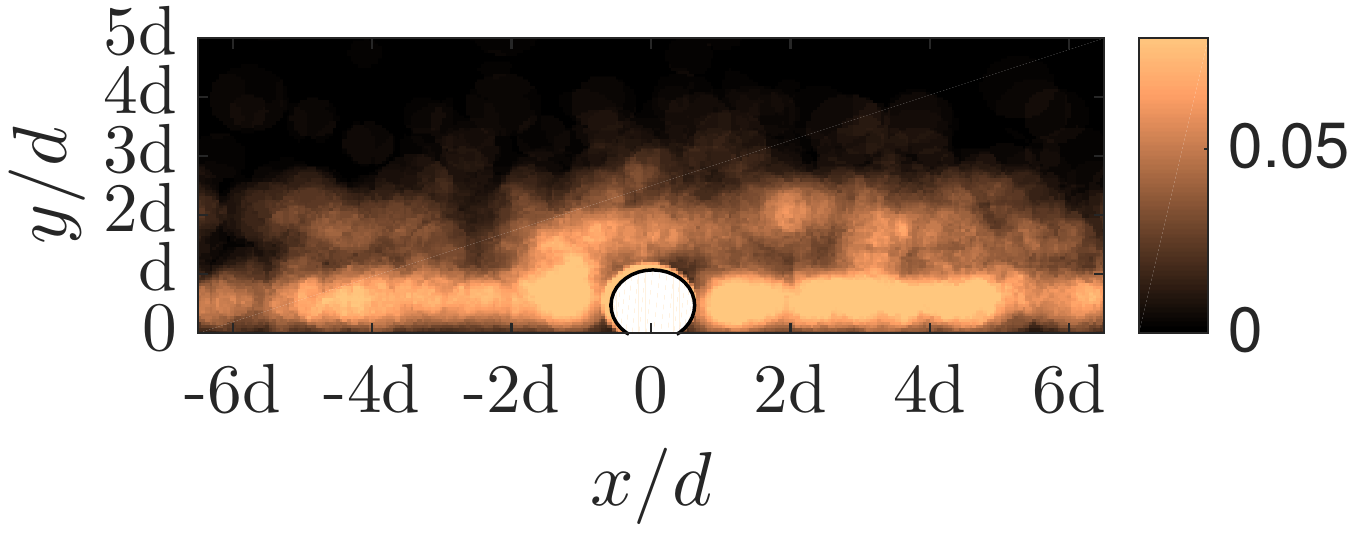}
  \includegraphics[width=0.24\linewidth]{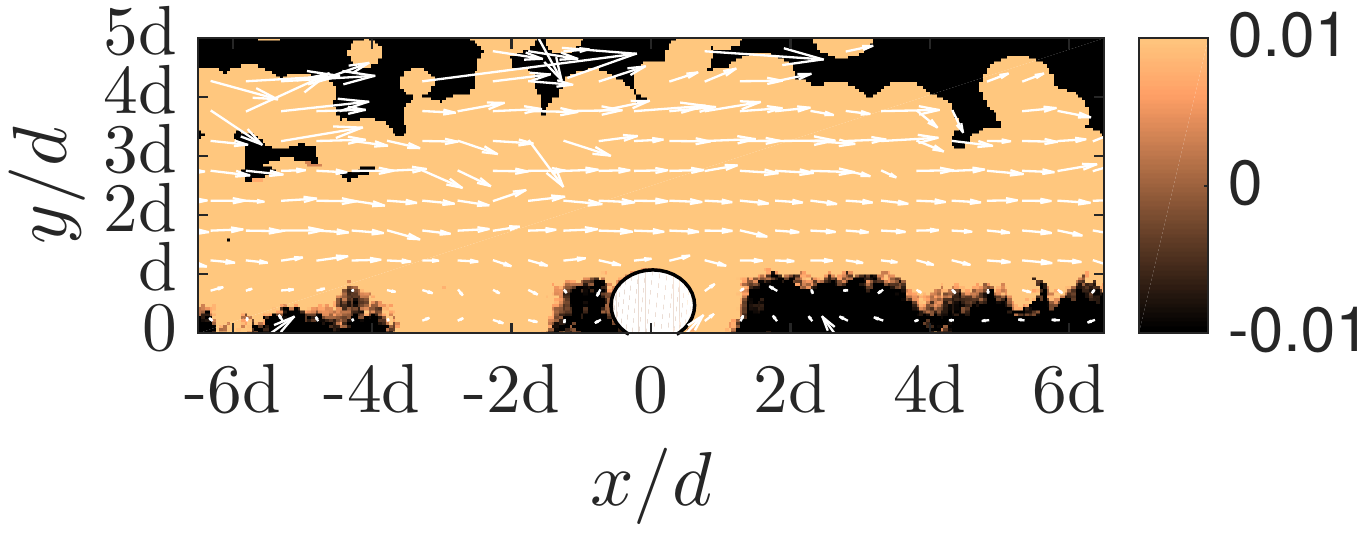}
  
    \includegraphics[width=0.24\linewidth]{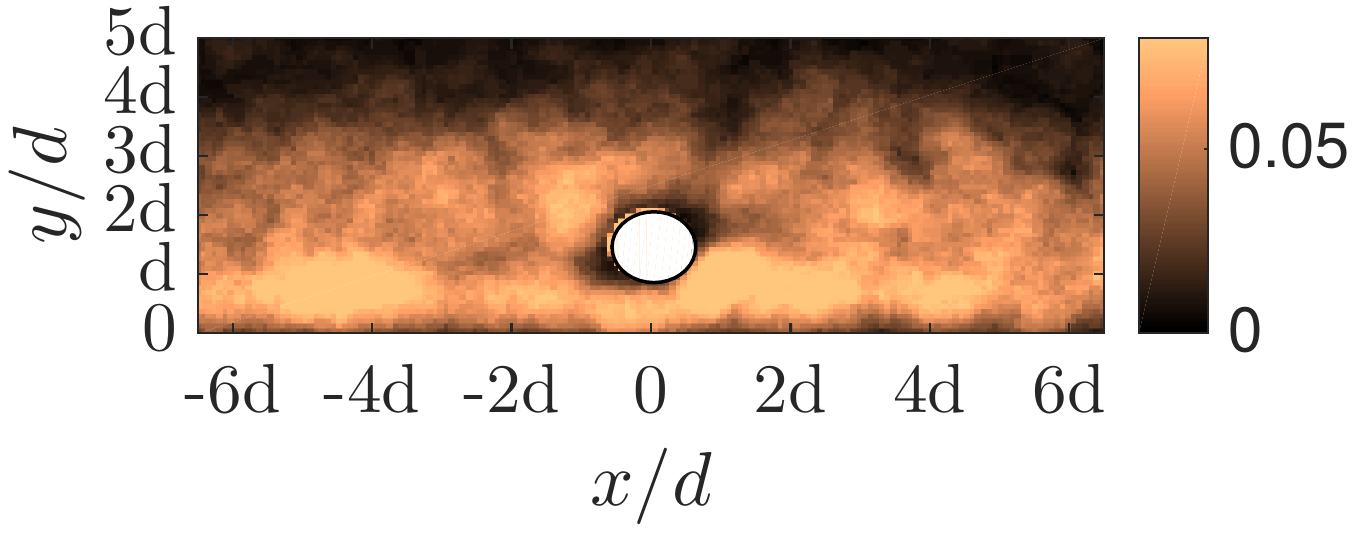}
    \includegraphics[width=0.24\linewidth]{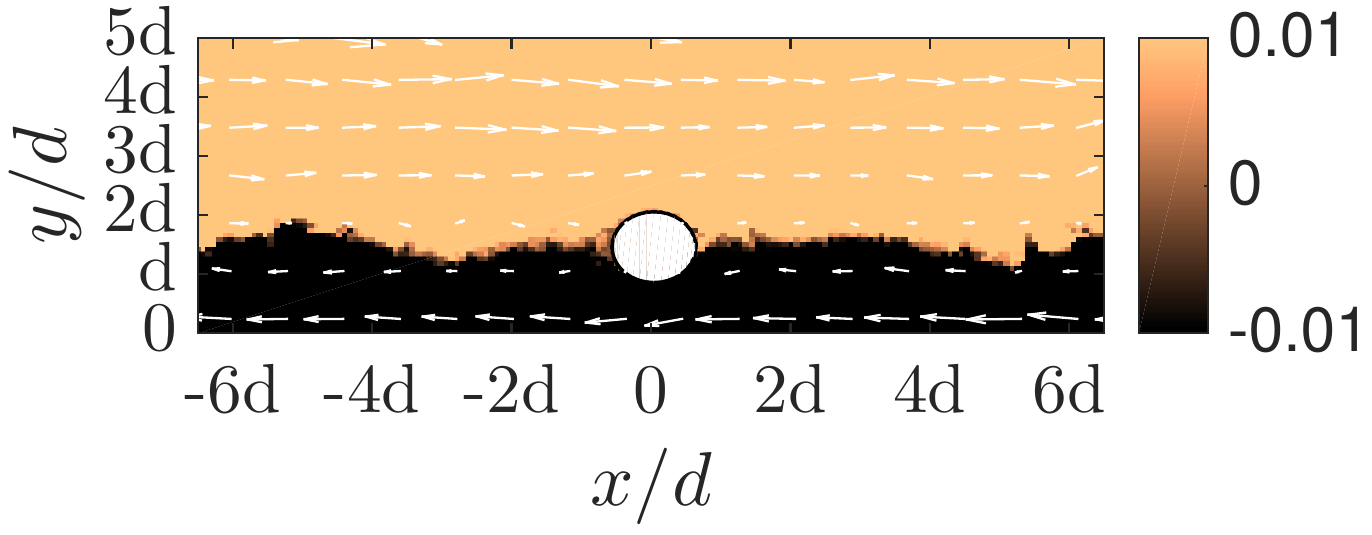}
  \includegraphics[width=0.24\linewidth]{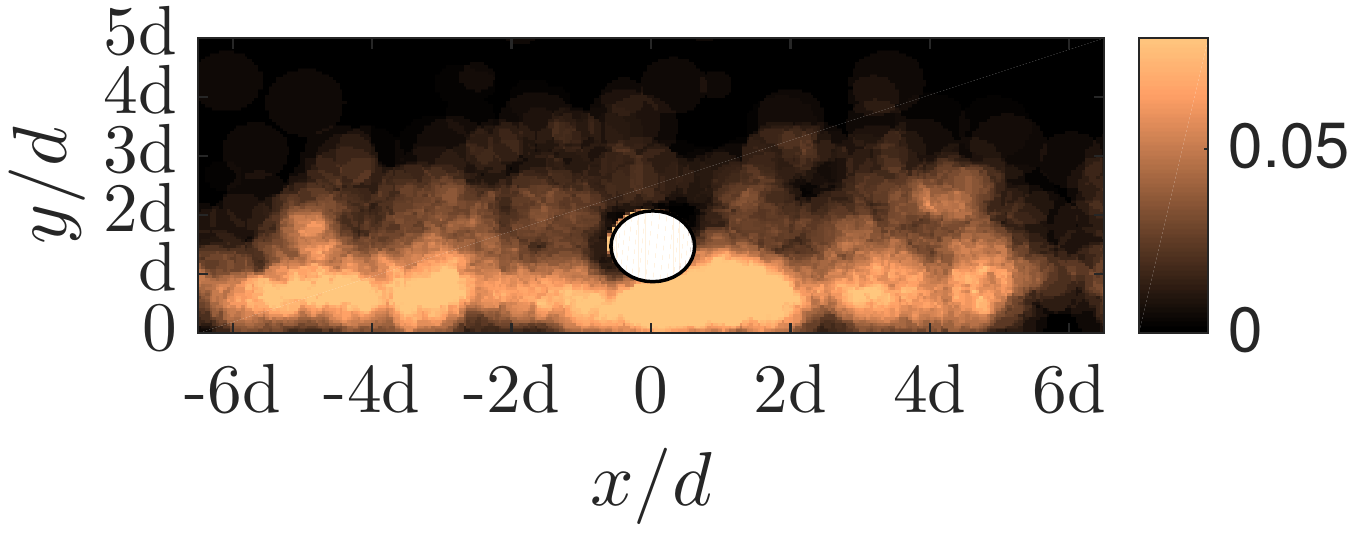}
  \includegraphics[width=0.24\linewidth]{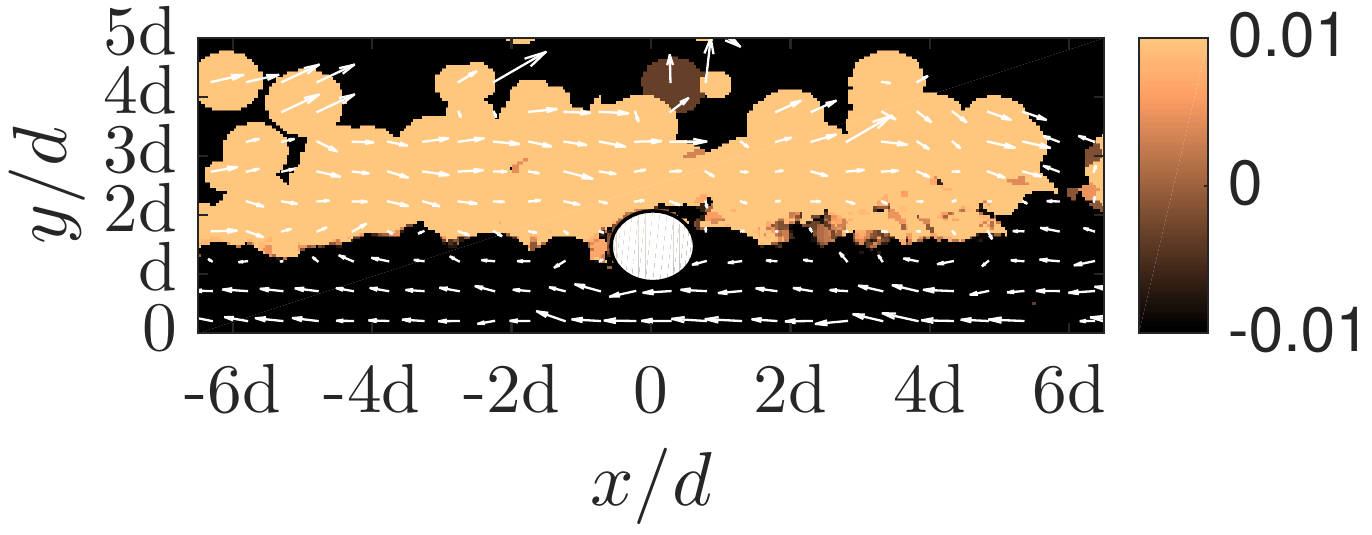}
  
      \includegraphics[width=0.24\linewidth]{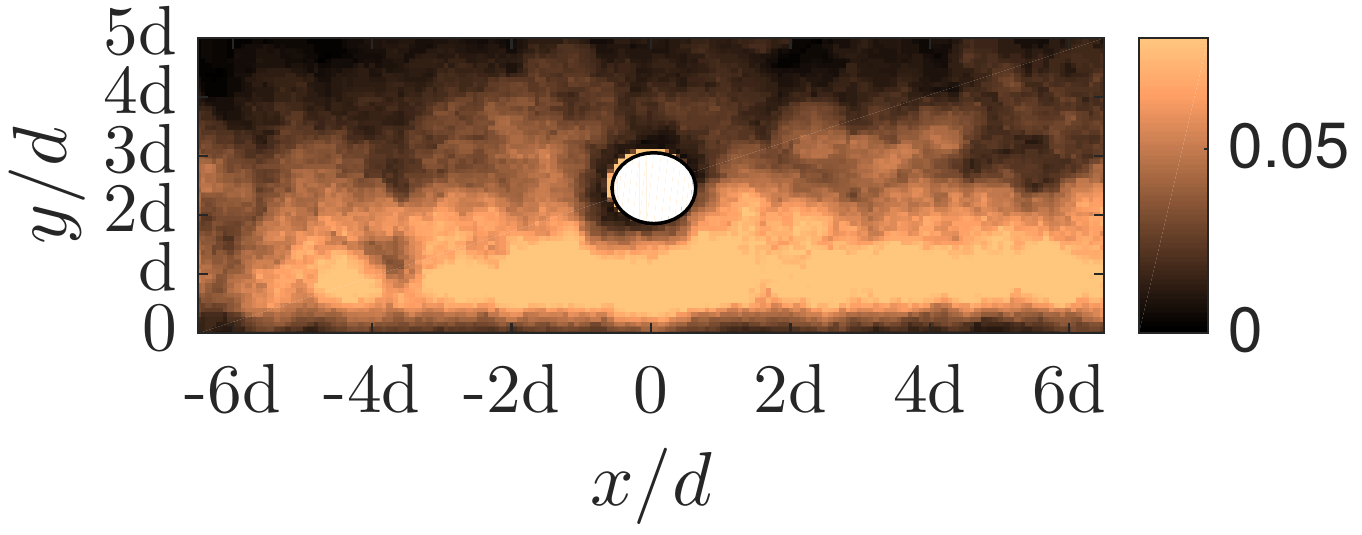}
    \includegraphics[width=0.24\linewidth]{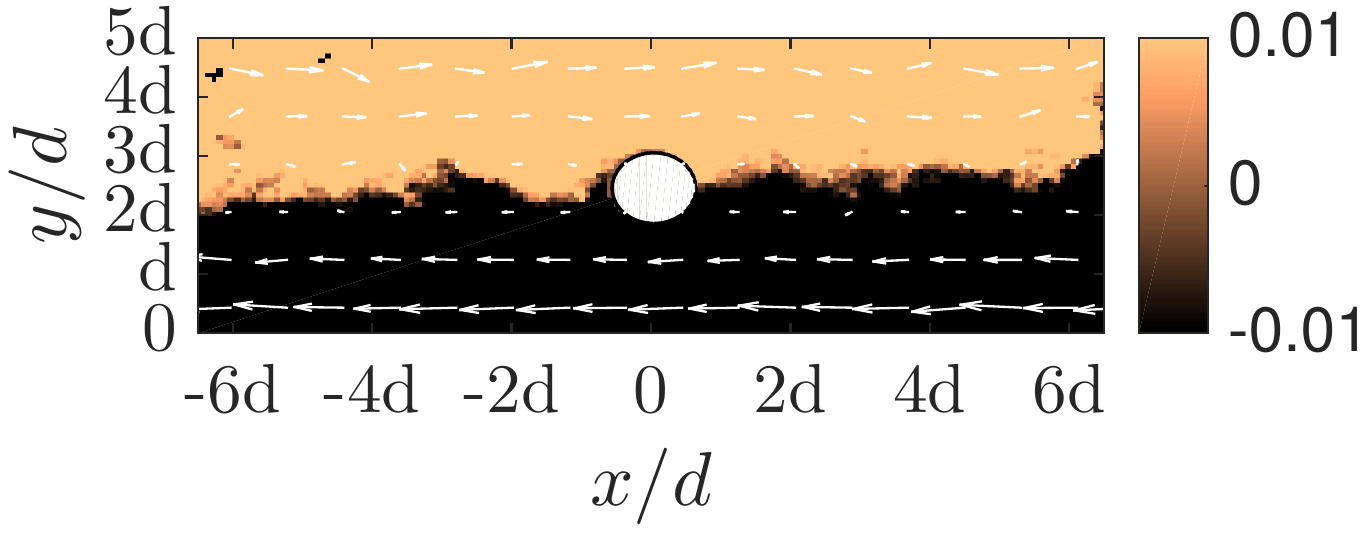}
  \includegraphics[width=0.24\linewidth]{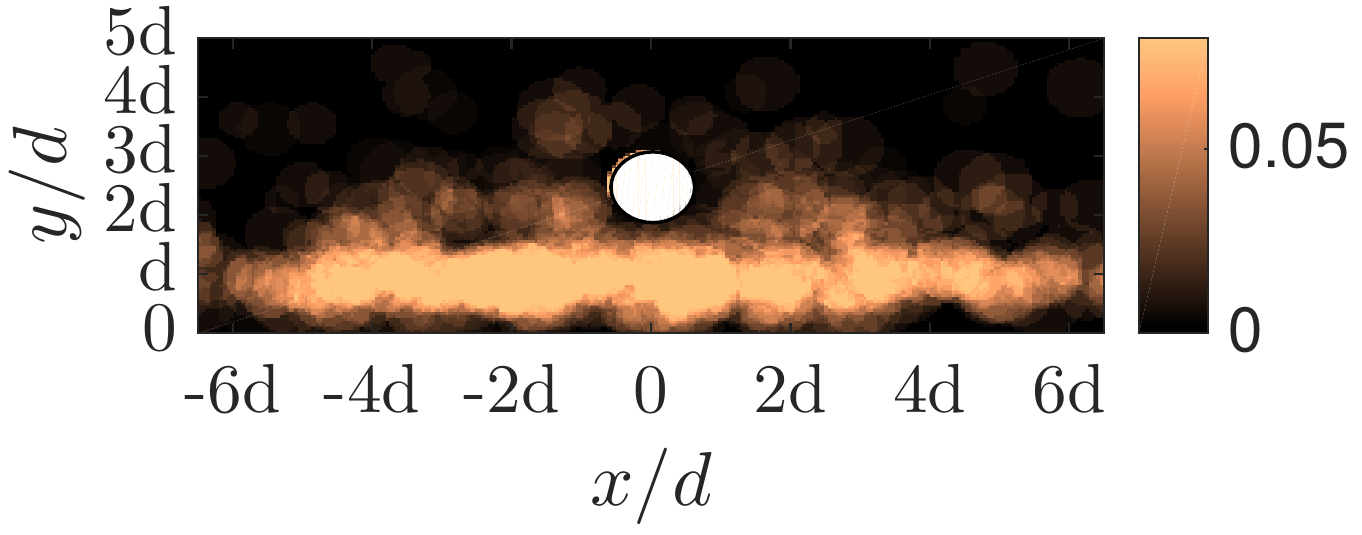}
  \includegraphics[width=0.24\linewidth]{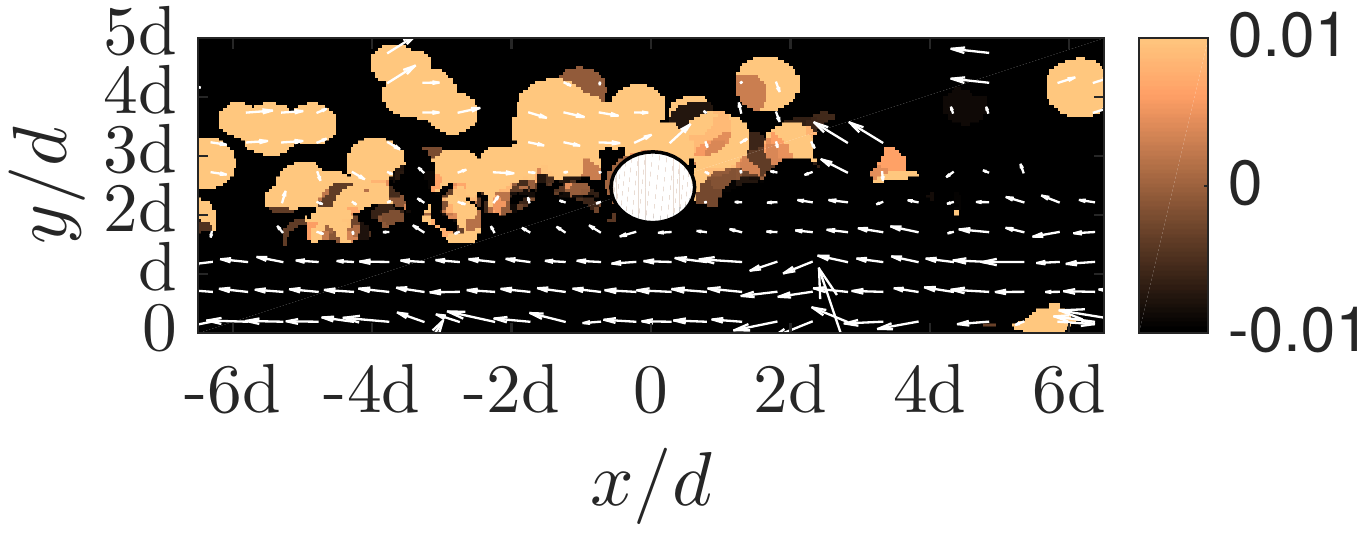}
\caption{Concentration and relative velocity distribution around a reference particle at $\Phi$ = 1\% for both SP and LP: Particle-centered distribution for SP \textit{(first-column)}, relative streamwise velocity for SP \textit{(second-column)},  particle distribution for LP \textit{(third-column)} and relative streamwise velocity for LP \textit{(fourth-column)}. The 3 rows of the above figure corresponds to 3 layers of particles: bottom near-wall layer \textit{(first-row)}, second layer above the wall \textit{(second-row)} and third layer above the wall \textit{(third-row)}. The white arrows in the second and fourth columns represent the net relative velocity $U_{Rel}\hat{x} + V_{Rel}\hat{y}$.}
\label{fig:Distributon function: 1p}  
\end{figure}

\begin{figure}
  \includegraphics[width=0.24\linewidth]{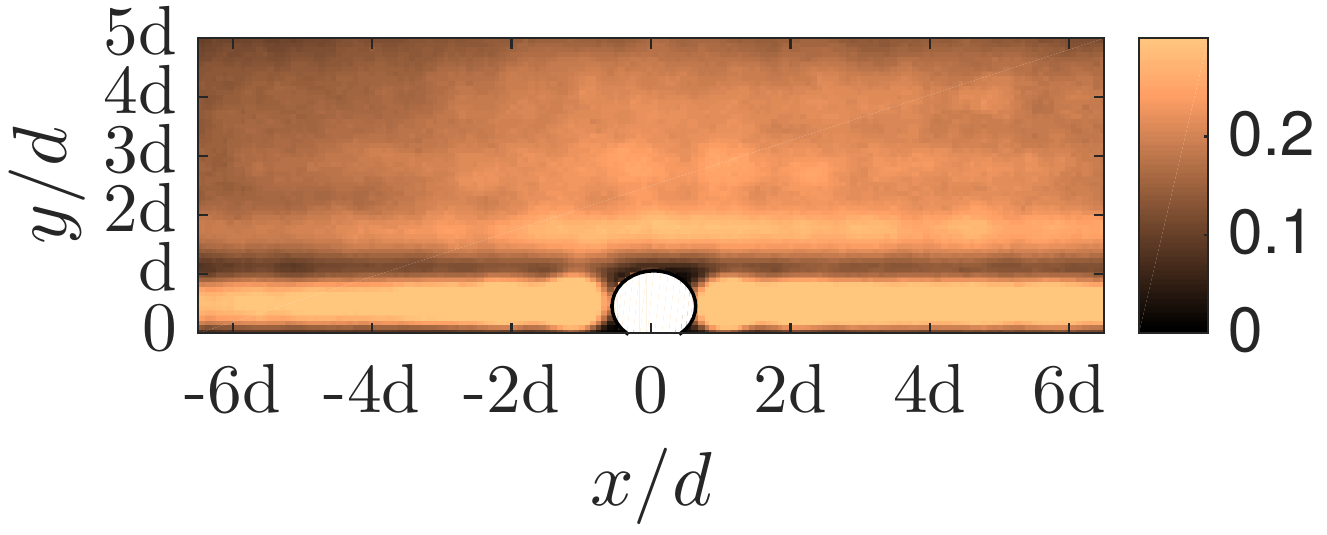}
    \includegraphics[width=0.24\linewidth]{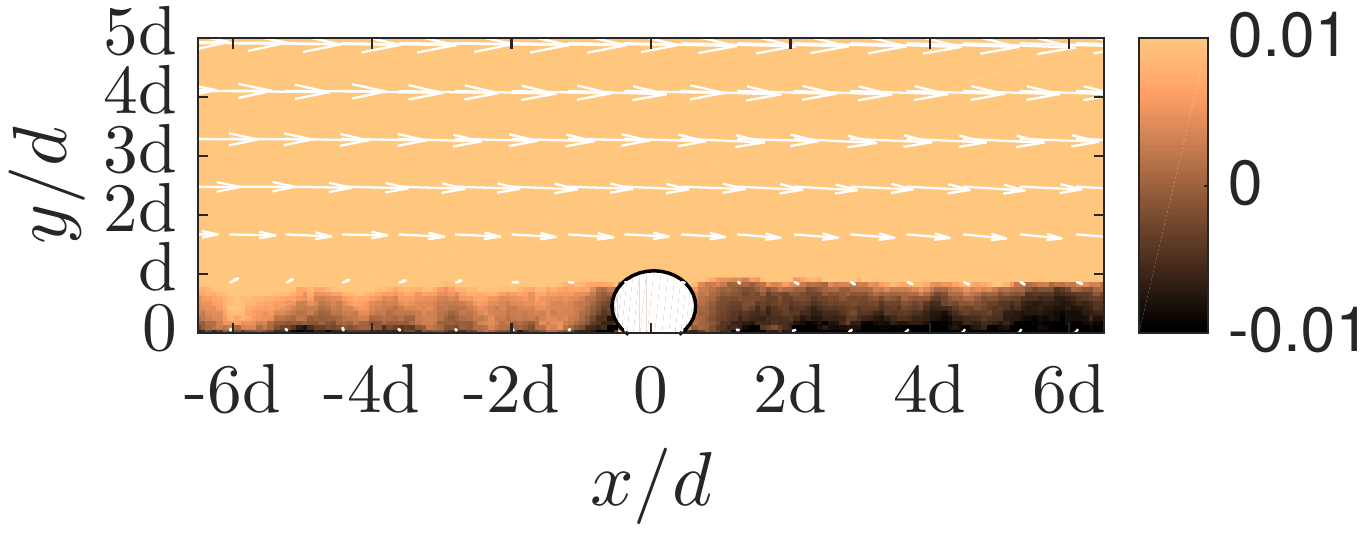}
  \includegraphics[width=0.24\linewidth]{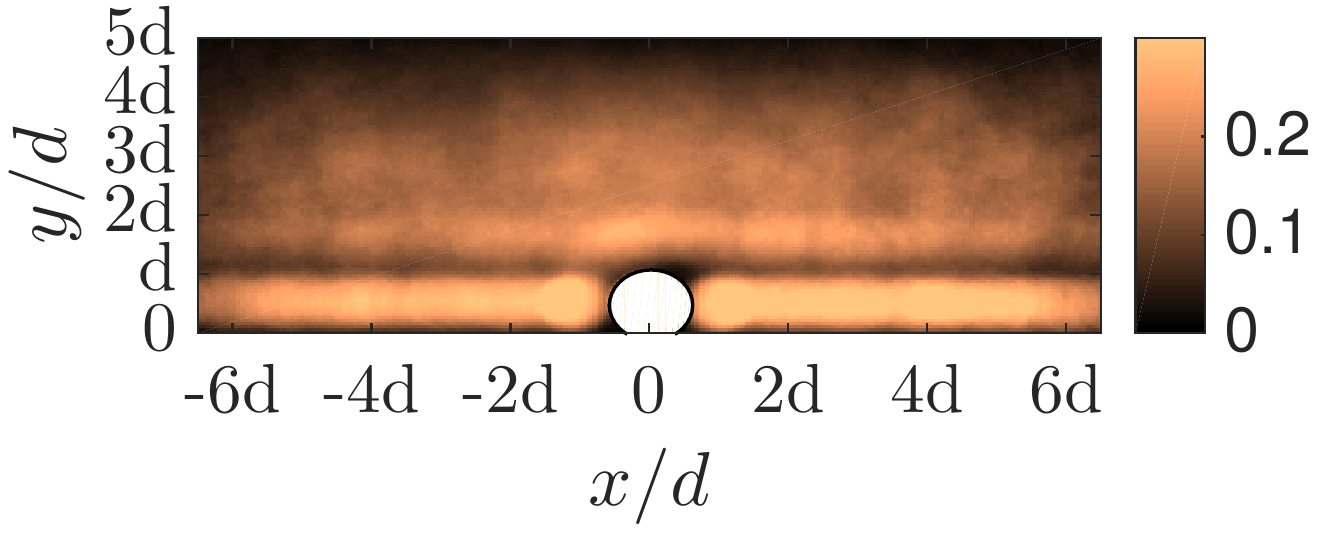}
  \includegraphics[width=0.24\linewidth]{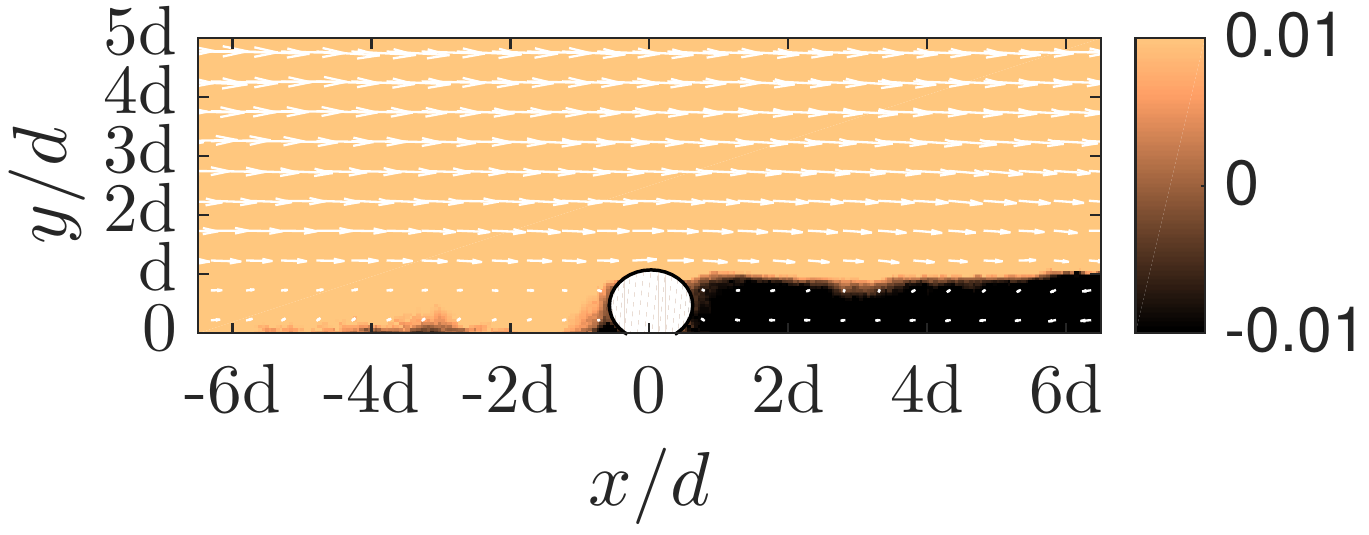}
  
    \includegraphics[width=0.24\linewidth]{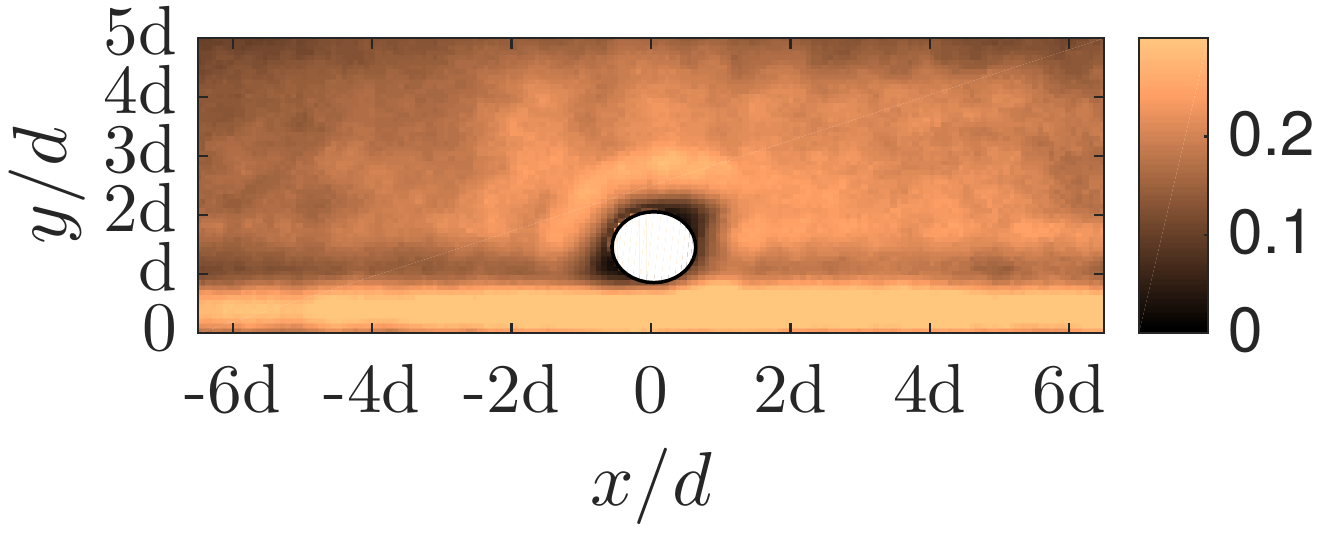}
    \includegraphics[width=0.24\linewidth]{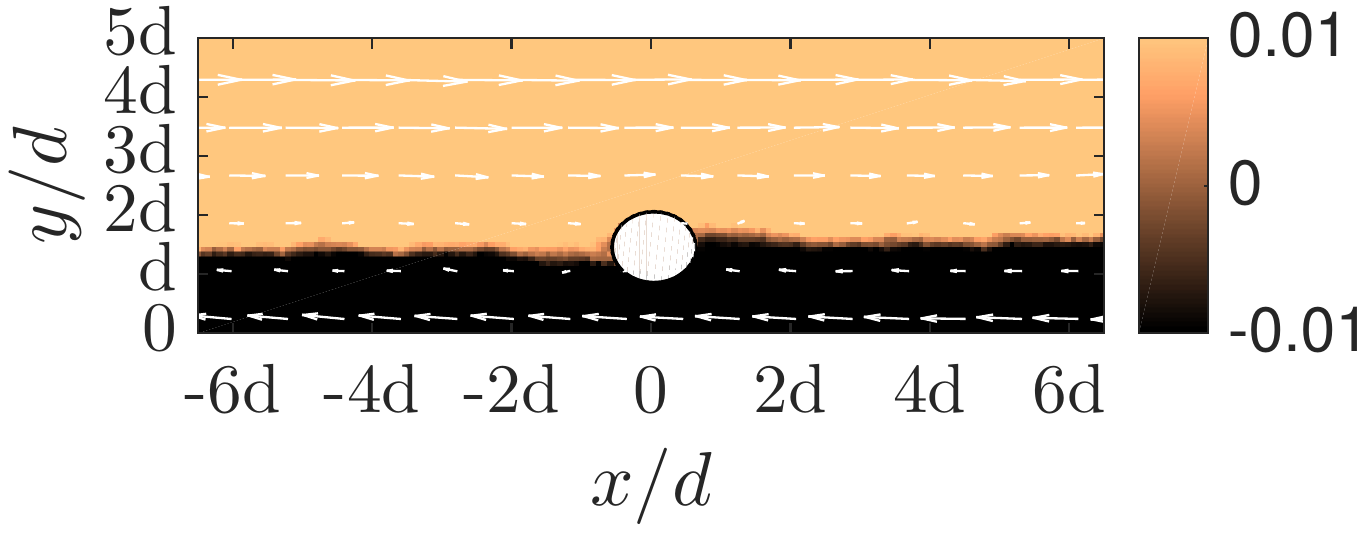}
  \includegraphics[width=0.24\linewidth]{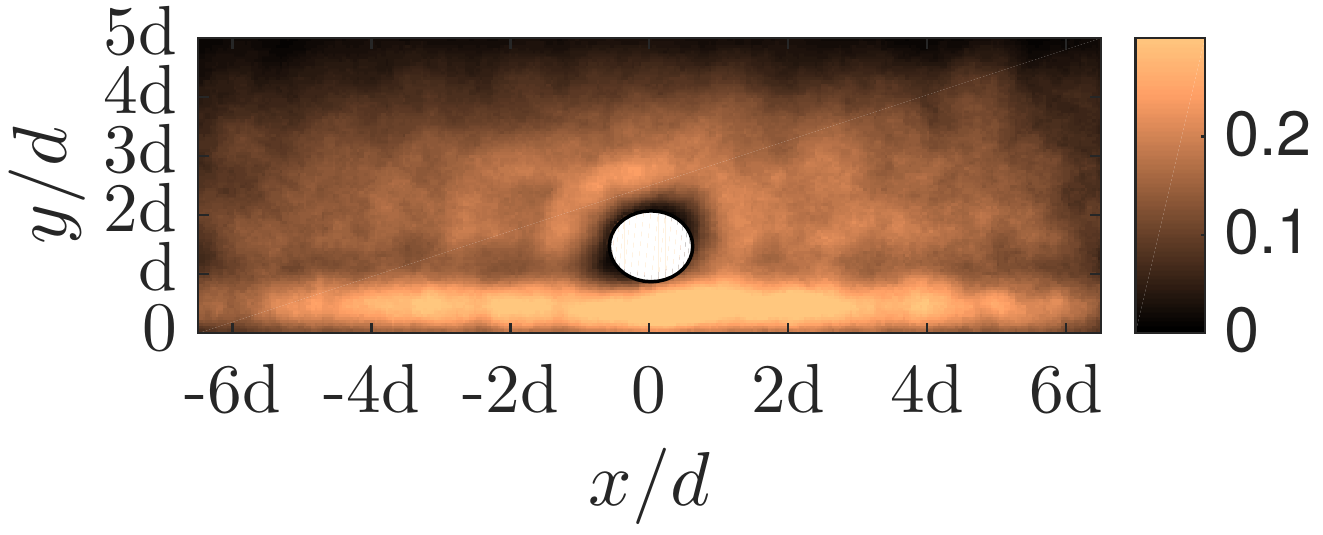}
  \includegraphics[width=0.24\linewidth]{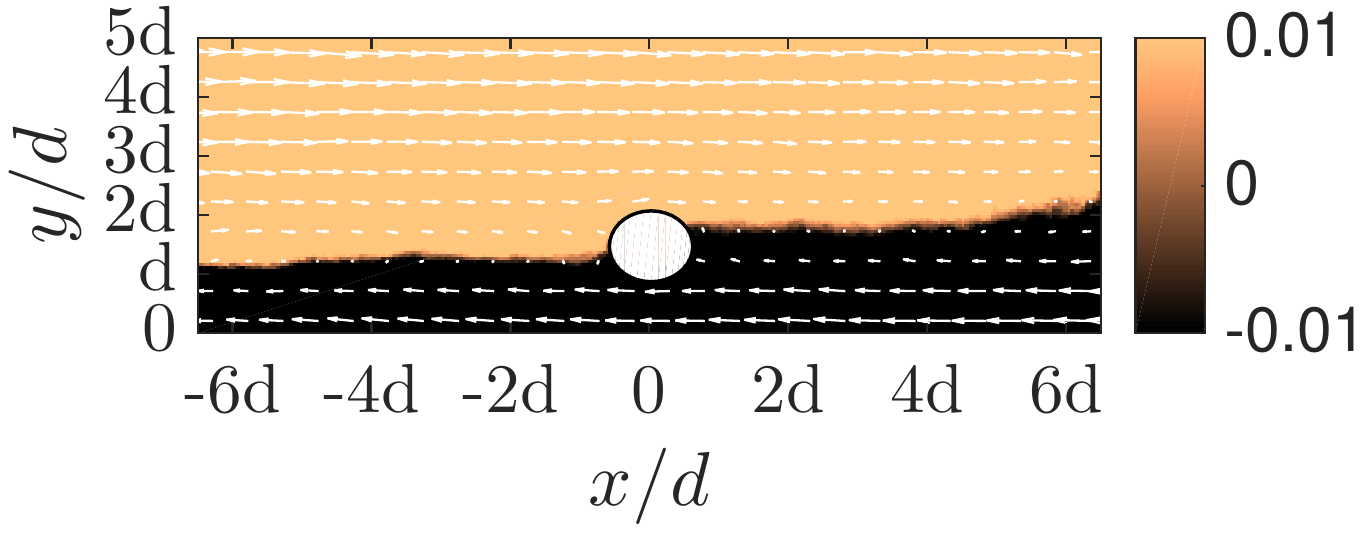}
  
      \includegraphics[width=0.24\linewidth]{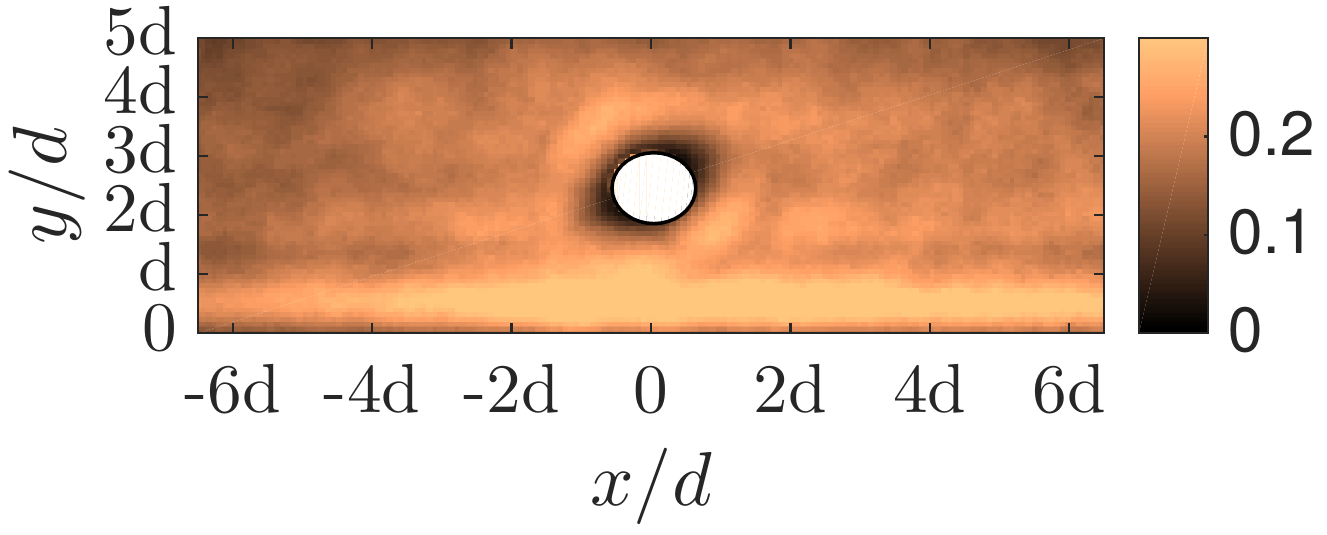}
    \includegraphics[width=0.24\linewidth]{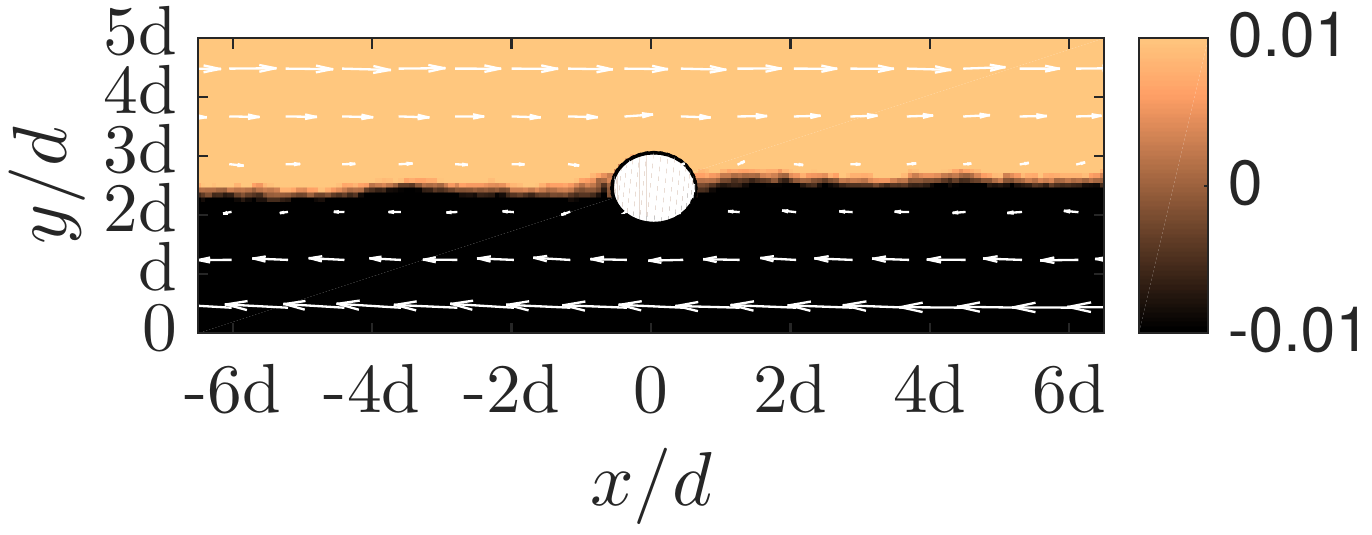}
  \includegraphics[width=0.24\linewidth]{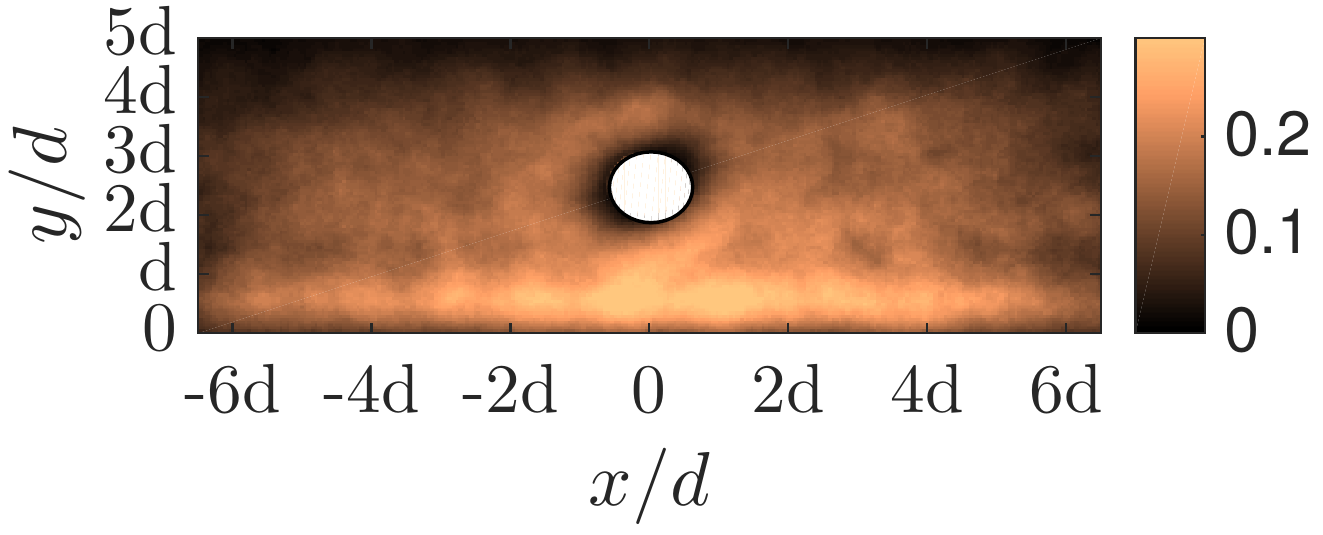}
  \includegraphics[width=0.24\linewidth]{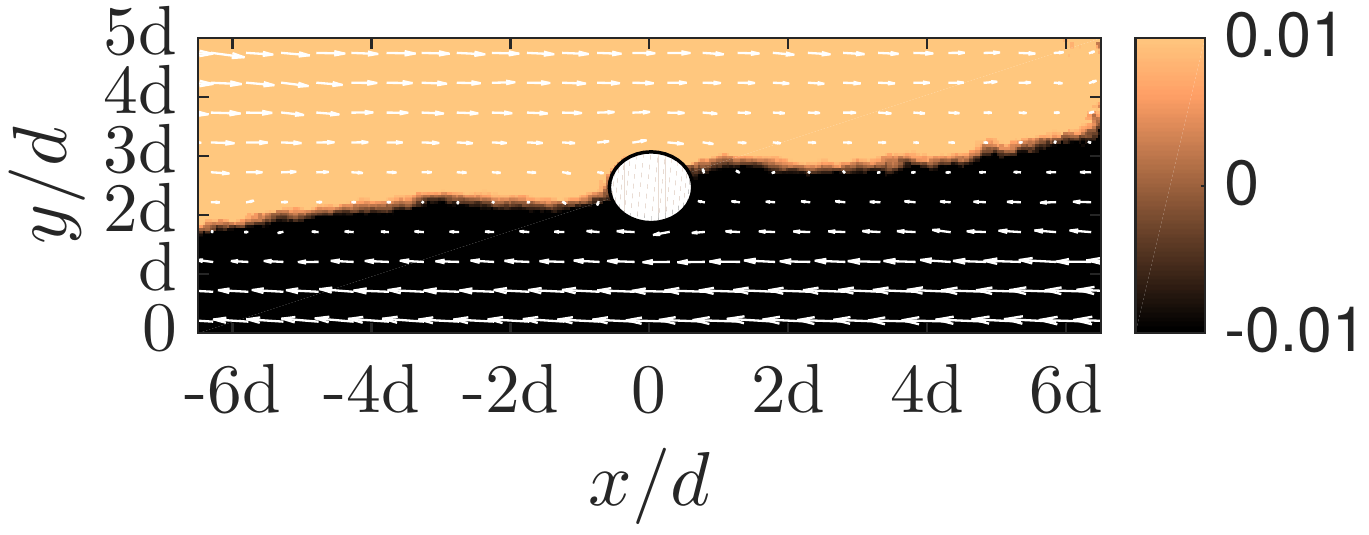}
\caption{Particle-centered distribution and relative velocity field at $\Phi$ = 5\% for both SP and LP: Details of the caption are same as in figure \ref{fig:Distributon function: 1p}.}
\label{fig:Distributon function: 5p}  
\end{figure}

The ability to track individual particles and their neighbors, within the area of observation, made it possible to extract particle-centered statistics. \sagar{This enables us to investigate  the dynamics of neighboring particles surrounding a reference particle.} Accordingly, the particle-centered distribution function and relative velocity are shown in figure \ref{fig:Distributon function: 1p} and \ref{fig:Distributon function: 5p} for the lowest and highest $\Phi$, respectively. The flow is from the left to the right. Since the flow is non-homogeneous in the wall-normal direction, these distribution functions are plotted for three elevations $y$, centered at $y_i = d_p/2, 3d_p/2$ and $5d_p/2$ i.e. the first, second and third particle layers respectively, corresponding to the three rows in figure \ref{fig:Distributon function: 1p} and \ref{fig:Distributon function: 5p}. In the $x-y$ measurement plane, the particle-centered distribution function for each elevation $y_i$ is defined as: 
\[
\sum_{n=1}^{N_i} M_i(x-x_p,y-y_p)/ N_i, \quad \forall\ y_p \in (y_i-d_p/2,y_i+d_p/2). 
\]
Here $M_i(x-x_p,y-y_p)$ is the mask matrix centered around a reference particle whose centroid is located at $(x_p,y_p)$. $M_i(x-x_p,y-y_p)$ has a value equal to 1 in the solid phase and 0 in the fluid phase. The averaging is performed over all $N_i$ reference particles whose centroid's wall-normal position $y_p$ lies inside the region $(y_i-d_p/2,y_i+d_p/2)$ i.e. in a band that is one particle diameter thick, and over all PIV images. The particle-centered relative velocity distribution is similarly calculated by subtracting the velocity of the reference particle from all the neighboring particles' velocity.

In figure \ref{fig:Distributon function: 1p} and \ref{fig:Distributon function: 5p}, the first two columns correspond to the distribution function and the streamwise relative velocity for the smaller particles, SP. Similarly, the last two columns correspond to the larger particles, LP. The distances are scaled by the corresponding particle diameter so, the last two columns are actually representing a larger physical area as compared to the first two columns due to the larger size of LP. 

From the distribution function for the bottom layer at  $\Phi$ = 1\% (first-row, first-column of figure \ref{fig:Distributon function: 1p}), it is evident that, on average, there is a region of high concentration immediately behind and in front of the reference particle. There is also a noticeably higher concentration in the top-left part. A small region with very low particle concentration appears to the bottom-left and top-right neighborhood of the reference particle. This particle-depleted region is perhaps more strongly visible in the second layer of SP (second-row, first-column of figure \ref{fig:Distributon function: 1p}). This preferential alignment of the particle-rich and particle-depleted region is due to particle inertia, quantified by the particle Reynolds number $\propto \dot{\gamma} d_p^2/\nu_f$, where $\dot{\gamma}$ is the mean shear rate. \citet{picano2013shear} showed that larger particle inertia can lead to larger excluded volume around the particle and thus, shear-induced thickening. The effects of particle inertia, now defined based on the relative slip-velocity ($U_f-U_p$), manifests in the relative velocity distribution (second and fourth column in figure \ref{fig:Distributon function: 1p}), especially for the bottom layer, where particles behind the reference particles are \textit{drafted} towards it and the velocity distribution is reminiscent of the wake behind a solid body. The larger noise in the statistics of LP is due to their smaller number at a given $\Phi$.

The above trends for $\Phi$ = 1\% are, in general, preserved at higher $\Phi$ = 5\% (see first and third columns of figure \ref{fig:Distributon function: 5p}). Here the particle-rich and particle-depleted regions can be clearly observed for the second and third layers (second and third-row of figure \ref{fig:Distributon function: 5p}). Also the wake behind the reference particle is prominently seen across all three particle layers (see second and fourth columns of figure \ref{fig:Distributon function: 5p}), especially for LP, which could be due to their larger inertia. For the bottom layer of particles, in contrast to the case for lower $\Phi$ = 1\%, the distribution function displays higher symmetry about the vertical line $x/d$ = 0, especially for SP. This is most likely because of the significantly higher particle concentration in the bottom layer leading to more frequent collisions and hence, a more uniform concentration.

\subsubsection{Particle cluster formation}
\label{Particle cluster formation}

\begin{figure}
  (a) \includegraphics[width=0.45\linewidth]{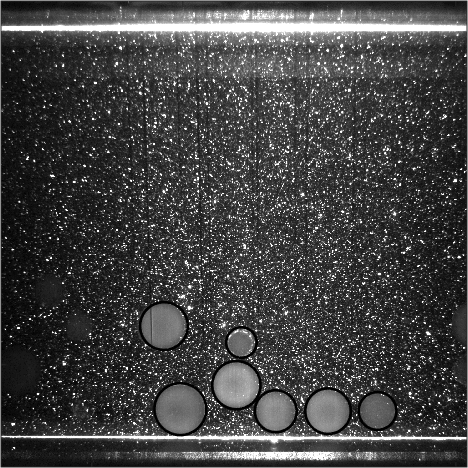}
  (b) \includegraphics[width=0.45\linewidth]{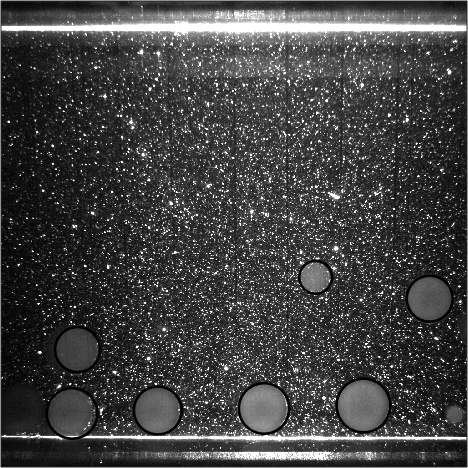}
\caption{(a) Particles occasionally moving in a train-like cluster and (b) particles moving with nearly constant relative spacing. These instances correspond to LP at $\Phi$ = 3\%.}
\label{fig:Cluster: 3p}  
\end{figure}

\begin{figure}
  (a) \includegraphics[width=0.45\linewidth]{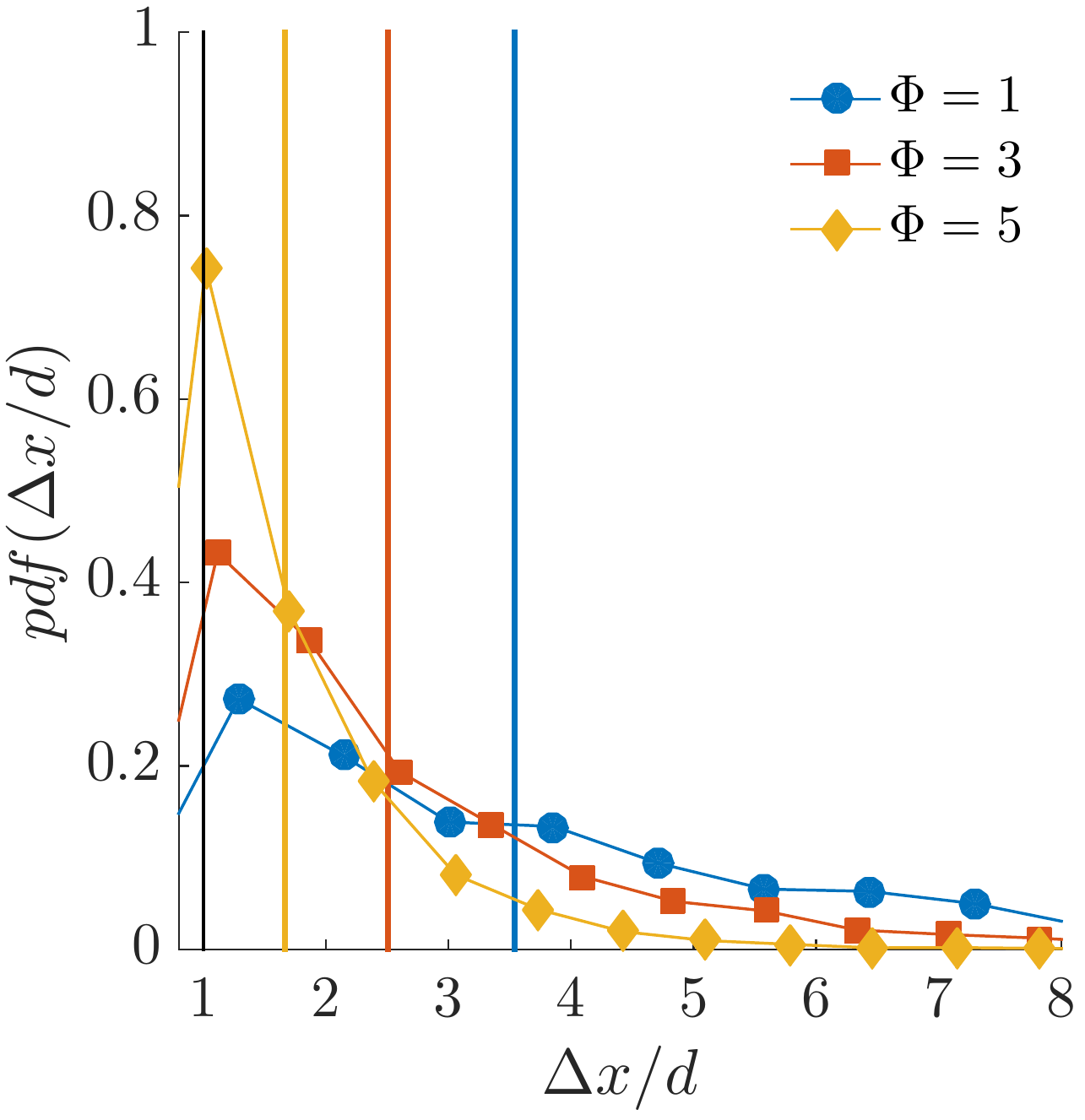}
   (b) \includegraphics[width=0.45\linewidth]{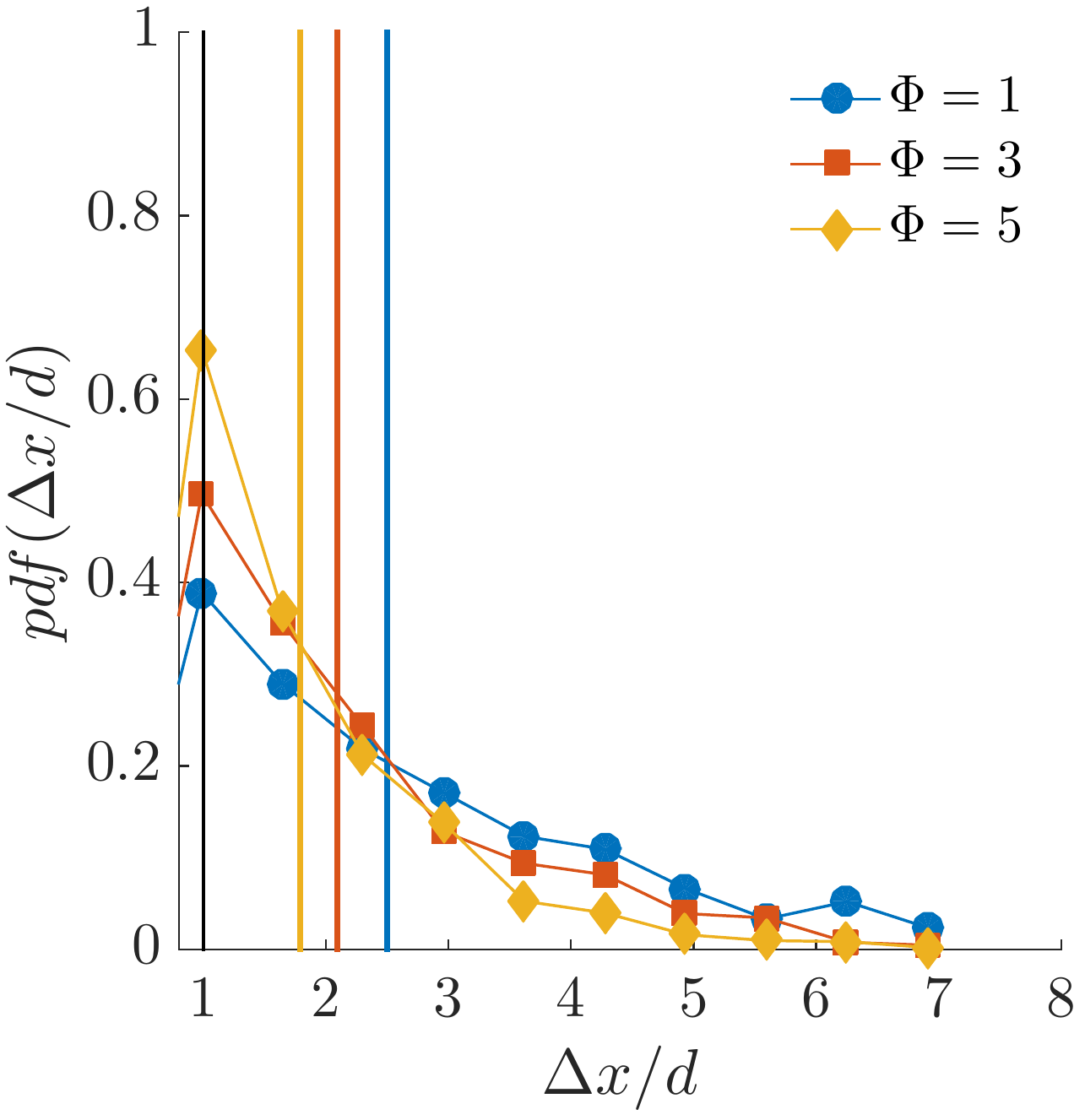}
\caption{Probability distribution function of the center-to-center separation distance between neighboring particles in the bottom near-wall layer for: (a) smaller particles SP (b) and larger particles LP. The vertical lines (same color corresponds to the same volume fraction) corresponds to the average spacing denoting a scenario where the particles would move with a constant relative spacing. The black vertical line corresponds to a separation distance of one particle diameter corresponding to the scenario where all neighboring particles are in contact and move as a continuous train.}
\label{fig:Relative distance pdf}  
\end{figure}

The particle-centered distribution function shown in figures \ref{fig:Distributon function: 1p} and \ref{fig:Distributon function: 5p} reflects the average spatial distribution of neighboring particles. Moreover, instantaneous visualizations of the particle distribution indicate the formation of occasional train-like clusters as shown in figure \ref{fig:Cluster: 3p} (a), along with instances where the spacing between neighboring particles is nearly uniform (figure \ref{fig:Cluster: 3p} (b)), and instances when the spacing between neighboring particles is larger than average (not shown). The probability distribution function $pdf$ of the center-to-center distance between neighboring particles provides evidence about the existence of such clusters. If the distance between neighboring particles remains nearly constant then the $pdf$ should display a peak around the average spacing. This average spacing must reduce with particle concentration, and approach a minimum value of one particle diameter when all particles move in contact as a packed bed at the maximum packing fraction. 

Figure \ref{fig:Relative distance pdf} (a) and (b) show the $pdf$ of the center-to-center distance between neighboring particles for the bottom layer of SP and LP. The $pdf$ has been normalized so that the area under the curve $\sum_{\Delta x/d} pdf * \Delta(\Delta x/d)$ equals 1. The average spacing, calculated as the mean value of the $pdf$: 
\[
\sum_{\Delta x/d} pdf * \Delta x/d * \Delta(\Delta x/d), \]
reduces with increasing $\Phi$ and is represented with a vertical line in figures \ref{fig:Relative distance pdf} (a) and (b). The peak of the $pdf$, or the modal value, is shifted towards distances smaller than the average spacing in the bottom near-wall layer, indicating a tendency for two or more particles to travel with smaller than average relative spacing between them i.e. train-like clusters exist. This cluster formation could be due to (i) the drafting motion behind the reference particle as seen in figure \ref{fig:Distributon function: 1p} and \ref{fig:Distributon function: 5p} where particles following the reference particle are attracted in its wake (see also \citet{matas2004trains}) and (ii) the tendency of particles to sample low speed streaks (as seen in \citet{kidanemariam2013direct,shao2012fully}). One can also see separation distances smaller than one particle diameter which is an effect caused by the partial illumination of the spherical particles by the thin laser light-sheet, discussed before, making the particles appear as smaller spheres.

\section{Conclusion}
\label{Conclusion}

We have considered the turbulent flow of a particle suspension in a square duct and have presented fluid-particle velocity and concentration statistics using RIM-PIV experiments for two different particle sizes (14.5 and 9 in terms of ratio between the duct height and particle diameter) and volume fractions $\Phi$ from 1 to 5\%. The pressure drop increases with $\Phi$ but, within the error-bars, it is insensitive to the particle size for the range of $\Phi$ and Reynolds numbers $Re_{2H}$ considered in this study. 
However, the fluid turbulent velocity statistics are considerably different for the two particle sizes; the effects with respect to the single phase flow being more pronounced for smaller particles owing to their larger number. We therefore suggest that two competing mechanisms are active in these flows, where the total pressure drop is given by the viscous stress at the wall and by the friction caused by the particles sliding on the wall. Thus, on one side, smaller particles are characterised by a lower wall viscous stress as they form a less permeable bed; on the other side, a more packed bed (smaller particles) is associated with higher drag and more contact points with the bottom wall. 
In particular, the mean velocity profiles look fuller for large particles and fluctuations are more intense for smaller particles.
These two effects appear to balance yielding a similar global pressure drop.

Moreoever, the secondary flow intensity is found to increase in the presence of particles. Particle centered distributions have identified regions with high and low concentration 
in the vicinity of the reference particle. 
This excluded volume manifests due to the finite inertia at the particle scale. Also, relative velocity distribution have shown the 
existence of wake-like regions behind the reference particle in the near-wall region. Trailing particles attracted in this wake zone along with particle segregation in low speed 
regions might be the reason for the observed particle clustering. 

Almost excellent quantitative agreement is seen between the turbulent statistics obtained from fully resolved DNS and experiments for the case with small particles SP at $\Phi$ = 1\% thus, corroborating both methods in such a complex flow setting. \sagar{Good agreement between simulations and experiments despite not exactly matching values for collision parameter (e.g. coefficient of friction and restitution) indicate that, in turbulent flows, the results are not significantly affected by these variables}. We hope that these high-resolution velocity and concentration measurements may serve as experimental dataset to provide new insights into the modification of turbulence induced by the presence of a mobile sediment bed, in the same spirit as \citet{revil2016turbulence} and \citet{ni2018stresses}.

\begin{acknowledgements}
This work was supported by the European Research Council Grant No.\ ERC-2013-CoG-616186, TRITOS, from the Swedish Research Council (VR), through the Outstanding Young Researcher Award to LB.
\end{acknowledgements}

\bibliography{prf.bib}

\end{document}